\documentclass[fleqn,usenatbib]{mnras}

\usepackage{newtxtext,newtxmath}

\usepackage[T1]{fontenc}

\DeclareRobustCommand{\VAN}[3]{#2}
\let\VANthebibliography\thebibliography
\def\thebibliography{\DeclareRobustCommand{\VAN}[3]{##3}\VANthebibliography}

\usepackage{graphicx}	
\usepackage{amsmath}	
\usepackage{pdflscape}
\usepackage[dvipsnames]{xcolor}
\usepackage{array}
\usepackage{afterpage}

\newcolumntype{C}[1]{>{\centering\arraybackslash}p{#1}}

\DeclareRobustCommand{\dataurl}{\url{https://osf.io/q8x76/?view_only=3dc4faf80aae40e186ec358ad261d490}}

\DeclareRobustCommand{\cgunit}{\mathrm{mag}}
\DeclareRobustCommand{\pz}{\emph{pz}}
\DeclareRobustCommand{\sz}{\emph{sz}}
\DeclareRobustCommand{\angstrom}{\mbox{\normalfont\AA}}

\DeclareRobustCommand{\reply}[1]{\textcolor{purple}{#1}}

\DeclareRobustCommand{\firefly}{\textsc{firefly}}

\title[Colour gradients of low-redshift galaxies]{Colour gradients of low-redshift galaxies in the DESI Legacy Imaging Survey}

\author[Liao et al.]{
Li-Wen Liao,$^{1,2}$\thanks{E-mail: liwen@gapp.nthu.edu.tw} 
Andrew P. Cooper,$^{1,2,3}$
\\
$^{1}$Institute of Astronomy and Department of Physics, National Tsing Hua University, Hsinchu 30013, Taiwan\\
$^{2}$Center for Informatics and Computation in Astronomy, National Tsing Hua University, Hsinchu 30013, Taiwan\\
$^{3}$Physics Division, National Center for Theoretical Sciences, Taipei 10617, Taiwan\\
}

\date{Accepted XXX. Received YYY; in original form ZZZ}

\pubyear{2021}

\begin{document}
\label{firstpage}
\pagerange{\pageref{firstpage}--\pageref{lastpage}}
\maketitle

\begin{abstract}
Radial colour gradients within galaxies arise from gradients of stellar age, metallicity and dust reddening. Large samples of colour gradients from wide-area imaging surveys can complement smaller integral-field spectroscopy datasets and can be used to constrain galaxy formation models. Here we measure colour gradients for low-redshift galaxies ($z<0.1$) using photometry from the DESI Legacy Imaging Survey DR9. Our sample comprises $\sim93,000$ galaxies with spectroscopic redshifts and $\sim574,000$ galaxies with photometric redshifts. We focus on gradients across a radial range 0.5$R_{\mathrm{eff}}$ to $R_{\mathrm{eff}}$, which corresponds to the inner disk of typical late type systems at low redshift. This region has been the focus of previous statistical  studies of colour gradients and has recently been explored by spectroscopic surveys such as MaNGA.
We find the colour gradients of most galaxies in our sample are negative (redder towards the centre), consistent with the literature.
We investigate empirical relationships between colour gradient, average $g-r$ and $r-z$ colour, $M_r$, $M_\star$, and sSFR.
Trends of gradient strength with $M_r$ ($M_\star$) show an inflection around $M_r\sim-21$  ($\log_{10} \, M_\star/\mathrm{M_\odot}\sim10.5$).
Below this mass, colour gradients become steeper with increasing $M_\star$, whereas colour gradients in more massive galaxies become shallower. We find that positive gradients (bluer stars at smaller radii) are typical for galaxies of $M_{\star}\sim10^{8}\,\mathrm{M_\odot}$. We compare our results to age and metallicity gradients in two datasets derived from fits of different stellar population libraries to MaNGA spectra, but find no clear consensus explanation for the trends we observe. Both MaNGA datasets seem to imply a significant contribution from dust reddening, in particular, to explain the flatness of colour gradients along the red sequence.
\end{abstract}




\section{Introduction}
\label{sec:intro} 

Photometry and spectroscopy for essentially magnitude-limited samples of millions of low-redshift galaxies have enabled the study of large-scale trends in the statistics of fundamental galaxy observables like magnitude, colour, size and morphology \citep[e.g.][]{Blanton2009PhysicalGalaxies}. The ease with which those trends can be compared to predictions from numerical simulations has led to rapid progress in understanding galaxy formation in the $\Lambda$CDM cosmogony and put strong constraints on theoretical models. The precision and depth offered by future all-sky imaging surveys such as the Rubin Observatory's LSST \citep{Ivezic2019LSST:Products} will strengthen those constraints and provide new ways to distinguish between theories. In this paper we use the DESI Legacy Imaging Survey \citep{Dey2019OverviewSurveys} to explore one simple but potentially powerful observable that is readily accessible to future surveys but not yet routinely compared to models: galaxy colour gradients.

Radial gradients in galaxy colour arise from underlying age and metallicity gradients \citep{Larson1974DynamicalGalaxies,Tinsley1978ChemicalDisks}. The distributions of different stellar populations within galaxies are very likely to be influenced by the same factors that determine their masses and average colours, such as the assembly history of their dark matter halos. Most simply, late-type galaxies, which form through a quiescent process of gradual gas condensation\reply{,} moderated by supernova-driven outflows, are expected to grow `inside out'. This mode of growth should, in general, produce stellar age gradients that give rise to negative colour gradients (redder cores and bluer outskirts). However, relationships between the strength of these gradients and other readily observable properties may not be trivial, and may encode useful information about the galaxy formation process. A better understanding of these relationships important in the context of modern galaxy formation models. The current generation of cosmological simulations can reproduce many low-redshift distributions of galaxy properties once the adjustable parameters of their sub-resolution components are calibrated (for example) to the galaxy stellar mass function and size distribution \citep[e.g.][]{Vogelsberger2014IntroducingUniverse,Crain2015EAGLE,Schaye2015EAGLE,Lacey2016AFormation,Nelson2018FirstBimodality,Pillepich2018FirstGalaxies}. The outstanding differences between models now concern how the balance between mass inflow, star formation and outflow (the baryon cycle) evolves over time and is moderated by the growth of dark matter halos. These can be constrained by observations at higher redshift, but also by the distribution of different stellar populations within present-day galaxies \citep[e.g.][]{Trayford2019ResolvedRelationsEagle}.

In this paper, we compute colour gradients for galaxies in the DESI Legacy Imaging Survey (Legacy Survey, LS), which provides deep ($r=23.4$) imaging over $\sim$14,000 deg$^2$. We construct a new homogeneous catalogue of optical galaxy colour gradient measurements at low redshift, over as wide a range of stellar mass and colour as possible, in order to explore large-scale trends in these gradients with other galaxy properties (such as averaged colour, $M_r$, $M_\star$, and sSFR) and to facilitate statistical comparisons with galaxy formation models. Our sample selection and our definition of the colour gradient are intended to be easy to reproduce and to compare with previous surveys and simulations. Our sample therefore provides a baseline for more extensive studies of galaxy colour profiles with all-sky imaging surveys in future. Moreover, by using photometric redshifts, we can explore the behaviour of colour gradient trends at significantly lower stellar masses. The DESI Bright Galaxy Survey is currently conducting a deep spectroscopic survey drawn from the same sample of galaxies as this work \citep{Ruiz-Macias2020, Hahn2022BGS}.

Results from previous statistical studies of colour gradients in the low redshift galaxy population, including
\citet[][]{Peletier:1990,DeJong1996Near-infraredGalaxies,LaBarbera:2005,Tortora:2010} and \citet{Gonzalez:2011}, are broadly consistent with the current consensus understanding of galaxy formation (we discuss this in more detail in Section~\ref{sec:discussion}). However, the effects of age, metallicity and dust on the broadband colours of stellar populations are notoriously degenerate, which greatly limits more detailed inferences about galaxy formation based on resolved photometric data alone \citep[although see][]{Abdurrouf2022Pixedfit}. In contrast, fits of population synthesis templates to spatially resolved spectroscopy can provide much stronger constraints on the underlying stellar populations, and hence allow trends in the distribution of age, metallicity and dust content to be measured more directly. Several recent surveys have pursed this approach by carrying out integral field (IFU) spectroscopy for large representative samples of low redshift galaxies, including CALIFA \citep{Sanchez2012Califa}, MaNGA \citep{Bundy2015Manga} and SAMI \citep{Croom2021Sami}.

Our work is motivated in part by the possibility that, at least in the near-term, large photometric studies can complement these recent advances in resolved spectroscopy, even in the brighter central regions of galaxies. Photometry can provide much larger samples that can be used to extrapolate inferences based on smaller spectroscopic datasets. Resolved colour distributions are a straightforward and robust output from standard imaging survey pipelines, such as the LS catalogue we use here. Future imaging surveys, such as the Rubin Observatory's Legacy Survey of Space and Time \citep[LSST;][]{Ivezic2019LSST:Products} should significantly extend both the redshift and mass range over which they can be measured. If the underlying relationships between colour and stellar population gradients can be tightly constrained by detailed IFU studies of $\lesssim10,000$ galaxies, photometric colour measurements for millions of galaxies could be used to reduce statistical uncertainties on large-scale trends, to explore trends with other properties (such as AGN activity, environment or redshift) and to reveal rare outlier populations. Photometry can be carried out at much higher spatial resolution and extended beyond the range of surface brightness probed by the IFU surveys to study fainter structure. All these extrapolations, however, require a solid grounding in the relationship between stellar populations and photometric observables at higher signal-to-noise.

Colour measurements also provide an alternative way to constrain simulations. Inferring ages and metallicities from spectra is a complex process, which involves assumptions, for example, regarding the choice of stellar population templates, dust extinction and star formation histories. This is reflected in the current variety of different techniques and template sets used for such inferences. Strictly, values inferred from fits to spectra should not be compared directly to simulated ages and metallicities, but to the results of fits to mock spectra computed from the simulations and convolved with the same observational uncertainties as the real data \citep[e.g.][]{Nanni2022EAGLEMockMaNGA}.

Of course, it is also possible to synthesise broadband colours from simulated ages and metallicities and compare them to (much simpler) photometric observations (this is sometimes called `forward-modelling' e.g. \citealt{Trcka2022TNGLF}). We believe that this approach is both useful as a cross-check, given present uncertainties in spectroscopic stellar population fits\footnote{A similar issue arises in the (now routine) procedure of constraining cosmological simulations to reproduce the galaxy stellar mass function: should the models be compared to stellar masses derived from observed luminosities (using spectral fits to determine mass-to-light ratios), or should mock luminosities be synthesised from the simulations and compared to observed luminosity functions? Ideally, both approaches should give consistent results, particularly when the same sets of stellar population templates are used; but this is not to be taken for granted \citep[e.g.][]{Mitchell2013StellarMass,Torrey:2015,Lacey2016AFormation,Trcka2022TNGLF}. We suggest similar considerations -- and cautions -- apply when comparing real and simulated resolved ages and metallicities, particularly if (as we suggest in Section~\ref{sec:manga}) different methods for fitting spectroscopic data have not converged on a unique explanation for the colour gradient trends we report here.}, and also practical, because photometric mock observables are routinely computed from simulations for other reasons (for example, reproducing survey selection functions in mock catalogues). In the case of colour gradients, the current generation of state-of-the-art simulations already provide mock images suitable for this purpose \citep[e.g.][]{Torrey:2015,Trayford2015Eagle,  Rodriguez-Gomez2019TheObservations}, although the treatment of dust extinction remains a significant complicating factor \citep[e.g.][]{Trayford2020Dust,Camps2022Dust}. We hope that the up-to-date colour gradient dataset we provide here will aid further comparisons to mock images from simulations.

In Section~\ref{sec:LS}, we describe how we select galaxies from the LS and how we compute the colour gradients.
In Section~\ref{sec:sz sample} and \ref{sec:pz sample}, we present our measurements for samples selected by spectroscopic and photometric redshift. We discuss the role of colour gradients in constraining galaxy formation theories in Section~\ref{sec:discussion}, including a comparison to results from MaNGA (\ref{sec:manga}) and remarks on blue-to-red colour gradients in low mass galaxies (\ref{sec:dw}). We conclude in Section~\ref{sec:conclusion}. We assume a flat $\Lambda$CDM cosmology, $\Omega_{\mathrm{m}} + \Omega_{\Lambda}= 1$, with matter density parameter $\Omega_\mathrm{m}=0.3$ and Hubble constant $H_0=100h \, \mathrm{km \, s^{-1} \, Mpc^{-1}}$ with $h=0.7$ \citep{Spergel2003Parameters}.

\begin{figure*}
	\includegraphics[width=2\columnwidth, trim=90 20 0 0]{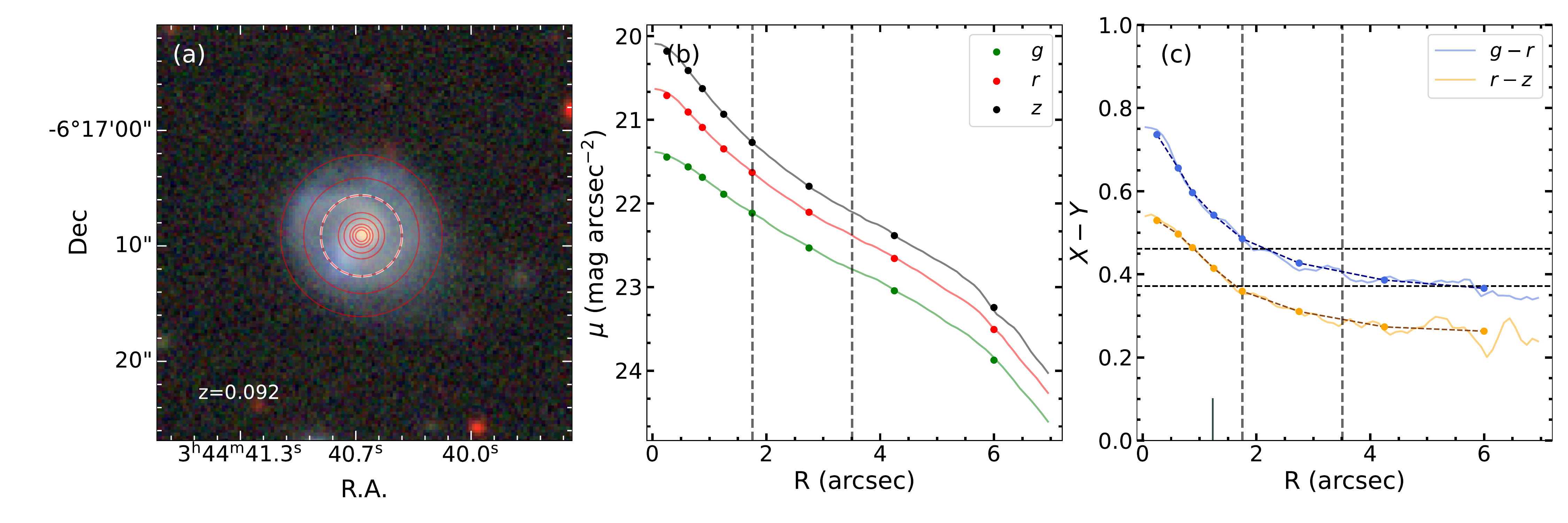}
    \caption{
    (a) $grz$ co-added image, (b) surface brightness profile, and (c) colour profile of an example $r=17$ galaxy in the Legacy Survey at redshift $z=0.092$.
    The colour gradients of this galaxy are $\nabla_{g-r}=-0.15$ and $\nabla_{r-z}=-0.13$.
    Red circles in panel (a) indicate the LS aperture radii. 
    The white dashed circle shows the effective radius, $R_\mathrm{eff}$.
    catalogueue aperture surface brightnesses (dots) in panel (b) are consistent with profiles derived directly from the images (solid lines). The vertical dashed lines in panels (b) and (c) mark $R_\mathrm{eff}/2$ and $R_\mathrm{eff}$, between which we measure colour gradients. 
    Colour profiles derived from interpolation of the catalogue aperture fluxes (dashed lines) in panel (c) are close to those obtained from direct measurements, demonstrating that the catalogue apertures are sufficient for our analysis.
    A short vertical bar shows the scale of the PSF for this source. Horizontal dashed lines mark the average $g-r$ and $r-z$ colour.
    }
    \label{fig:apertures_on_img_blue}
\end{figure*}

\begin{figure*}
	\includegraphics[width=2\columnwidth, trim=90 20 0 0]{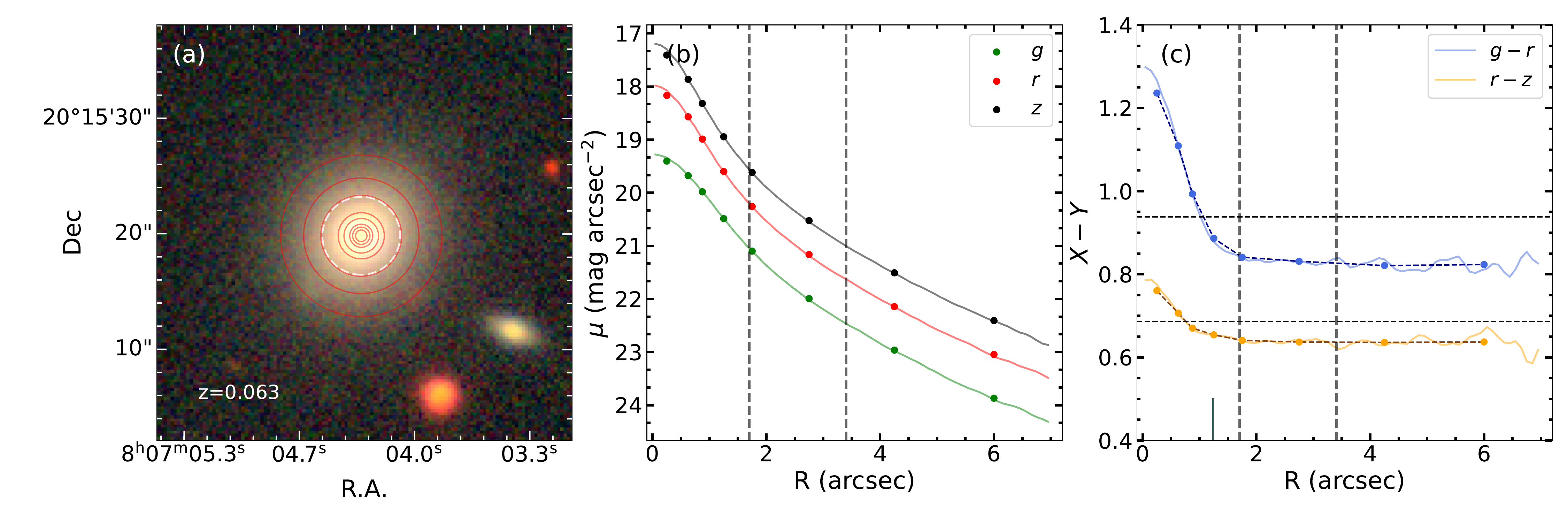}
    \caption{
    (a) $grz$ co-added image, (b) surface brightness profile, and (c) colour profile of a red $r=15.7$ galaxy in the Legacy Survey at redshift $z=0.063$.
    The colour gradients of this galaxy are $\nabla_{g-r}=-0.04$ and $\nabla_{r-z}=-0.01$.
    Labels are the same as those in Fig.~\ref{fig:apertures_on_img_blue}
    }
    \label{fig:apertures_on_img_red}
\end{figure*}

\section{The Legacy Survey}
\label{sec:LS}

The DESI Legacy Imaging Surveys cover approximately $14,000\,\mathrm{deg}^2$ of the northern hemisphere sky (approximately $-18\degr < \delta < +84\degr$ in celestial coordinates and $\lvert b\rvert > 18\degr$ in Galactic coordinates) in three optical bands ($g$, $r$, and $z$).
The observations were performed with DECam at the CTIO 4m  \citep[DECaLs;][]{Flaugher2015THECAMERA}, an upgraded MOSAIC camera at the KPNO 4m \citep[MzLS, Mayall $z$-band Legacy Survey;][]{Dey2016Mosaic3:Telescope}, and the 90Prime camera at the Steward Observatory 2.3m telescope \citep[BASS, Beijing-Arizona Sky Survey;][]{Zou2017ProjectSurvey}.
The Legacy Survey is approximately $2 \, \mathrm{mag}$ deeper than the Sloan Digital Sky Survey (SDSS) with $5\sigma$ point-source detection limits of $g = 24.0$, $r = 23.4$, and $z=22.5$ AB mag \citep{Dey2019OverviewSurveys}. The plate scale of the LS images is 0.262 $\mathrm{arcsec \, pix^{-1}}$ and the typical seeing $\sim1 \, \mathrm{arcsec}$ .
The Legacy Survey data releases provide single-exposure and co-added images processed with the NOAO Community Pipeline, and source catalogues generated from the processed images using {\it{the Tractor}}, a forward-modeling approach to source extraction \citep{Lang2016TheMeasurement}. We use the 9$^{\mathrm{th}}$ data release of the survey (LS DR9).

For each source, the LS catalogue provides a sky position, fluxes in three bands, a source type based on the best-fit model of the surface brightness profile (PSF, exponential, de Vaucouleurs) and an effective radius. These properties are derived from the most likely model as determined by {\it{the Tractor}}. The catalogues also provide fluxes measured in circular apertures with radii [0.5, 0.75, 1.0, 1.5, 2.0, 3.5, 5.0, 7.0] arcsec around each source (see Fig.~\ref{fig:apertures_on_img_blue}a). 

The results we present in this paper are obtained directly from the catalogue aperture fluxes. Additional information could be recovered from detailed surface photometry of the original images, for example, by a finer radial spacing or  elliptical rather than circular apertures. However, our motivation is to make comparisons to simulations and other datasets as straightforward and robust as possible by minimizing the number of additional reduction steps beyond those routinely applied to large surveys (and mock catalogues). Fig.~\ref{fig:apertures_on_img_blue} (panels b,c) shows that the aperture measurements are representative of more finely-resolved (circularly symmetric) profiles.

\subsection{Sample selection}

\begin{figure*}
    \centering
    \includegraphics[width=0.9\textwidth,trim=30 15 0 0]{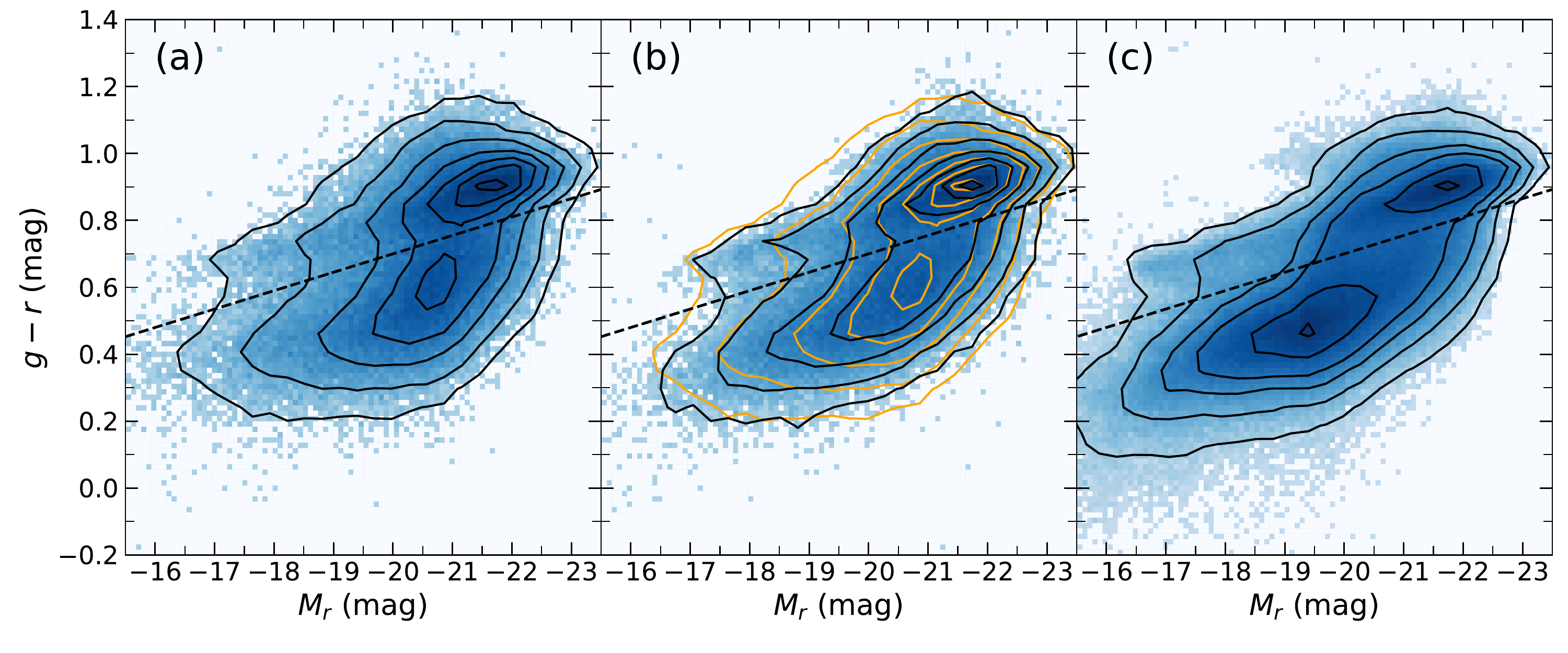}
    \caption{Colour magnitude diagrams for (a) our sample based on SDSS spectra (\sz\/);
    (b) \sz\/ sample but using photometric redshifts from \citet{Zhou2020TheRedshifts}.
    and (c) our full photometric redshift sample (\pz\/), limited to $r<19$.
    The orange contours overlaid on panel (b) repeat the \textit{sz} CMD from panel (a).
    The dashed line indicates our fiducial separation between red sequence (RS) and blue cloud (BC) galaxies, $(g-r) = -0.055M_r-0.4$.}
    \label{fig:cmd}
\end{figure*}

\begin{figure}
    \centering
    \includegraphics[trim=30 25 0 0 ,width=0.8\columnwidth]{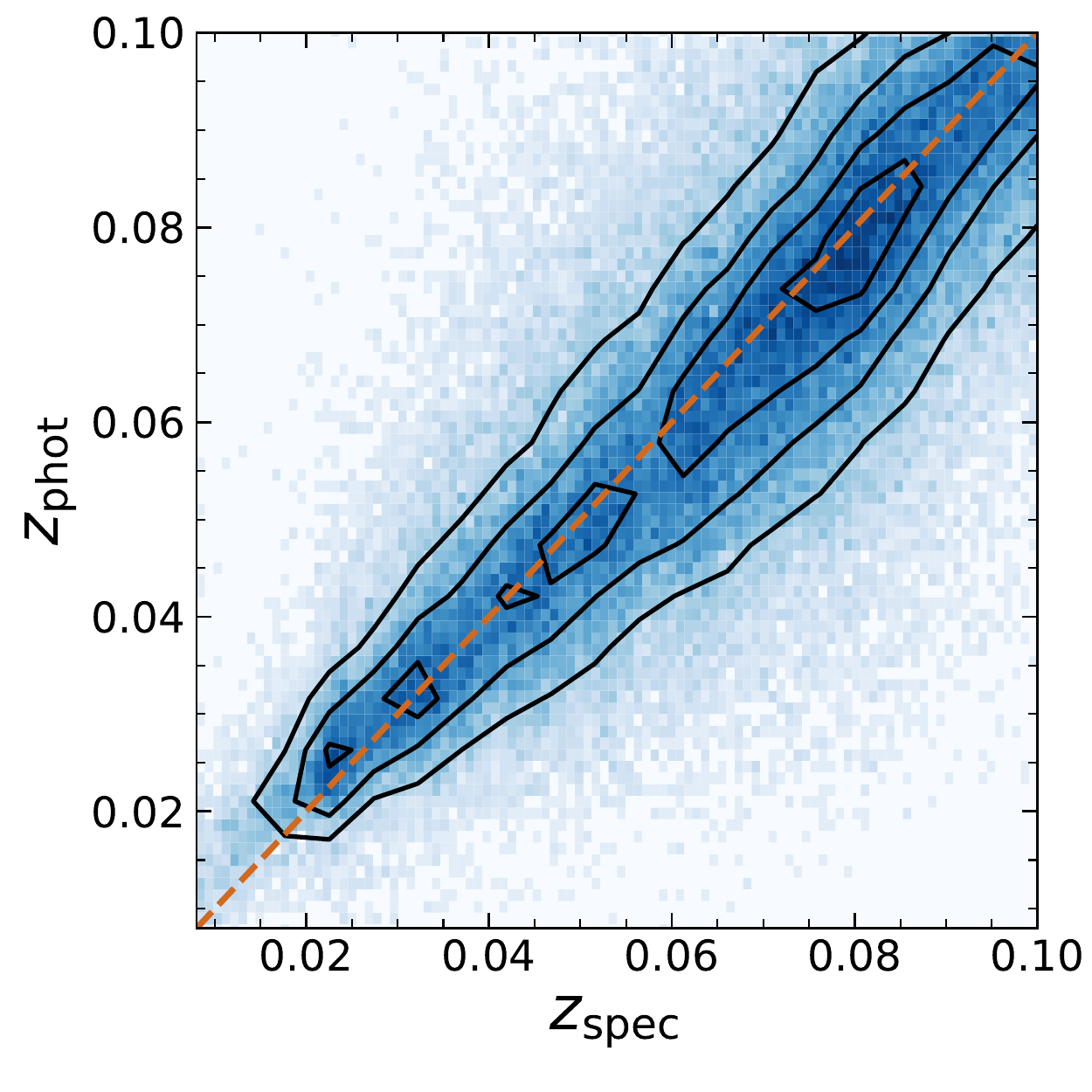}
    \caption{Spectroscopic redshift versus photometric redshift for galaxies in our \pz\/ sample.
    The contours enclose 20, 40, 60, and 80\% of the sample.
    The median photometric redshift error ($\sigma_\mathrm{photz}$) of the \sz\/ sample is $\sim21\%$. 
    The dashed line shows the $1:1$ relation.}
    \label{fig:pz_sz}
\end{figure}

We select two samples of galaxies from LS as described below, one based on SDSS spectroscopic redshifts (\sz\/) and the other on photometric redshifts derived from the LS (\pz\/). Table.~\ref{tab:sample} summarizes the properties of our two samples. 

\begin{table}
 \caption{Summary of our two LS galaxy selections. From left to right, columns give the redshift range, magnitude limit, effective radius range and sample size.}
 \label{tab:sample}
 \begin{tabular}{lcccc}
  \hline
   & $z$ & $r \, \mathrm{(mag)}$ & $R_\mathrm{eff} \, \mathrm{(arcsec)}$ & $N$\\
  \hline
  \sz\/ sample & 0.008-0.1 & $r < 17.77$ & $3\arcsec-6\arcsec$ & 93,680\\
  \sz\/ (gswlc) & & & & 81,085 \\
  \sz\/ (red) & & & & 37,990 \\
  \sz\/ (blue) & & & & 43,095 \\
  \sz\/ (low sSFR) & & & & 45,831 \\
  \sz\/ (high sSFR) & & & & 35,254 \\
  \hline
  \pz\/ sample & 0.008-0.1 & $r < 19$ & $3\arcsec-6\arcsec$ & 574,493\\
  \pz\/ (red) & & & & 195,408\\
  \pz\/ (blue) & & & & 379,085\\
  \hline
 \end{tabular}
\end{table}

\subsubsection{Sample with spectroscopic redshifts (\sz\/ sample) \label{sec:szsample}}
Our primary sample is derived from the subset of LS sources with spectra from the SDSS. LS DR9 provides a catalogue that is row-by-row matched to the SDSS DR16 spectroscopic catalog\footnote{\url{https://www.legacysurvey.org/dr9/files/}.}.
In this spectroscopic sample, we only include sources that meet the following criteria
to select high quality spectra with no duplicates \citep{Eisenstein2011SDSS-III:Systems}: \\
\indent{\fontfamily{cmtt}\selectfont ZWARNING = 0 or 16}, \\
\indent{\fontfamily{cmtt}\selectfont PLATEQUALITY = "good"}, \\
\indent{\fontfamily{cmtt}\selectfont SPECPRIMARY > 0}.\\
We also impose the following criteria on LS photometry, following the target selection quality cuts proposed for the DESI Bright Galaxy Survey by \citet{Ruiz-Macias2020}: \\
\indent{\fontfamily{cmtt}\selectfont NOBS\_x > 0}, \\
\indent{\fontfamily{cmtt}\selectfont FRACMASKED\_x < 0.4}, \\
\indent{\fontfamily{cmtt}\selectfont FRACIN\_x > 0.3}, \\
\indent{\fontfamily{cmtt}\selectfont FRACFLUX\_x < 5}, \\
\indent{\fontfamily{cmtt}\selectfont FLUX\_IVAR\_x > 0}, \\
where $x$ corresponds to the $g$, $r$, and $z$ bands.
{\fontfamily{cmtt}\selectfont NOBS\_x > 0} makes sure that images exist for the corresponding bands. The values used for image quality cuts, {\fontfamily{cmtt}\selectfont FRACMASKED\_x, FRACFLUX\_x} and {\fontfamily{cmtt}\selectfont FRACIN\_x,} are profile-weighted integrals over the pixels to which the model of the source contributes.
{\fontfamily{cmtt}\selectfont FRACMASKED\_x} is the profile-weighted fraction of pixels masked in all observations of the object, {\fontfamily{cmtt}\selectfont FRACFLUX\_x} is the profile-weighted fraction of the flux from other sources divided by the total flux, and {\fontfamily{cmtt}\selectfont FRACIN\_x} is the fraction of model flux within the group of pixels (`blob') identified with the source by the Legacy Survey source-detection algorithm. The  {\fontfamily{cmtt}\selectfont FRACFLUX\_x} cut selects for isolated objects, while the {\fontfamily{cmtt}\selectfont FRACMASKED\_x} and {\fontfamily{cmtt}\selectfont FRACIN\_x} cuts select for well-constrained model fits. The requirement of positive {\fontfamily{cmtt}\selectfont FLUX\_IVAR\_x} removes a very small number of spurious sources.
We also exclude sources that are extreme outliers in average colour, by requiring $-1 < X-Y < 4$ for both $g-r$ and $r-z$. 
To separate stars and galaxies, we first remove sources having \texttt{TYPE = PSF} in LS DR9.
Following \citet{Ruiz-Macias2020}, we further exclude sources which appear in the {\it{Gaia}} catalogue and meet the criterion $G-rr < 0.6$, where $G$ is the {\it{Gaia}} $G$-band magnitude and $rr$ is the raw $r$-band magnitude from the LS DR9 (without correction for Galactic extinction). This latter step eliminates non-PSF sources in the \textit{Tractor} catalogue for which the {\it{Gaia}} point-source magnitude is close to the \textit{Tractor} model magnitude. Such sources are likely to be stars.

The magnitudes we quote throughout this paper are corrected for Galactic dust extinction as
\begin{equation}
m_{x} = 22.5-2.5\log_{10} (F_{x}/\tau_{x}),
\end{equation}
with $F_{x}$ the Legacy Survey band $x$ flux of the source in nanomaggies, and $\tau_x$ the corresponding Milky Way transmission at the position of the source ({\fontfamily{cmtt}\selectfont MW\_TRANSMISSION} in the {\it{Tractor}} catalog, for which a value of 1 represents a fully transparent line of sight and 0 a fully opaque line of sight). 
We do not apply any K-corrections or corrections for internal extinction (we discuss the effect of dust below and in Appendix \ref{appendix:dust}).

We restrict our \sz\/ sample to the nominal  $r < 17.77$. magnitude limit of the SDSS main galaxy sample \citep{Abazajian2008TheSurvey}. 
We impose additional constraints of $0.008 < z < 0.1$ and $3\arcsec < R_\mathrm{eff} < 6\arcsec$ to reflect the limits imposed by our use of LS catalogue aperture photometry and the finite PSF size (see Section \ref{sec:color gradient}).
Our final \textit{sz} sample contains $\sim93,000$ galaxies.
Fig.~\ref{fig:cmd}a shows the colour-magnitude diagram (CMD) of this sample. 
In addition, we match these galaxies to the GSWLC-X2 catalogue \citep{Salim2016GALEX-SDSS-WISEGalaxies,Salim2018DustAnalogs}, which provides stellar masses ($M_\star$) and star-formation rates (SFR) based on SED fitting (using the CIGALE code with \citet{BC2003} templates and assuming a Chabrier IMF) to UV, optical and mid-IR photometry from GALEX, SDSS and WISE, respectively; the GALEX photometry was corrected for blending effects. The SED fitting was performed on the UV and optical photometry using a novel approach to include the total IR luminosity as a constraint; see \citet{Salim2018DustAnalogs} for details. The fits involve assumptions regarding the range of metallicity ($\gtrsim -1$) form of the star formation history (double exponential), treatment of stellar and nebular emission and dust attenuation law deemed appropriate for the bulk of low-redshift ($z<0.3$) galaxies in the SDSS catalogue. The footprint of GSWLC-X2 is limited by the availability of sufficient deep GALEX data (although the galaxies included are not themselves required to be detected in GALEX). Approximately $81,000$ galaxies in our sample have both $M_\star$ and SFR measurements in GSWLC-X2.

\subsubsection{Sample with photometric redshifts (\pz\/ sample) \label{sec:pzsample}}

For comparison to our primary sample, we select a much larger sample based on  photometric redshifts ({\fontfamily{cmtt}\selectfont z\_phot\_median}) computed from LS photometry by \citet{Zhou2020TheRedshifts}. This catalogue provides accurate photometric redshifst for objects with magnitude $z<21$.
\citet{Zhou2020TheRedshifts} developed a new method for estimating photometric redshift using the Lagacy Survey DECam and WISE photometry, which they applied to the LS DR9 data. 
They report a normalized median absolute deviation (NMAD, $\sigma_\mathrm{NMAD} = 1.48\times\mathrm{median}(\lvert \Delta z \rvert / (1+z_\mathrm{spec}))$) of $\sim 0.013$, and an outlier rate (fraction of outliers with $\lvert \Delta z \rvert > 0.1 \times (1+z_\mathrm{spec})$) of $\sim1.5$ per cent, where $\Delta z = z_\mathrm{phot}-z_\mathrm{spec}$.
Fig.~\ref{fig:pz_sz} shows spectroscopic redshift versus photometric redshift for galaxies in our \sz\/ sample that have a corresponding spectroscopic redshift. 
The $\sigma_\mathrm{NMAD}$ is $\sim0.010$, similar to the value reported by \citet{Zhou2020TheRedshifts}.

With photometric redshifts, we can extend our selection to fainter galaxies.
For this sample, we apply the same quality cuts, Galactic extinction correction, star galaxy separation, redshift and size criteria described in in Sec.~\ref{sec:szsample}.
We apply an apparent magnitude cut at $r < 19$ for this sample, because galaxies fainter than this are typically not well enough resolved in the LS photometry to measure colour gradients.
This limit is equivalent to an absolute magnitude of $M_r = -19.3$ at the upper redshift limit our our sample. 
At this magnitude, the size limits we impose correspond to a range of effective surface brightness from 22.6 to 24.1 $\mathrm{mag\, arcsec^{-2}}$.

Our \pz\/ sample contains approximately 0.5 million galaxies, and consequently provides much better statistics in comparison to the \sz\/ sample, at the cost of greater redshift (hence absolute magnitude) uncertainty for individual galaxies.
The CMD of this sample is shown in Fig.~\ref{fig:cmd}c. The differences between this panel and Fig.~\ref{fig:cmd}a are dominated by the different composition of the samples: compared the \sz\/ sample, the \pz\/ sample contains more faint and blue galaxies.
In Fig.~\ref{fig:cmd}b, to isolate the effects of the different redshift estimates, we construct the CMD of galaxies in the \sz\/ sample using their $z_\mathrm{phot}$ rather than $z_\mathrm{spec}$. The $z_\mathrm{spec}$ CMD contours from Fig.~\ref{fig:cmd}a is repeated for comparison (in orange). In this panel only, we do not restrict to $z_\mathrm{phot} < 0.1$, such that every \sz\/ galaxy is included at the distance implied by its $z_\mathrm{phot}$.

The two sets of contours in Fig.~\ref{fig:cmd}b are almost identical, which implies that redshift errors do not dominate the differences between the CMDs of the \sz\/ and \pz\/ samples.
The most noticeable difference is in the upper envelope of the red sequence ($-20 < M_{r} < -18$), where the outermost contours shift brighter by $\lesssim0.5$ mag. This implies $z_\mathrm{phot}$ measurements for the fainter, very red galaxies in this region are systematically larger than $z_\mathrm{spec}$. Since this region of the CMD is known to contain dusty star-forming galaxies, we speculate that systematic bias in $z_\mathrm{phot}$ may be due to strong dust reddening.

We obtain a very similar CMD if we simply restrict the \pz\/ sample to the apparent magnitude limit of the \sz\/ sample, $r<17.77$. In that case, the difference in the distribution at the faint end of the blue cloud becomes more pronounced. We believe this is mostly because the \pz\/ sample includes many more galaxies in this part of the diagram, but Fig.~\ref{fig:cmd}b suggests it could also be because $z_\mathrm{phot}$ systematically underestimates $z_\mathrm{spec}$ for some of those galaxies. Overall, we conclude that scatter between $z_\mathrm{phot}$ and $z_\mathrm{spec}$ does not distort the CMD of our \pz\/ sample significantly.

We have also compared the magnitude and redshift distributions of the two samples with and without the cuts on $R_{\mathrm{eff}}$. For the \sz\/ sample, these distributions are not significantly different.
For the \pz\/ sample, the size cuts do not bias the $M_r$ distribution up to $z\sim0.06$. At higher redshift, the least luminous galaxies are under-represented in the size-limited \pz\/ sample, in comparison to the flux-limited parent sample.

\subsection{Measurement of colour gradients \label{sec:color gradient}}

Figs. \ref{fig:apertures_on_img_blue} and \ref{fig:apertures_on_img_red} illustrate our definition of the colour gradient and our approach to measuring these gradients from LS photometry. Typical examples of red and blue galaxies from our sample are shown in each panel (a), together with the LS apertures. 

To compute colour gradients, we use the LS catalogue aperture fluxes and effective radius ($R_\mathrm{eff}$) of each galaxy to construct a surface brightness profile (Fig. \ref{fig:apertures_on_img_blue} and \ref{fig:apertures_on_img_red} panel (b)) and a colour profile (panel (c)) for bands $X$ and $Y$. We obtain colours at $R_1$ (inner radius) and $R_2$ (outer radius) through linear interpolation of the aperture values (in magnitudes).
We compute the colour gradient as 
\begin{equation}
    \nabla_{X-Y} = \delta(X-Y)/\delta(R/R_\mathrm{eff}),
\end{equation}
defined as the slope of $X-Y$  colour versus $R/R_\mathrm{eff}$.
With this definition, a positive colour gradient means that the galaxy is bluer at the centre and a negative colour gradient means the galaxy is redder at the centre. Here we choose $R_1 = 0.5 R_\mathrm{eff}$ and $R_2 = R_\mathrm{eff}$.

Since we use fixed aperture radii to calculate colour gradients, we cannot derive gradients for galaxies with $R_\mathrm{eff} < 0.5\arcsec$ or $R_\mathrm{eff} > 6\arcsec$. The colour gradients of galaxies with small $R_\mathrm{eff}$ may be affected by the PSF. The median FWHM scale of the PSF is approximately 1.5\arcsec, so we further restrict our sample to galaxies with sizes in the range $3\arcsec < R_\mathrm{eff}<6\arcsec$.
We obtain an error on the colour gradient through standard propagation of the aperture flux errors.
The median $1\sigma$ errors on our colour gradients are $\sigma\sim0.024\, \cgunit$ for $\nabla_{g-r}$ and $\sigma\sim0.027\, \cgunit$ for $\nabla_{r-z}$.

\subsection{1/$V_\mathrm{max}$ weighting}

We apply a $1/V_\mathrm{max}$ weighting \citep{Schmidt1968SpaceSources} to correct our statistical results for the incompleteness of intrinsically faint galaxies in our flux-limited sample. For each galaxy, the corresponding maximum luminosity distance ($d_{L,\mathrm{max}}$) is given by
\begin{equation}
    d_{L,\mathrm{max}} = 10^{-0.2(M_r-m_\mathrm{lim}-5)},
\end{equation}
where $M_r$ is the absolute magnitude and $m_\mathrm{lim}$ is the limiting apparent magnitude ($r=17.77$ for our \sz\/ sample and $r=19$ for our \pz\/ sample).
We obtain the corresponding maximum comoving distance ($d_{c,\mathrm{max}}$) at which each source can be observed, and convert this to a maximum volume as
\begin{equation}
    V_\mathrm{max} = \frac{4\pi}{3}
    (d_c(z_\mathrm{max})^3-d_c(z_\mathrm{min})^3),
\end{equation}
where $z_\mathrm{min}$ is 0.008.
Since we apply a redshift cut at $z=0.1$, we set $d_{c,\mathrm{max}}$ to $d_c(z_\mathrm{max}=0.1)$ if $d_{c\mathrm{,max}} > d_c(z=0.1)$.

\begin{figure}
    \centering
    \includegraphics[trim=0 0 0 0,width=0.8\columnwidth]{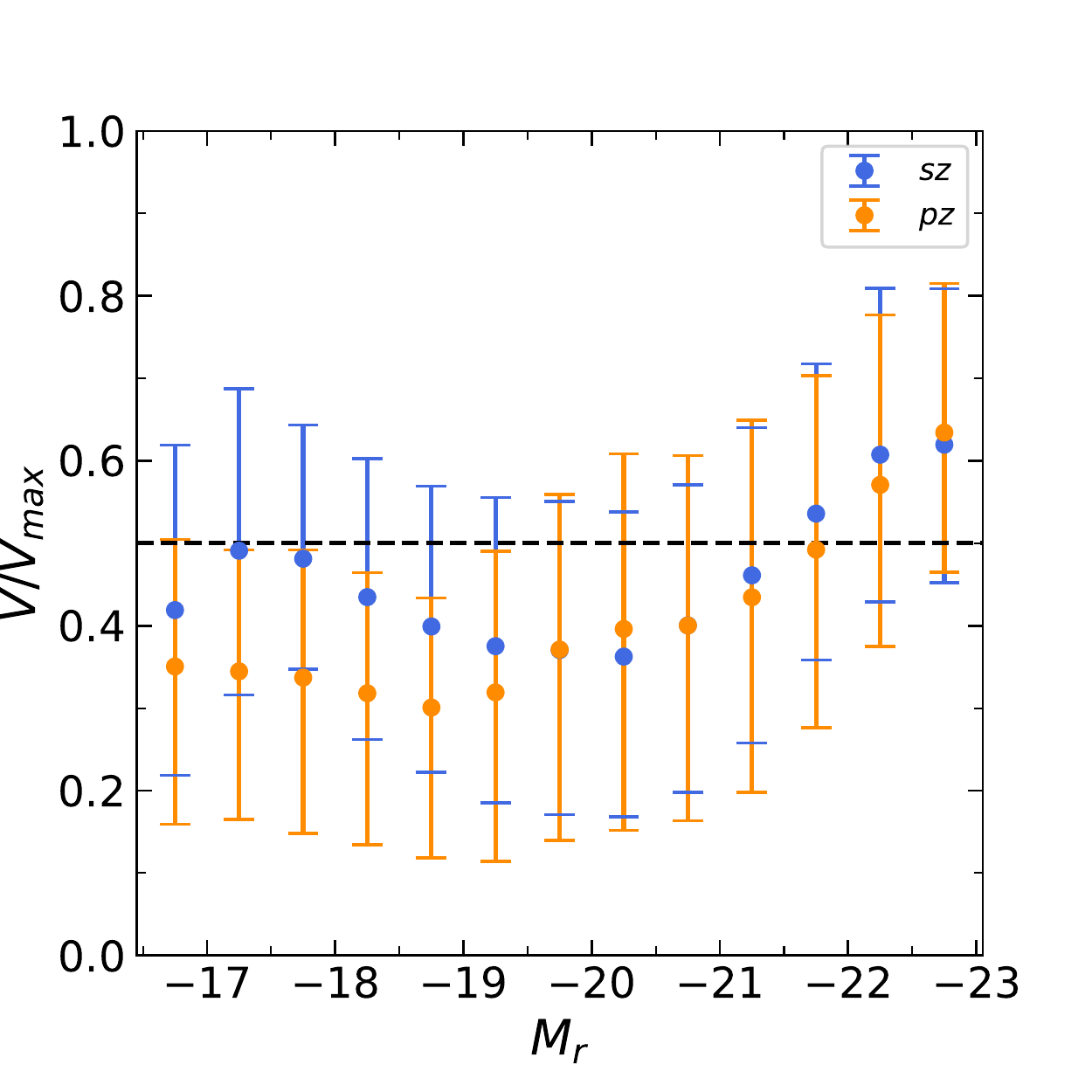}
    \caption{$V/V_\mathrm{max}$ as a function of $M_r$ for our \sz\/ (blue) and \pz\/ (orange) samples. The error bars indicate the 25-75\textsuperscript{th} percentile range.}
    \label{fig:VVmax}
\end{figure}

Fig.~\ref{fig:VVmax} shows the mean of $V/V_\mathrm{max}$ in bins of absolute magnitude. Most magnitude bins have $\langle V/V_\mathrm{max} \rangle \approx0.5$, implying that the distribution of galaxies is approximately uniform throughout the survey volume \citep{Schmidt1968SpaceSources}. 
$V/V_\mathrm{max}$ is larger than 0.5 for bright galaxies in both the \sz\/ and \pz\/ samples. In a purely magnitude limited selection this could be caused by clustering, which increases the counts at large distances and decreases them at small distances, relative to a uniform distribution. However the minimum and maximum angular size limits we impose on our sample may also contribute to systematic deviations of $V/V_\mathrm{max}$. For intrinsically bright galaxies (with large physical size) the upper limit on angular size will further reduce the counts at small distances and hence have the same effect on $V/V_\mathrm{max}$ as clustering. For intrinsically faint galaxies (with small physical size) we expect the opposite effect: a  reduction in counts at larger distances leading to lower $V/V_\mathrm{max}$. A reduction at faint magnitudes is only apparent in our \pz\/ sample. Since the \pz\/ sample has a fainter apparent magnitude limit than the \sz\/ sample, it contains more intrinsically faint galaxies with large distances, at which a greater fraction will be removed by our minimum angular size cut.

\section{Results for spectroscopic redshift sample}
\label{sec:sz sample}

\begin{figure}
    \centering
    \includegraphics[width=0.8\columnwidth, trim=0 0 0 0]{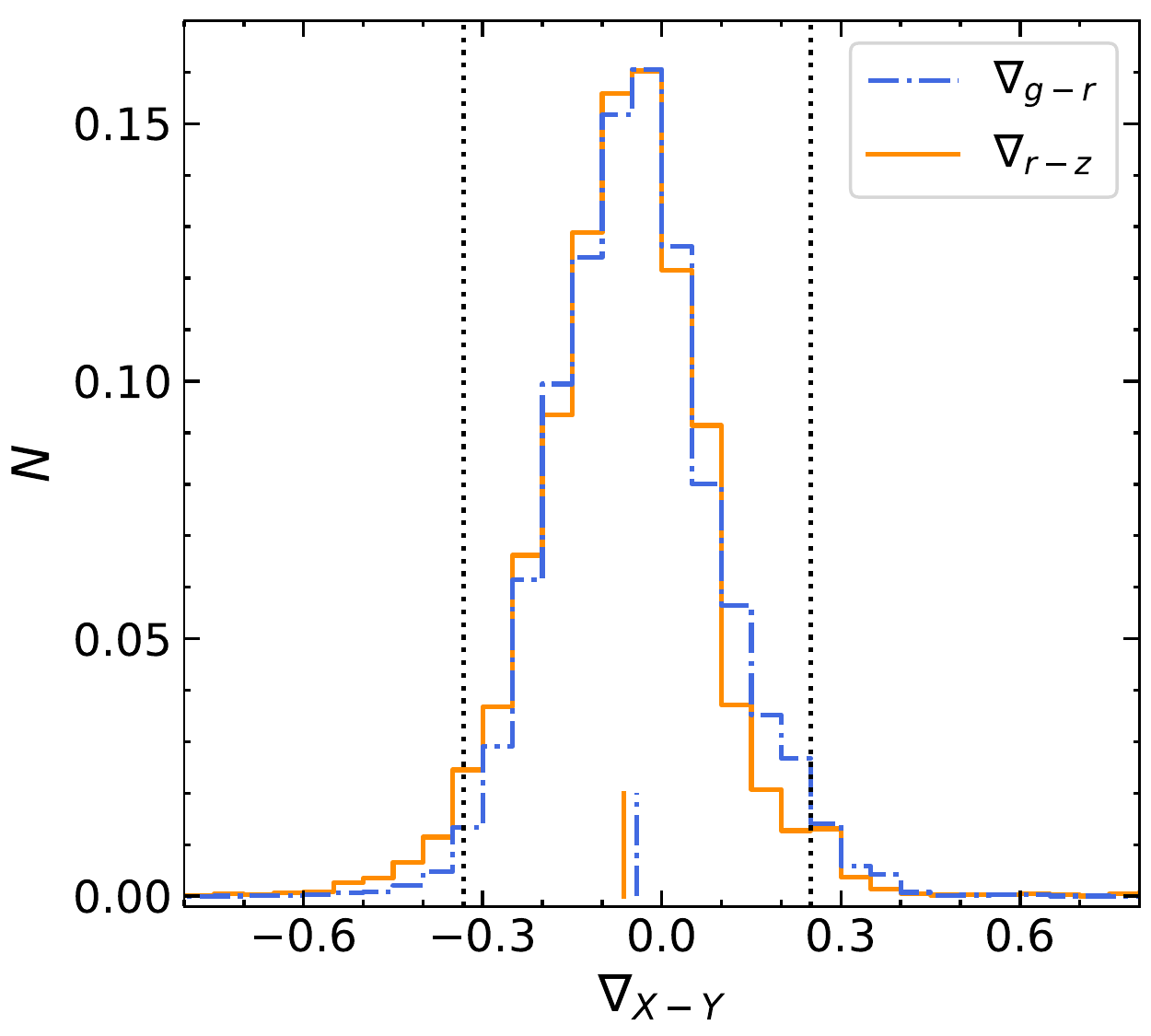}
    \caption{$1/V_\mathrm{max}$-weighted PDF of $\nabla_{g-r}$ (blue dot-dashed line) and $\nabla_{r-z}$ (orange solid line) for galaxies in our \textit{sz} sample.
    Short vertical bars of the same colour and style indicate mean colour gradients $\nabla_{g-r}=-0.041$ and $\nabla_{r-z}=-0.064$. The standard deviations of these distributions are $\sigma_{g-r}=0.15$ and $\sigma_{r-z}=0.17$.
    Vertical dashed lines mark $\pm2\sigma$.}
    \label{fig:fig2cghist}
\end{figure}

\begin{figure*}
    \centering
    \includegraphics[width=0.7\textwidth, trim=0 10 0 0]{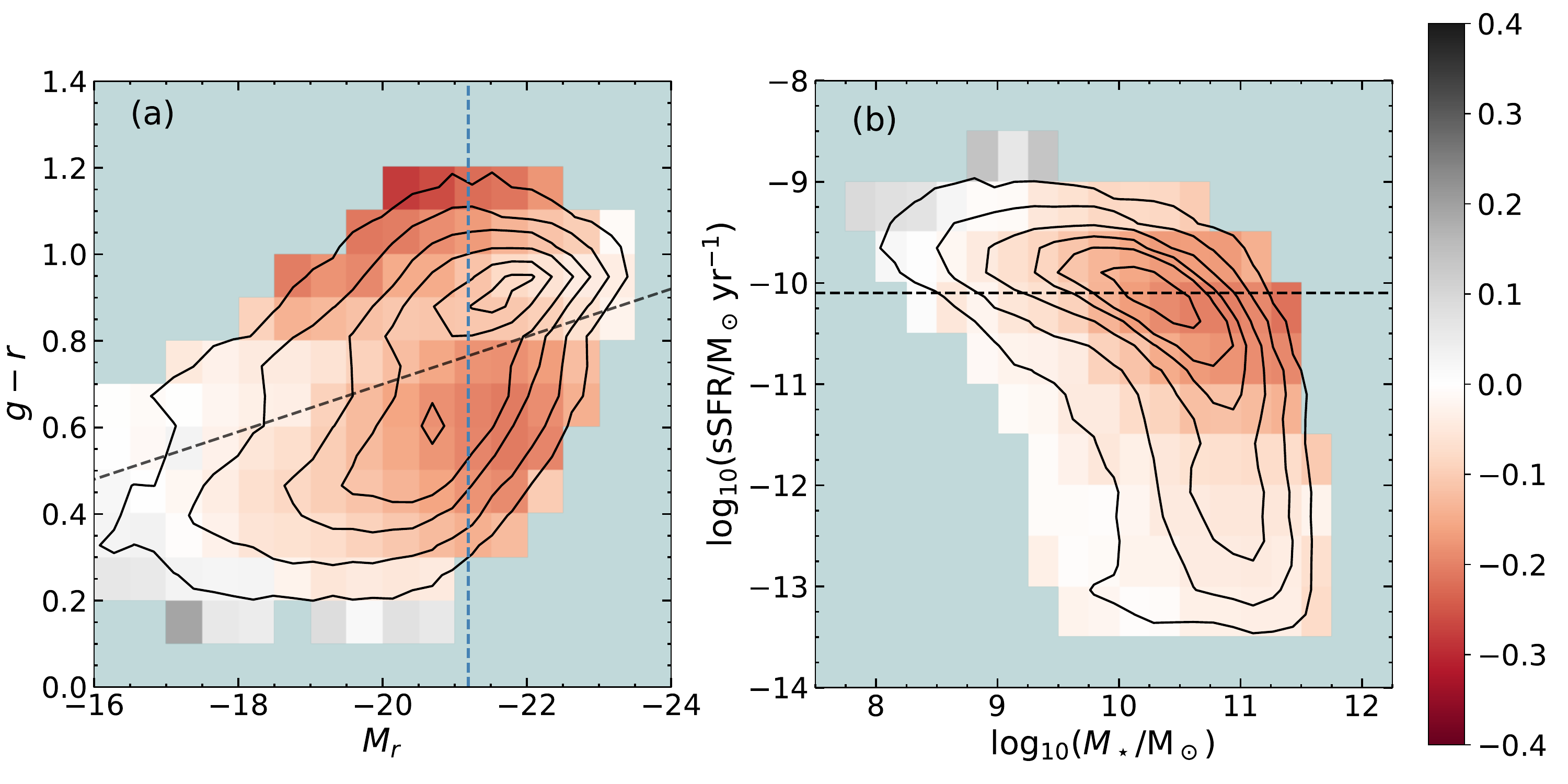}
    \caption{Average $\nabla_{g-r}$ (colour scale) binned in the plane of (a) $M_r, g-r$ and (b) $M_\star$, sSFR. The contours show the joint density distribution of galaxies in these quantities and the black dashed lines show our separation between (a) red and blue galaxies and (b) low- and high-sSFR galaxies.
    The blue dashed line in panel (a) indicates the inflection point shown in Fig.~\ref{fig:cg_optcolor}c (see Table~\ref{tab:fitresults}).}
    \label{fig:binned_map}
\end{figure*}

Fig.~\ref{fig:fig2cghist} shows the distributions of the colour gradients $\nabla_{g-r}$ and $\nabla_{r-z}$ for galaxies in our \sz\/ sample. The two distributions are similar overall. The mean values of $\nabla_{g-r}$ and $\nabla_{r-z}$ are $-0.041$ and $-0.064$, respectively.
In these optical bands, galaxies tend to be redder at their centres. 
The corresponding standard deviations are $\sim0.15$ and $0.17$, both of which are much larger than typical errors on individual colour gradients, 0.024 for $\nabla_{g-r}$ and 0.027 for $\nabla_{r-z}$.
Negative mean colour gradients are consistent with previous studies \citep{Peletier:1990,LaBarbera:2005,Liu2009TheSurvey, Gonzalez:2011,Chanjcc:2016,Kennedy:2016}. 
Slightly more negative values of $\nabla_{r-z}$ compared to $\nabla_{g-r}$ are consistent with galaxies having smaller effective radii at longer wavelengths \citep[e.g.][]{Kennedy2016GalaxyDecompositions}. 
Since we use the same $R_\mathrm{eff}$ in all bands, the flux from $z$-band is expected to be more concentrated at the centre, resulting in a steeper colour gradient.

We further examine galaxies having extreme colour gradients, defined as more than 3 standard deviations from the mean. 
In total, we have around 600 galaxies ($0.6\%$) with colour gradients outside 3$\sigma$. 
There are 4 times more galaxies in our sample with extreme negative $\nabla_{g-r}$ than extreme positive gradients. 
Many galaxies with extreme positive gradients are blue galaxies blended with a bright red star. Galaxies with extreme negative gradients are more varied; some are blends with blue stars, but others are dusty spiral galaxies (see Appendix~\ref{appendix:extcg}).

\subsection{Colour gradient as a function of average colour and $M_r$}
\label{sec:cg_colorMr}

Fig.~\ref{fig:binned_map}a shows how the colour gradient varies across the colour magnitude diagram by plotting the mean gradient in small bins of $M_r$ and $g-r$. This figure shows that colour gradients have a complex dependence on both magnitude and global colour. The mean gradient in most bins is negative (red pixels), as for the population as a whole (Fig.~\ref{fig:fig2cghist}). However, Fig.~\ref{fig:binned_map}a also shows clear systematic variation of colour gradients across the population.

For blue cloud galaxies (below the black dashed line in Fig.~\ref{fig:binned_map}a), we see a trend from positive to negative colour gradients as galaxies become brighter. For typical red sequence galaxies, gradients are flatter and do not show strong variation parallel to the red sequence. However, bins corresponding to extremely red colours ($\sim0.3$~dex above the red sequence for $-21 < M_{r} < -19$) show the steepest negative colour gradients of all. This region of the CMD is dominated by star-forming galaxies with high dust\footnote{Without considerable further work to estimate resolved extinction gradients, we cannot correct the colour gradients themselves for reddening. We expect dust contributes to the observed scatter in gradients, but does not drive the mean values shown on this diagram \citep[e.g.][]{DeJong1996Near-infraredGalaxies} with the possible exception of extremely dusty galaxies. However, see also Sec.~\ref{sec:manga}.} extinction \citep[e.g.][ see also Appendix~\ref{appendix:dust}]{Tempel2011GalaxyDR7,Xiao2012DustGalaxies}.

\begin{figure*}
    \centering
    \includegraphics[width=0.75\textwidth,trim=0 10 0 0]{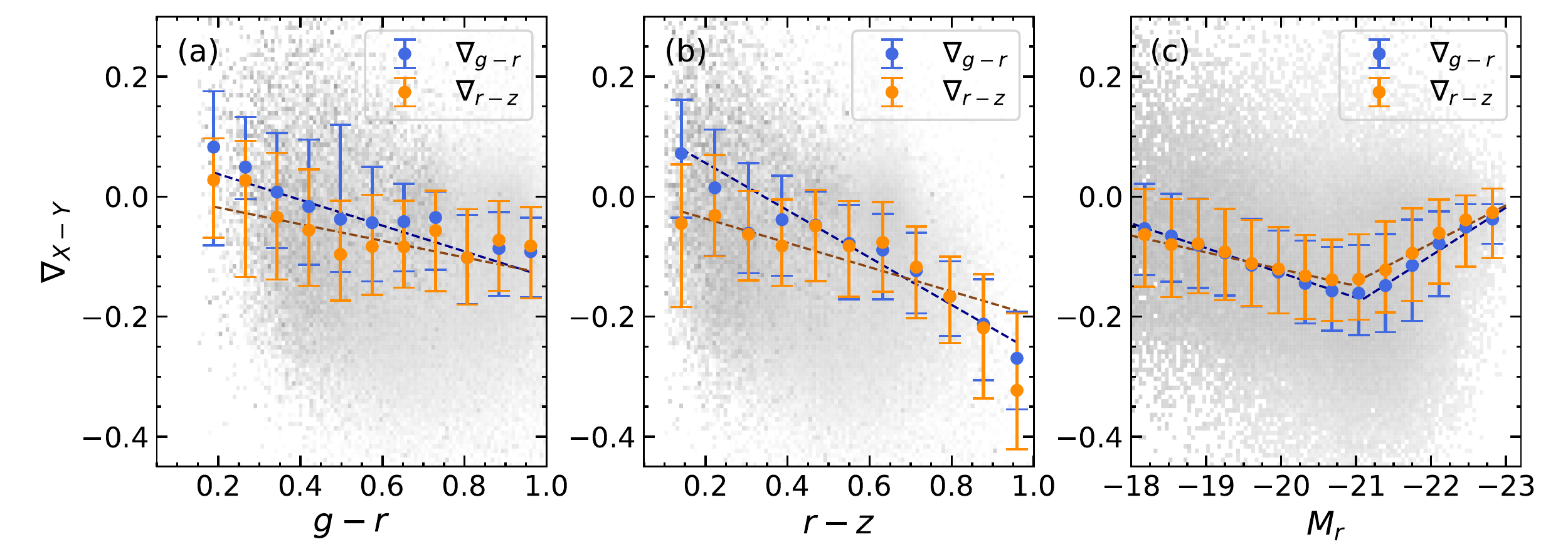}
    \caption{Colour gradient as a function of (a) $g-r$, (b) $r-z$, and (c) $M_r$. 
    Colour gradients derived from $g-r$ and $r-z$ are shown in blue and orange, respectively.
    Means of binned colour gradients are shown as points, with error bars indicating the 25–75 percentile range.
    We fit straight lines to the data (dashed blue and brown lines fit blue and orange points, respectively).
    For $M_r$, we fit a broken linear function; the parameters of the fit are given in Table~\ref{tab:fitresults}.
    The grey-scale map shows the density of galaxies in our sample (for $\nabla_{g-r}$ only).
    }
    \label{fig:cg_optcolor}
\end{figure*}

In Fig.~\ref{fig:cg_optcolor} we plot the average colour gradient as a function of average colour and absolute (LS model) magnitude, $M_r$. Trends of colour gradients with other properties are often presented this way in the literature, although some of the complexity seen in Fig.~\ref{fig:binned_map}a is less apparent in this one-dimensional representation. 
In the first and second panels of Fig.~\ref{fig:cg_optcolor} we show that gradients are negative for all but the bluest galaxies and become steeper as galaxies become redder overall. 
This is consistent with previous studies based on similar optical colours 
\citep[e.g.][]{Park2005MorphologySpace, Suh2010DemographyGradients, Tortora:2010}. The trend with magnitude shows more complex behaviour.
Starting from the faint end, colour gradients become steeper with increasing $M_r$ but show an inflection around $M_r\sim-21$, becoming shallower for brighter galaxies.
This has also been observed in previous studies, for example \citet[][]{Tortora:2010}, who report an inflection in the trend of SDSS $g-i$ gradient with SDSS $r$ band magnitude at $M_{r,\mathrm{SDSS}} \sim -20$. The $\sim1$ mag difference with our result could be due to different sample selections, choices of colour and definitions of colour gradient (the difference between the LS and SDSS $r$-band filters is relatively small). In Section~\ref{sec:cg_sm_sfr} we show the corresponding inflection point in $\nabla$ vs. $M_\star$ agrees well with that reported by \citet{Tortora:2010}.

In Table~\ref{tab:fitresults} we report linear least-squares fits to these trends,
\begin{equation}
\label{eq:linefit}
    \nabla_{X-Y} = a\times P + b,
\end{equation} 
where $P$ is the variable of interest. The trends of gradient with average colour (both $g-r$ and $r-z$) are described well by a single slope, $a \approx -0.3$.  To describe the trend with $M_r$, we use two linear functions constrained to be equal at an inflection point. We find the best-fit inflection point to be $M_{r,i}=-21$. For the fainter subset of our sample, $M_r > M_{r,i}$, the trends in $\nabla_{g-r}$ and $\nabla_{r-z}$ have slightly different slopes, $a\sim0.04$ and $0.03$ respectively.
For the brighter part, $M_r < M_{r,i}$, the slope is $a\sim-0.07$ for both colours.

\begin{figure}
    \centering
    \includegraphics[width=1.0\columnwidth, trim=20 20 0 0]{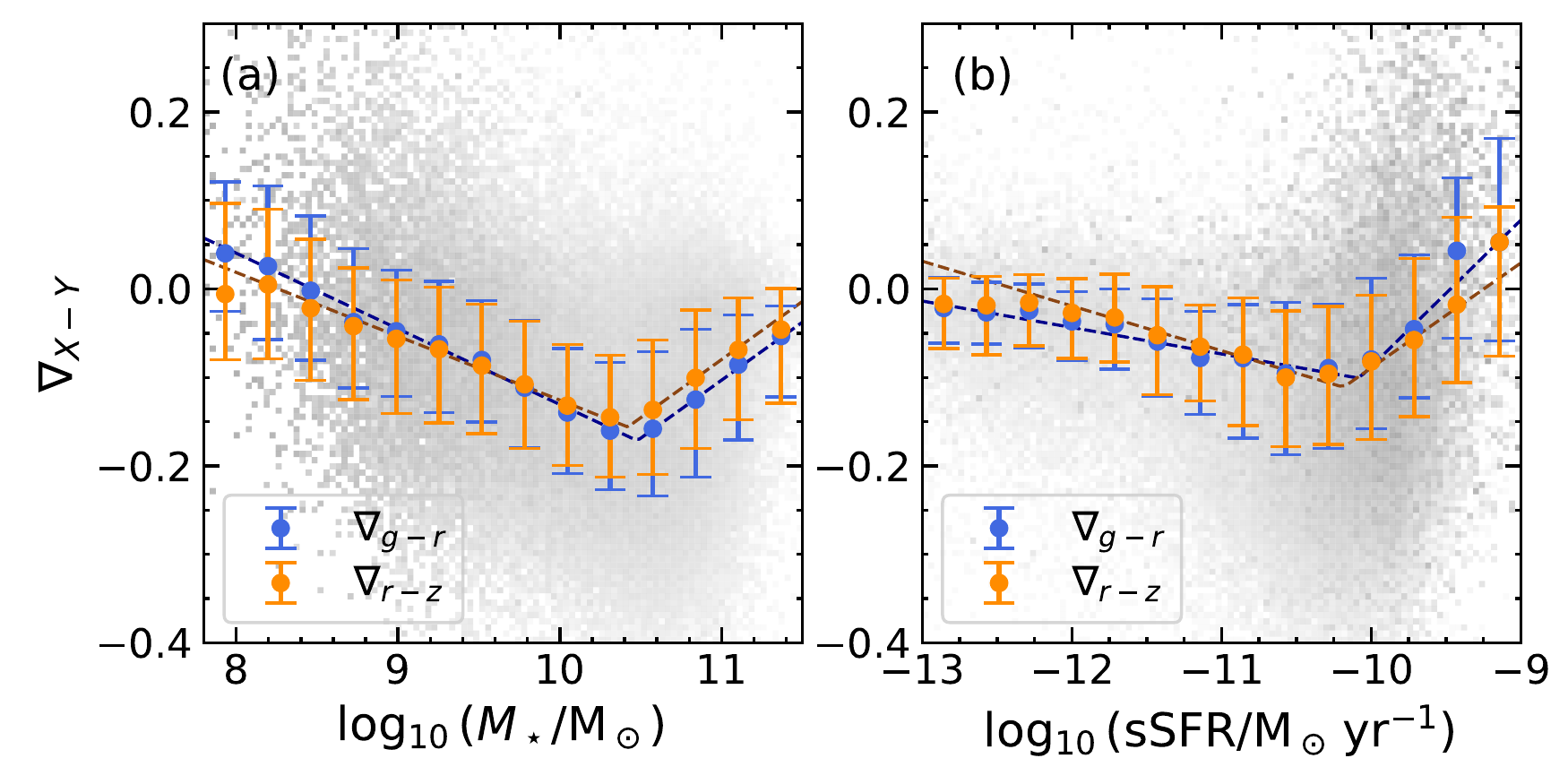}
    \caption{Similar to Fig.~\ref{fig:cg_optcolor}, here showing colour gradients as a function of (a) $M_\star$ and (b) sSFR. The error bars indicate the 25–75 percentile range.
    We fit both relations with a broken linear function. The best fit parameters are given in Table.~\ref{tab:fitresults}.}
    \label{fig:cg_sm_sfr}
\end{figure}

The trends shown in Figs.~\ref{fig:binned_map}a and \ref{fig:cg_optcolor} are qualitatively consistent with expectations from the literature, in particular observations of age and metallicity gradients from spatially resolved spectroscopy in much smaller samples of galaxies (see further discussion in Section~\ref{sec:discussion}). Most work to date attributes steep negative colour gradients in the blue cloud to `inside out' formation: star-forming gas is accumulated slowly with increasing angular momentum, giving rise to strong age gradients in the most massive (earliest forming) galaxies. The corresponding colour gradient can be reinforced by a declining metallicity gradient, driven by decreasing star formation rates and dilution of the star-forming ISM by the inflow of metal-poor gas. As we discuss below, it is not immediately clear whether these steeper gradients arise from age gradients in a single dynamical component (i.e. a galactic disc), or from the transition between two or more components with  different origins \citep[e.g.\ between bulge, bar and disc;][]{Kennedy2016GalaxyDecompositions}. 
We find fainter field dwarf galaxies ($M_r \gtrsim -17$) have flat gradients, implying more uniform stellar populations, consistent with physically small systems dominated by a small number of recent star formation events. Those with the bluest colours have slightly positive gradients.

We find colour gradients to be much flatter along the red sequence. The contrast with the blue cloud is particularly striking for galaxies brighter than $M_{r,i}\approx-21$ (among which red sequence galaxies dominate in number and hence drive the trend in Fig.~\ref{fig:binned_map}a). Flatter gradients on the red sequence (and decreasing scatter in gradient at a fixed magnitude) may reflect both a smaller range of (old) ages for the stars in those galaxies and the increasing importance of growth through gas-poor mergers. Mergers can flatten gradients by mixing stellar populations \citep[e.g.][]{White1980MixingMergers,Bell2005DryGalaxies, Kim2013Optical-NearGalaxies}.

These explanations may not be unique, or sufficient to explain the quantitative behaviour we see in Fig.~\ref{fig:binned_map}a. In Section~\ref{sec:discussion}, we discuss the possible origins of these trends in more detail and the implications of our results for galaxy formation models. We first present analogous trends with other galaxy properties and explore subsets of our sample.

\subsection{Colour gradient as a function of $M_\star$ and sSFR}
\label{sec:cg_sm_sfr}

Fig.~\ref{fig:binned_map}b shows how the mean colour gradient changes across the $M_\star$--sSFR diagram, for the subset of our sample matched to the catalogue of \citet{Salim2016GALEX-SDSS-WISEGalaxies,Salim2018DustAnalogs}.
Again, the mean gradient in most bins is negative.
We find a clear trend of steeper colour gradients with increasing $M_{\star}$ for galaxies with high sSFR ($\log_{10}\, (\mathrm{sSFR/M_\odot \, yr^{-1}}) > -10.1$, above the black dashed line). 
This is consistent with the results shown in Fig.~\ref{fig:binned_map}a, assuming most high-sSFR galaxies are in the blue cloud.
For low-sSFR galaxies\footnote{We note that \citet{Salim2016GALEX-SDSS-WISEGalaxies} caution that sSFR measurements below $-11.5$ to $-12$ dex (dependent on the depth of the UV data available for a given galaxy) can only be considered as upper limits.}, the colour gradient appears to correlate with sSFR.This is clearest at higher masses where the range of sSFR is greatest. The colour gradient becomes flatter as sSFR decreases.
The value we choose to separate high- and low-sSFR galaxies is different from the more conventional $\log_{10} \, \mathrm{sSFR} = -11 \, \mathrm{M_\odot \, yr^{-1}}$ (roughly one order of magnitude below the star formation `main sequence'). Our value corresponds to the infection point of the $\nabla$--sSFR relation shown in Fig.~\ref{fig:cg_sm_sfr} and therefore represents a clear division in our sample.

In Fig.~\ref{fig:cg_sm_sfr} we plot colour gradient against $M_\star$ and sSFR separately, analogous to Fig.~\ref{fig:cg_optcolor}.
The lefthand panel shows the trend with $M_\star$, which is similar to that with $M_r$. 
The colour gradients become steeper with mass for $M_\star < M_{\star,i}$, and flatter with mass for $M_\star > M_{\star,i}$.
As above, we fit this relation with a broken linear function, finding $\log_{10}\,M_{\star,i}/\mathrm{M_{\sun}}\sim10.5$.
We find $\nabla_{g-r}$ and $\nabla_{r-z}$ slopes of [$-0.09$,  $-0.07$] $\mathrm{dex}^{-1}$ respectively for the low-mass subset and [$0.13$, $0.13$] $\mathrm{dex}^{-1}$ respectively for the high-mass subset.
The inflection at $M_\star \sim10^{10.5} \, \mathrm{M_\odot}$ is similar to that found by \citet{Tortora:2010} and is consistent with standard expectations for galaxy formation in a $\Lambda$CDM cosmogony (see discussion in Sec.~\ref{sec:origin_cg}).

The righthand panel of Fig.~\ref{fig:cg_sm_sfr} shows trends with sSFR. Colour gradients are flat at very low sSFR, become steeper (more negative) with increasing sSFR up to an inflection point, above which they flatten again, becoming positive for the highest sSFR galaxies in our sample. The scatter is small at low sSFR and large at high sSFR. We find the inflection point to be $\log_{10}(\mathrm{sSFR}/\mathrm{M_{\odot}\,yr^{-1}}) = -10.1$ and the slopes of the $\nabla_{g-r}$ and $\nabla_{r-z}$ trends with sSFR to be [$-0.03$, $-0.05$] for low-sSFR subset and [$0.17$, $0.12$] for high-sSFR subset, respectively.
These results follow from those shown in previous figures, if we simply assume that the lowest sSFRs correspond to massive red galaxies (flat gradients) and the highest sSFR  to blue dwarf galaxies (increasing steep positive gradients with decreasing mass).
However, we caution that systematic uncertainties in sSFR are large below $\mathrm{sSFR} < -11.5 \, \mathrm{M_\odot} \, \mathrm{yr}^{-1}$  \citep[see][]{Salim2016GALEX-SDSS-WISEGalaxies}.

As expected, trends in colour gradient with $M_\star$ and sSFR emphasize those seen in $M_r$ and $g-r$. At high masses, the slope of the relation is similar when plotted against mass or magnitude, but the slope of the relation for low mass galaxies is significantly steeper as a function of stellar mass. This is not surprising; less massive galaxies have a wide range of colours, hence $r$-band mass-to-light ratios, so any trend with $M_{\star}$ will be weakened when plotted against $M_{r}$. The flat or slightly positive gradients for dwarf galaxies stand out more clearly in  Fig.~\ref{fig:cg_sm_sfr}, as do the extremely steep gradients of galaxies with the highest sSFR (which comprise both low-mass blue dwarfs and the more massive `dusty discs' that lie above the red sequence in Fig.~\ref{fig:binned_map}a).

\subsection{Colour gradients for red and blue galaxies}
\label{sec:cg_red_blue}

Fig.~\ref{fig:binned_map} implies that red and blue galaxies (or quiescent and star-forming galaxies) will show different trends in colour gradient with mass or luminosity.
Hence, we now examine the trends shown in Fig.~\ref{fig:cg_optcolor} for red and blue galaxies separately.

\begin{figure}
    \centering
    \includegraphics[width=\columnwidth, trim=0 40 0 30 
    0]{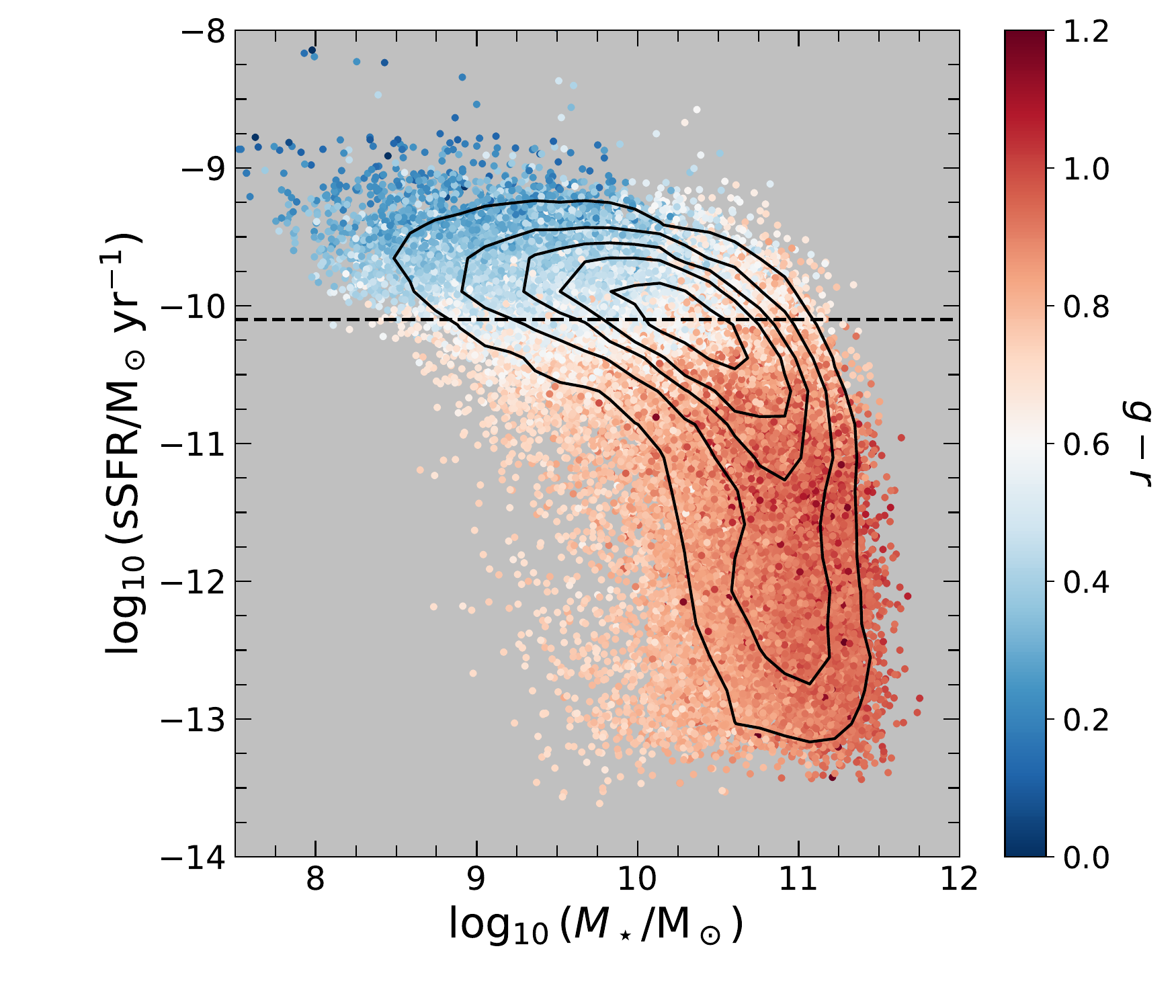}
    \caption{$M_\star-$sSFR relation of our sample.
    The contours show the density of galaxies (indicating the star-formation main sequence) and the colours of individual points indicate $g-r$.
    The dashed line at $\mathrm{sSFR}=10^{-10.1} \, \mathrm{M_\odot} \, \mathrm{yr}^{-1}$ separates high- and low-sSFR galaxies.}
    \label{fig:sfms}
\end{figure}

We divide our sample using two different methods: a CMD locus, shown by the black dashed line in Fig.~\ref{fig:cmd}, and a specific star formation rate threshold, $\log_{10} \, (\mathrm{sSFR}/\mathrm{M_\odot}\, \mathrm{yr}^{-1}) \sim-10.1$. This threshold is determined by the inflection point shown in Fig.~\ref{fig:cg_sm_sfr}. 
We refer to the two subsets of galaxies separated by the CMD locus as `red' and `blue', and to the subsets separated by the $M_{\star}$--sSFR locus as `low' and `high' sSFR. 
Fig.~\ref{fig:sfms} shows the relationship between these two selections by plotting galaxies in the $M_\star$--sSFR diagram, with the colour of each point corresponding to $g-r$.
We find that our separation at $\log_{10} \, (\mathrm{sSFR}/\mathrm{M_\odot}\, \mathrm{yr}^{-1}) = -10.1$ roughly approximates a color of $g-r\sim0.6$ and cuts through the peak of density distribution on this diagram.
The global $g-r$ colour of high-sSFR galaxies (those above the black dashed line in Fig.~\ref{fig:sfms}) becomes bluer with increasing sSFR and does not depend strongly on mass, while the $g-r$ colour of low-sSFR galaxies becomes redder with increasing $M_\star$ and does not depend strongly on sSFR.

\begin{figure*}
    \centering
    \includegraphics[width=0.75\textwidth, trim=0 0 0 0]{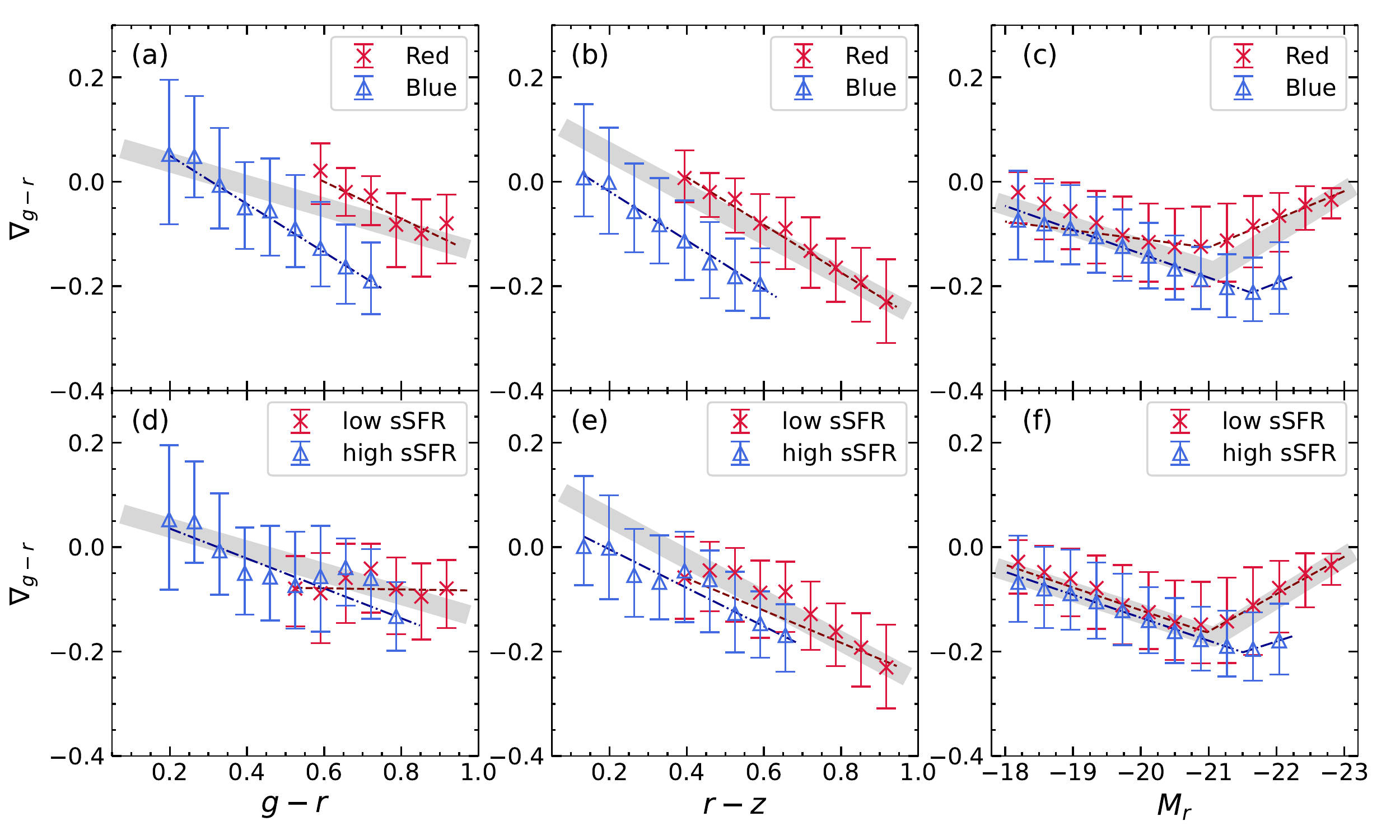}
    \caption{Color gradients in our \pz\/ sample as a function of $g-r$, $r-z$, and $M_r$ for red/blue galaxies (upper panels) and high-/low-sSFR galaxies (lower panels).
    The red/low-sSFR galaxies are labelled with red crosses and the blue/high-sSFR galaxies with blue triangles. The error bars indicate the 25–75 percentile range.
    The red dashed and blue dashed-dotted lines correspond to our fits for red/low-sSFR and blue/high-sSFR galaxies respectively.
    The thick grey line shows our fit to the corresponding $\nabla_{g-r}$ relations for the \sz\/ sample. Best-fit parameter are given in Table~\ref{tab:fitresults}.}
    \label{fig:cggr_color_rsbc_sz}
\end{figure*}

Figs.~\ref{fig:cggr_color_rsbc_sz} plots $\nabla_{g-r}$ against $g-r$, $r-z$, and $M_r$ for red/blue galaxies and low-/high-sSFR galaxies. We only present results for $\nabla_{g-r}$ in following sections because results for $\nabla_{r-z}$ are very similar. The fitted slopes are listed in Table~\ref{tab:fitresults}. Both subsets have trends similar to those seen in the full \sz\/ sample, showing steeper colour gradients for redder galaxies and an inflection in the trend with $M_r$. The thick grey lines in Fig.~\ref{fig:cggr_color_rsbc_sz} show the corresponding trends for the full \sz\/ sample. These mostly overlap the trends for the red/low-sSFR subset, which dominates at the bright end of the sample and shows similar behaviour to the blue/high-sSFR subset at the faint end.

Trends of steeper gradients with redder $g-r$ and $r-z$ global colour are apparent for both the colour-selected and sSFR-selected subsets, but are much clearer and more distinct for the colour-selected subset (compare panels (a) and (d) of Fig.~\ref{fig:cggr_color_rsbc_sz}). This suggests colour-dependent trends are not driven by differences in star formation rate alone. Red galaxies do not always correspond to low-sSFR galaxies and vice versa. In particular, in the `green valley' colour range $0.5 < g-r < 0.7$, a significant fraction of galaxies may be in transition between active star formation and quiescence. These results agree with Fig.~\ref{fig:binned_map}a, including our interpretation of the reddest galaxies as late-type systems with very strong dust reddening.

As in the full sample, gradients for all subsets become steeper with increasing luminosity up to $M_r\sim-21$, above which the trend reverses (Fig.~\ref{fig:cggr_color_rsbc_sz}c and f). However, there are some differences in this behaviour between the subsets. The inflections in the blue/high-sSFR subsets occur at a brighter magnitudes than those in the red/low-sSFR subsets (panels (c) and (f) in Fig.~\ref{fig:cggr_color_rsbc_sz}). This can also be seen in Fig.~\ref{fig:binned_map}a; the gradients for blue galaxies become monotonically steeper with magnitude at a fixed colour, whereas the steepest gradients for red galaxies occur around the inflection point.
In addition, brighter than their inflection points, the trends for blue/high-sSFR galaxies ($a\approx-0.05$ and $-0.04$ $\mathrm{dex}^{-1}$ for $\nabla_{g-r}$) are less steep than those of red/low-sSFR galaxies ($a\approx-0.05$ and $-0.07$ $\mathrm{dex}^{-1}$).
Combined with the fact that the scatter of red/low-sSFR galaxies decreases toward the bright end, these results suggest that the brightest red/low-sSFR galaxies (the tip of the red sequence) have experienced more dry major mergers, which act to wash out colour gradients \citep[][]{Bell2005DryGalaxies}.
Very bright blue/high-sSFR galaxies are almost all close to the red sequence (Fig.~\ref{fig:binned_map}a). That their gradients still flatten with increasing brightness, but at a slower rate compared to redder galaxies, may imply that they have experienced relatively fewer gas-poor merging events. Saturation of passively-evolving populations with very different formation times towards similar (although metallicity-dependent) maximally red colours may also contribute to the flattening of colour gradients in this mass range.

\begin{figure}
    \centering
    \includegraphics[width=1.03\columnwidth, trim=18 50 0 0]{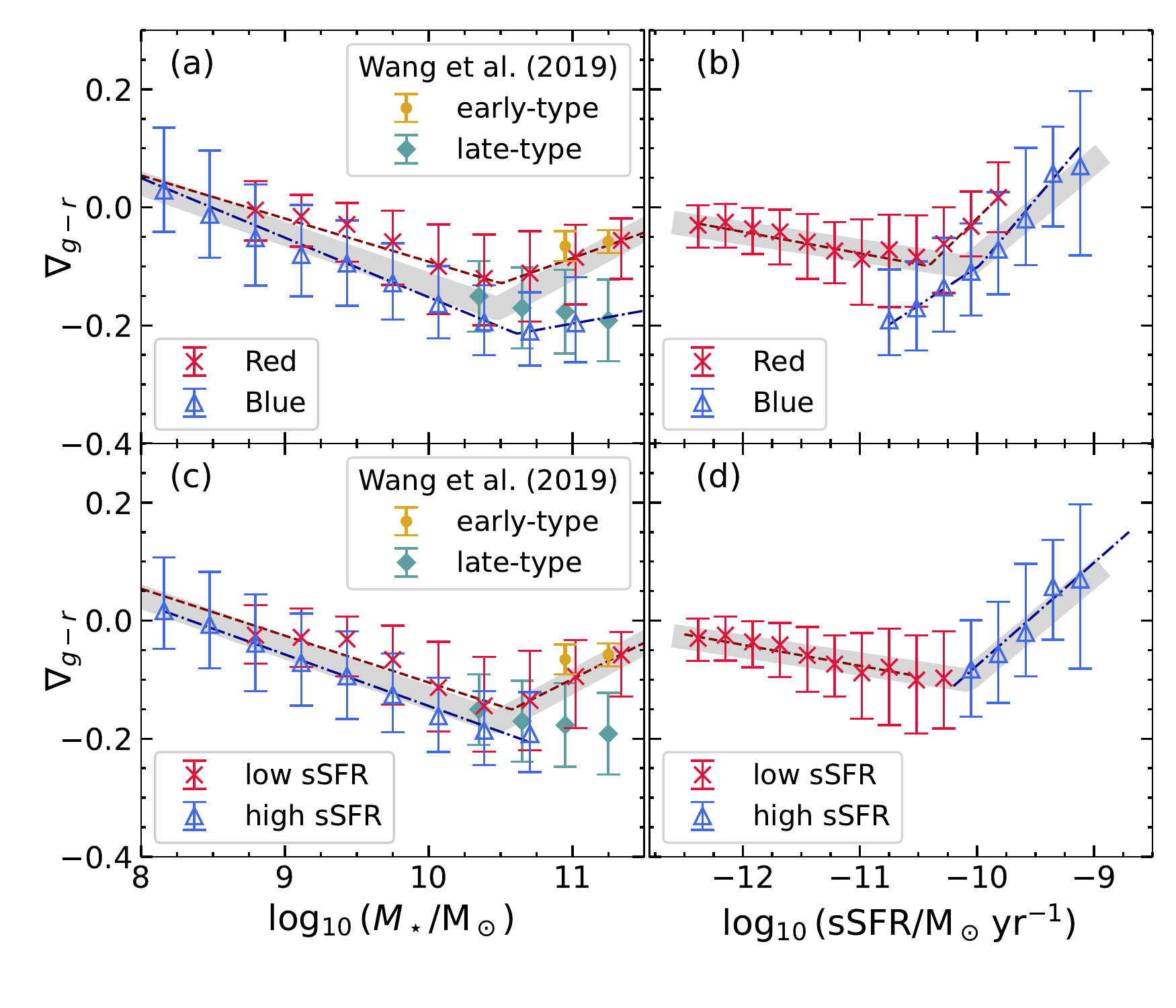}
    \caption{Colour gradients as a function of $M_\star$ and sSFR for red/blue galaxies (upper panels) and high-/low-sSFR galaxies (lower panels). The error bars indicate the 25–75 percentile range for LS galaxies.
    The yellow dots and light blue diamonds show  colour gradients we have computed from the stacked colour profiles in \citet[][]{WangWenting2019}, respectively for high and low concentration galaxies (error bars are estimates of the $1\sigma$ dispersion, see text).
    Other labels are as Fig.~\ref{fig:cggr_color_rsbc_sz}.}
    \label{fig:cggr_sm_rsbc_sz}
\end{figure}

Fig.~\ref{fig:cggr_sm_rsbc_sz} shows the correlations with $M_\star$ and sSFR for red/low-sSFR and blue/high-sSFR galaxies.
In general, the trends for these subsets are similar to those of the entire \sz\/ sample, with the gradients of low-luminosity galaxies becoming steeper and the gradients of high-luminosity galaxies becoming shallower as either variable increases.
Red galaxies dominate the trends with sSFR for the low-sSFR subset.
The scatter decreases toward lower star formation rates.
This is expected, assuming massive ellipticals dominate the sample with reddest colors and and lowest star formation rates. 
Panel (b) in Fig.~\ref{fig:cggr_sm_rsbc_sz} shows there are some  galaxies in our red subset with higher sSFR. 
Visual inspection of a random sample of these galaxies suggest that they comprise late-type galaxies with prominent bulges and disturbed disks, as well as galaxies with elliptical morphology.
These may be galaxies that have recently experienced a merger, but have not yet fully relaxed.
In contrast, images of randomly selected low-sSFR galaxies from our blue subset typically show a red central region surrounded by a disc with spiral structure.
These appear to be typical galaxies, perhaps at the onset of their quenching.

In Fig.~\ref{fig:cggr_sm_rsbc_sz}, we overlay gradient measurements derived from  colour profiles measured in stacked Hyper-SuprimeCam images by \citet[][]{WangWenting2019}, who separated their sample into early and late types by concentration index.
We use the relation of \citet[][]{NedkovaKalina2021} to estimate a median $R_{\mathrm{eff}}$ for each of \citeauthor{WangWenting2019}'s stellar mass bins.
We assume a 20 per cent intrinsic Gaussian dispersion in $R_{\mathrm{eff}}$ and propagate this to the dispersion of colour gradient at a given mass (adding in quaderature to the formal errors of the \citeauthor{WangWenting2019} colours), which we show with errorbars in Fig.~\ref{fig:cggr_sm_rsbc_sz}. The colour gradient trend is flat for early-type galaxies in \citet[][]{WangWenting2019} and steeper for their late type galaxies, consistent with the trends of our red and blue subsets.
For massive galaxies, the \citeauthor{WangWenting2019} data imply average colour gradients are flatter than our results suggest. This may be due to differences between the subsets of galaxies selected by colour or sSFR and those selected by concentration. The concentration criterion of \citeauthor{WangWenting2019} may isolate a more pure sample of galaxies with high Sersic index, particularly around the transition mass. The greater separation between the average gradients of their two subsets would then be consistent with the interpretation that mergers drive the trends we see in this regime.

\section{Results for photometric redshift sample}
\label{sec:pz sample}

\begin{figure}
    \centering
    \includegraphics[width=0.8\columnwidth, trim= 10 0 0 0]{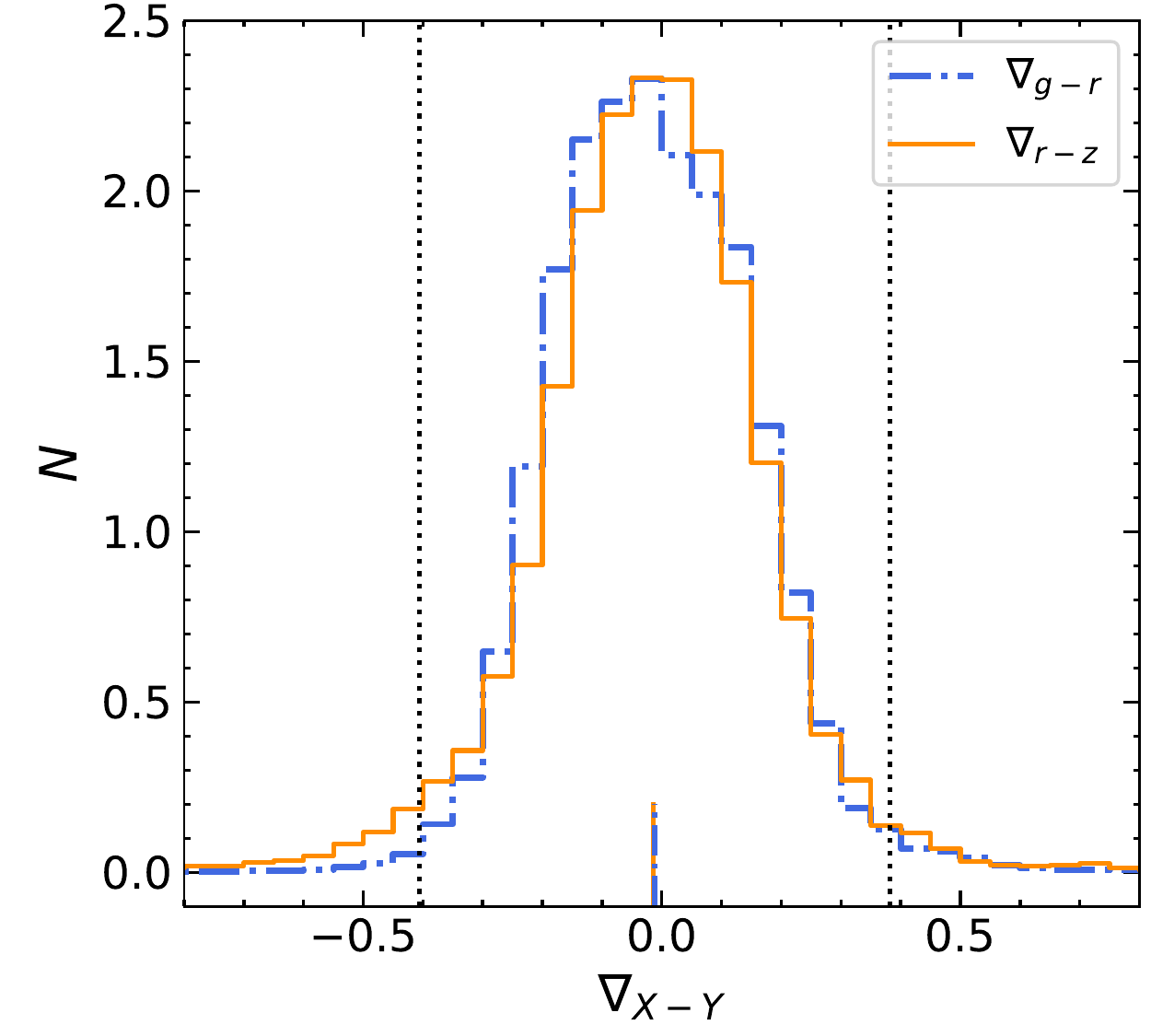}
    \caption{$1/V_{\mathrm{max}}$-weighted PDF of $\nabla_{g-r}$ (blue dot-dashed line) and $\nabla_{r-z}$ (orange solid line) for our \pz\/ sample. Short vertical lines of the same style and colour indicate mean gradients, $\nabla_{g-r} = -0.012$ and $\nabla_{r-z} = -0.014$.
    Standard deviations are $\sigma_{g-r}=0.20$, $\sigma_{r-z}=0.41$. The vertical dotted lines indicate $\pm2\sigma_{g-r}$.}
    \label{fig:cghist_pz}
\end{figure}

In this section, we repeat the analysis in Sec.~\ref{sec:sz sample} for our larger \pz\/ sample.
Fig.~\ref{fig:cghist_pz} shows the $V_{\mathrm{max}}$-weighted distribution of this sample.
The mean values of $\nabla_{g-r}$ and $\nabla_{r-z}$ are $-0.012$ and $-0.014$, respectively.
The negative mean gradient agrees with our result for the the \sz\/ sample; most galaxies are significantly redder at $R_{\mathrm{eff}}/2$ than at $R_{\mathrm{eff}}$.

\subsection{Colour gradient as a function of average colour and $M_r$}
\label{sec:cg_color_pz}

\begin{figure}
    \centering
    \includegraphics[width=0.87\columnwidth, trim=0 0 0 5, clip=True]{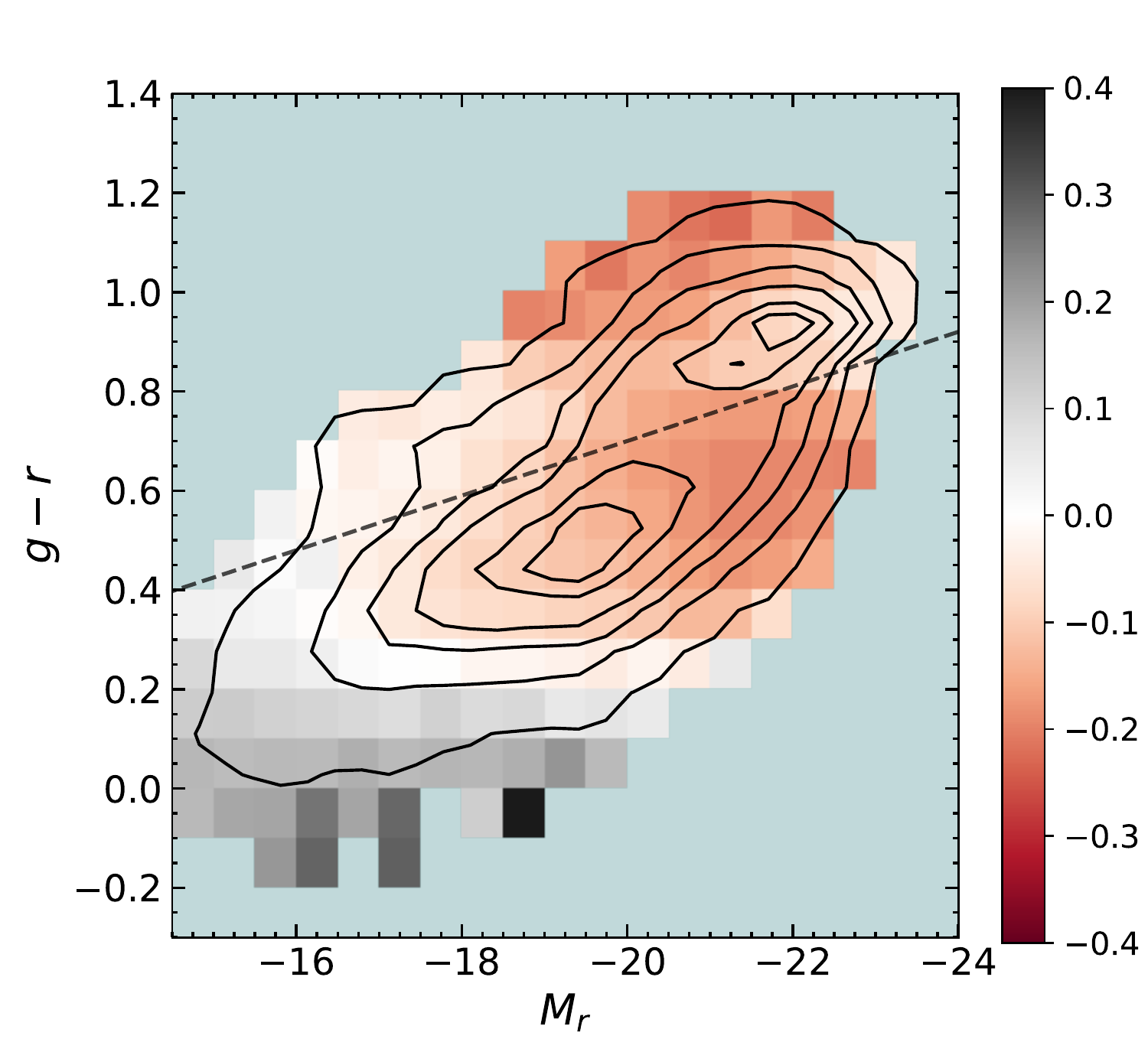}
    \caption{$\nabla_{g-r}$ in bins of $M_r$ and $g-r$ (colour scale) for our \pz\/ sample.
    The contours show the galaxy density and the black dashed line shows our separation between red and blue galaxies.
    The blue dashed line indicates the inflection point in our fit to the 1-dimensional trend of $\nabla_{g-r}$.}
    \label{fig:binned_cg_Mr_gr_photoz}
\end{figure}

Fig.~\ref{fig:binned_cg_Mr_gr_photoz} shows how the mean colour gradient varies across the colour magnitude diagram. 
The mean colour gradient in most bins is negative, as expected (Fig.~\ref{fig:cghist_pz}).
As in our \sz\/ sample, the colour gradient of blue cloud galaxies shows a significant trend, from positive gradients in faint and blue galaxies to negative gradients in bright and red galaxies. In this case the strong positive gradients of the faintest blue galaxies, which are present in greater numbers in this sample, are seen much more clearly. On the red sequence, the steepest colour gradients occur at $M_r\sim-21$ and $g-r\sim1.1$. The gradients are shallower (tending to flat) at the faint and bright ends of the red sequence.

\begin{figure*}
    \centering
    \includegraphics[width=0.75\textwidth, trim=0 20 0 0 ]{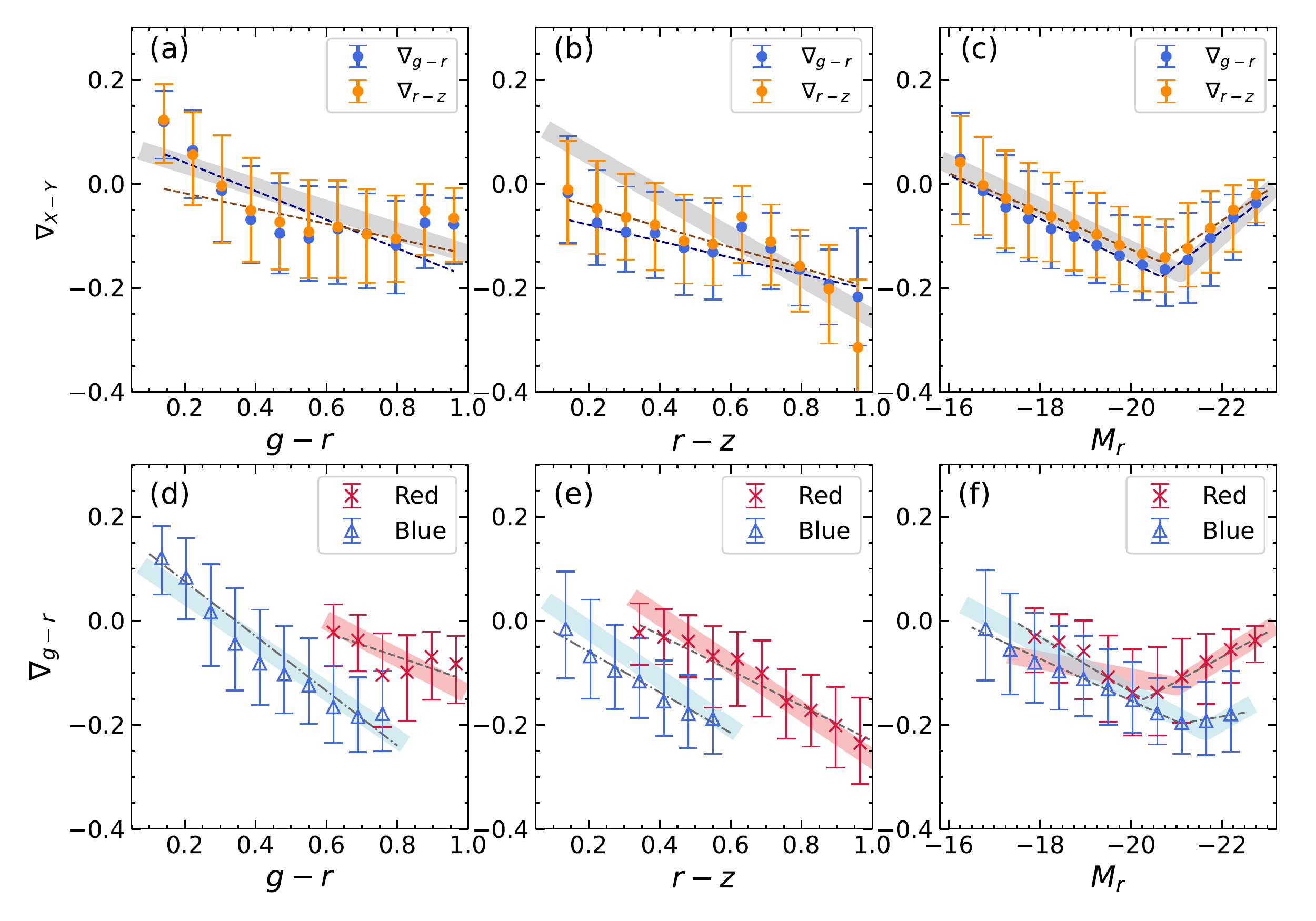}
    \caption{{\it{Upper panels}}: The equivalent of Fig.~\ref{fig:cg_optcolor} for the \pz\/ sample, showing colour gradients as a function of (a) $g-r$, (b) $r-z$ and (c) $M_r$. The thick grey lines show our fits to these trends (for $\nabla_{g-r}$ only) in the \sz\/ sample.
    {\it{Lower panels}}: The equivalent of Fig.~\ref{fig:cggr_color_rsbc_sz} in the \pz\/ sample, showing colour gradients as a function of (d) $g-r$, (e) $r-z$ and (f) $M_r$ for red and blue galaxies. The thick red and blue lines show the fits to red and blue galaxies in the \sz\/ sample. The error bars indicate the 25–75 percentile range.}
    \label{fig:cg_color_pz}
\end{figure*}

In Fig.~\ref{fig:cg_color_pz} we plot one-dimensional trends of colour gradient against $g-r$, $r-z$, and $M_r$ (upper panels).
The results are again broadly consistent with those for our spectroscopic redshift sample.
Compared to the \sz\/ sample, the trend with $g-r$ is less tight and the mean colour gradient at $g-r\sim0.5$ is steeper.
This is mainly due to the mixture of red and blue galaxies, which have distinct trends with $g-r$; more faint blue galaxies are included in our \pz\/ sample.
The slopes of the $g-r$ and $r-z$ relations are [$-0.28$, $-0.16$] for $\nabla_{g-r}$ and [$-0.15$, $-0.20$] for $\nabla_{r-z}$.
We fit a broken linear relation to the trend with $M_r$ and find the inflection point to be $M_r = -20.7$.
The gradients of the fainter subset become steeper with increasing magnitude, with a slope of $0.04$, whereas gradients at magnitudes brighter than the inflection become flatter, with a slope of $-0.07$. 
Both results are consistent with those from our \sz\/ sample.

\subsection{Colour gradients for red and blue galaxies}
\label{sec:cg_rsbc_pz}

Since we do not have star formation rate estimates for our \pz\/ sample, we can only separate red and blue galaxies using their colour and magnitude (Fig.~\ref{fig:cmd}c).
We use the same cut we applied to the \sz\/ sample in Fig.~\ref{fig:cmd} (see also Fig.~\ref{fig:binned_cg_Mr_gr_photoz}).
The number of blue galaxies is almost twice that of red galaxies, because more faint galaxies are included in the \pz\/ sample.
Fig.~\ref{fig:cg_color_pz}d-f show colour gradients as a function of $g-r$, $r-z$, and $M_r$ for red and blue galaxies.

The trends for red and blue galaxies do not show a significant difference between the \sz\/ and \pz\/ samples. The separate trends with $g-r$ and $r-z$ for red and blue galaxies are clearer those for the whole \pz\/ sample.
This supports our speculation above, that the curved relationships shown in Fig.~\ref{fig:cg_color_pz}a and b are due to a mixture of distinct trends for red and blue galaxies.
The gradients of blue galaxies are slightly steeper at a given color than those for red galaxies, but the slope of the relation is similar for both subsets.
We overlay the results of \sz\/ sample in Fig.~\ref{fig:cg_color_pz} as thick red and blue  lines, which demonstrate that the results for samples based on spectroscopic and photometric redshifts broadly agree.

Separating the broken trends of $\nabla_{g-r}$ and $\nabla_{r-z}$ with $M_r$ for red and blue galaxies shows that the inflection point for blue galaxies ([$-21.1$, $-21.1$]) is brighter than that for red galaxies ([$-20.3$, $-20.2$]), again consistent with our earlier results.
Fainter than the inflection, the trend for red galaxies (slope $0.05$) is slightly steeper than that for blue galaxies (slope $0.04$). Brighter than the inflection, as in the \sz\/ sample, we find red galaxies have significantly steeper gradients, and there is less galaxy-to-galaxy scatter. In this region the trends for red and blue galaxies have slopes $-0.05$ and $-0.02$ respectively. 

The highly consistent results we find in our comparison of the \pz\/ sample to the smaller \sz\/ sample indicate that modern photometric redshift techniques are accurate enough to recover the clear correlations between colour gradients and other galaxy properties that we have identified, at least in the range of magnitude where the two samples overlap. This gives us confidence that the extension of trend we see towards the fainter magnitude limit of the \pz\ sample is robust. Gradients flatten for dwarf galaxies and even become positive at the faint end of our sample ($M_{r}\sim-16$). We comment further on this interesting result in Section~\ref{sec:discussion}. 

\section{Discussion}
\label{sec:discussion}

In this section, we discuss our results in the context of previous work on stellar population gradients and recent results from the MaNGA survey, and the significance of our finding that the lowest mass galaxies in our sample typically have positive colour gradients.

\subsection{Colour gradients and galaxy formation}
\label{sec:origin_cg}

The consensus view of galaxy formation in the $\Lambda$CDM cosmogony is that low-mass galaxies form in through a `quiescent' process in which a reservoir of cold ISM gas builds up gradually through radiative cooling of a hydrostatic halo \citep{White1978CoreClustering} and is converted into stars at a steady rate of $\sim1$-$10\,\mathrm{M_{\sun}\,yr^{-1}}$, regulated by supernova feedback \citep{Larson1974DynamicalGalaxies,White1991GalaxyClustering,Kauffmann1993TheHaloes,Cole2002HierarchicalFormation}. In this regime, galaxies grow "inside out" as gas with higher angular momentum accumulates, and eventually exceeds the effective density threshold for instability to star formation, at ever-larger radii \citep[e.g.][]{Larson1976,Fall1980,Mo1998TheDiscs,DeLucia2007TheGalaxies, Dekel2009FormationSpheroids}. Observations imply the conversion of baryons into stars must be slow and inefficient in massive late-type galaxies \citep[e.g.][]{Mo1998TheDiscs,Guo2010HowHaloes,Aumer2013a}. Star formation in the outskirts of these galaxies is expected to be dominated by metal poor gas accreted from the cosmic web at late epochs, rather than by chemically enriched, initially low angular momentum gas evacuated from their central regions by supernova-driven winds \citep[e.g.][]{Yates2014DilutionRate}. The build-up of central stellar mass may stabilize  gas that remains (or subsequently flows to the centre) against further star formation \citep{Martig2009MorphologicalRed}. Over time, although central star formation may continue at a reduced level, the centres of galaxies become dominated by populations that formed early and rapidly, and are passively evolving at the present day. Occasional central starbursts may be triggered at late times by gas rich mergers and large-scale disk instabilities \citep[e.g.][]{Bower2006BreakingFormation,Guo2010FromCosmology}.
When galaxies reach a critical stellar mass \citep[$M_\star>10^{10.5} \, \mathrm{M_\odot}$; see][]{Kauffmann1996DetectionGalaxies,DeLucia2006TheGalaxies,Guo2008GalaxyCosmology}, now more often understood as a critical host halo mass \citep[e.g.][]{Bower2006BreakingFormation,Bower2017TheEnd}, feedback from AGN combines with intrinsically long radiative cooling times to suppress the inflow of gas from the halo. The result is a rapid and permanent decline in star formation activity, often referred to as `quenching' \citep[e.g][]{Pipino2006TheGalaxies,Croton2006TheGalaxies}. Above the transition mass, stellar mass growth is driven mostly by low mass ratio mergers between similarly old, passively-evolving galaxies \citep{DeLucia2006TheGalaxies, DeLucia2007TheGalaxies}.

This well-established picture makes readily testable predictions for metallicity and age gradients within galaxies, which are generally supported by recent observations.
For example, \citet[][]{Parikh2021SDSS-IVGalaxies} show that late-type galaxies tend to have negative metallicity and age gradients, which are consistent with inside-out formation scenarios \citep{Tinsley1978ChemicalDisks}.
Negative age and metallicity gradients for late-type galaxies are also reported in \citet{GonzalezDelgado2015TheGalaxies} and \citet[][]{2Goddard2017SDSS-IVType}.
For early-type galaxies, negative metallicity gradients have been reported by many studies \citep[e.g.][]{Davies1993Line-strengthGalaxies,Kobayashi1999GradientsGalaxies,Saglia:2000,Spolaor2009THEGalaxies,Tortora:2010,Parikh2021SDSS-IVGalaxies}.
In contrast to the consensus for late types, different studies come to divergent conclusions about typical age gradients in early-type galaxies  \citep[e.g.][]{Tortora:2010,Koleva2011AgeSequence,GonzalezDelgado2015TheGalaxies,Zibetti2020InsightsCALIFA,Parikh2021SDSS-IVGalaxies}.
These differences may be explained by the results of \citet[][]{Zibetti2020InsightsCALIFA}, who found U-shaped age profiles in early-type galaxies. They suggest that observed age gradients could range from positive to negative, depending on the range of radii over which they are measured.
The negative metallicity gradients and relatively flat age gradients in early types have been interpreted as a sign of "outside-in" formation \citep[e.g.][]{Parikh2021SDSS-IVGalaxies}, in which star formation is suppressed earlier or more effectively at larger radii \citep{Pipino2006TheGalaxies}. Mergers are increasingly important for stellar mass growth and the transition from disk to elliptical morphology at higher masses. High mass ratio mergers generally preserve radial trends in stellar populations, but violent relaxation in low mass ratio mergers may wash out gradients over several successive events \citep[e.g.][]{White1980MixingMergers,GaoLiang2004,Bell2005DryGalaxies,Kim2013Optical-NearGalaxies}. This effect has been discussed in the context of observations, for example, by \citet{Pastorello2014TheRadii} and \citet{Hirschmann2015TheRadii}.

At face value, the data we present here for a large sample of colour gradients in low-redshift galaxies appear consistent with the smaller samples of age and metallicity measurements cited in the previous paragraph, and hence with the current understanding of galaxy formation outlined at the start of this section. At the faint limit of our sample, field dwarf galaxies are expected to be young and associated with dark matter halos that are growing rapidly at the present day in relatively low density regions of the cosmic web \citep{Efstathiou1995, Tully:199612, Mo1998TheDiscs}. These galaxies are also expected to be uniformly metal poor, most simply because they are young. Both factors imply blue colours. If the bulk of the stellar mass in these galaxies has formed recently, their internal dispersion in age should be small, and their oldest and youngest stars should occupy roughly the same region of phase-space at similar density. Even if a significantly older population is present, the strong mass weighting towards recently formed stars should produce an approximately uniform colour distribution. There is therefore little reason to expect strong gradients in either age or metallicity within the half-mass radius in these galaxies. These simple expectations are consistent with our results, which show that bluer and less massive galaxies have systematically flatter colour gradients.

\begin{figure}
    \centering
    \includegraphics[width=\columnwidth, trim=0 15 0 0]{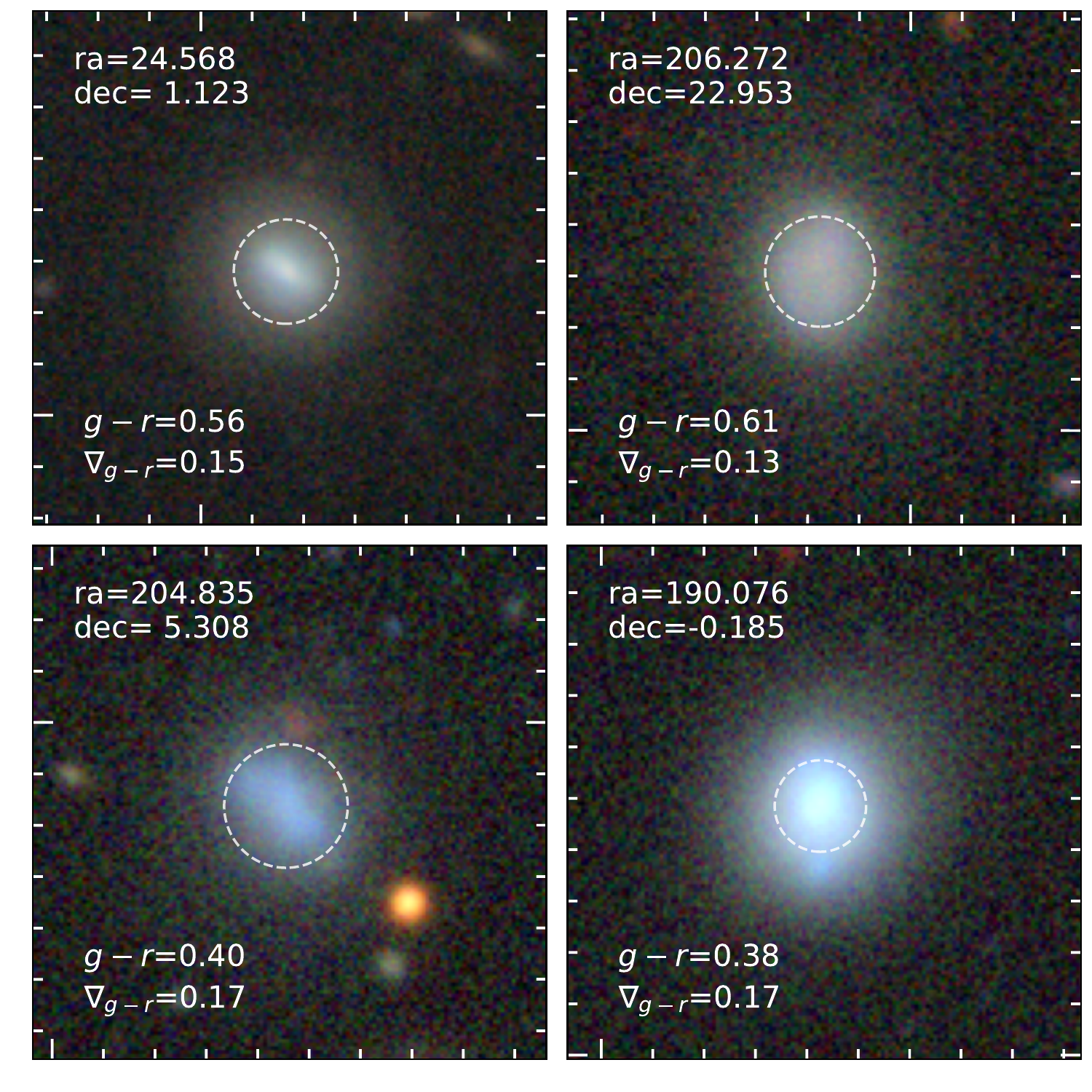}
    \caption{Examples of galaxies with positive colour gradients. The top row shows galaxies with redder average colour, and the bottom row galaxies with bluer average colour. All four galaxies have similar colour gradients. The white dashed circles indicate $R_{\mathrm{eff}}$. Each panel is $36\arcsec\times36\arcsec$.}
    \label{fig:poscg_ex}
\end{figure}
We also find a significant fraction of positive colour gradients, 
particularly at the lowest masses in our sample; our results based on photometric redshifts suggest that galaxies of $M_{r}\approx-16$ or $M_{\star} \lesssim 10^{8} \,\mathrm{M_{\odot}}$ are typically bluer at $R_{\mathrm{eff}}/2$ than at $R_{\mathrm{eff}}$. This is not readily apparent from images alone, partly because the gradients are relatively weak compared to those of massive late types, and partly because the colour scaling of typical three-colour images makes these galaxies appear appear blue or grey throughout. Some representative examples are showing in Fig.~\ref{fig:poscg_ex}. We discuss similar findings in previous work and broader context of this result in Section~\ref{sec:dw}.

For galaxies of intermediate mass, up to $M_\star < 10^{10.5} \, \mathrm{M_\odot}$, we find that colour gradients become steeper towards redder average colour and with increasing $M_\star$.
This is compatible with the well-established concept of inside-out formation driven by cooling and subject to a minimum gas density threshold for star formation, enhanced by the suppression of star formation by gas depletion and dynamical effects in the cores of galaxies  \citep{Tinsley1978ChemicalDisks}. Inside-out formation naturally implies steeper (negative) colour gradients for older and larger galaxies, arising from the combination of negative age and metallicity gradients. In $\Lambda$CDM models, higher stellar mass (for late type galaxies) is correlated with older ages, larger sizes and higher central metallicities \citep[e.g.][]{Mo1998TheDiscs, Shen2003size}, giving rise to a steepening trend of colour gradients with mass. The trend with average colour reflects the fact that star formation rates peak at earlier times for more massive galaxies associated with earlier dark matter halo collapse times. The precise slopes of these trends in a given band will depend on the details of cooling, star formation and feedback.

For very massive galaxies with $M_\star > 10^{10.5} \, \mathrm{M_\odot}$, the trend of colour gradient with $M_\star$ (and $M_r$) reverses, such that gradients are almost flat for the most massive systems. In this mass range, we suggest that both major mergers and passive evolution following quenching may contribute to this effect. Massive galaxies are expected to experience more low mass ratio merging events, giving rise both to rapid size growth and elliptical morphology \citep[e.g.][]{Naab:2009,Keenan2014Evolution1}.
A succession of gas-poor (`dry') major mergers could redistribute stellar populations on the scale of $R_{\mathrm{eff}}$ \citep[e.g.][]{White1980MixingMergers,Kim2013Optical-NearGalaxies}.
This explanation is consistent with our finding that the average gradient flattens more quickly for red galaxies than blue galaxies at high stellar masses (e.g.\ Fig.~\ref{fig:cggr_sm_rsbc_sz}), because the red population is dominated by ellipticals, which are thought be the result of dry major mergers. The decreasing scatter around the average gradient for high-mass galaxies supports this interpretation \citep[e.g.][]{DeLucia2006TheGalaxies}. It has also been suggested \citep[e.g.][]{Pipino2004} that, following the suppression of cooling (for example by AGN feedback, or environmental effects acting on satellite galaxies in clusters) star formation activity can persist longer in the centres of massive galaxies than in their outskirts, leading to an `outside in' progression of star formation that may be indirectly correlated with a transition to elliptical morphology. This scenario could also flatten the steep negative gradients established before the onset of AGN feedback. 

An alternative explanation for the trend at high mass is simply that, in a passively evolving galaxy, a fixed age gradient results in a flatter colour gradient for larger absolute age. For example, a disk with a steep age gradient (arising from the contrast between an older, passive central region and an active outer region) will have a steep negative (red to blue) colour gradient, as long as the star formation continues. If the outer region is quenched, the younger outer population will redden more rapidly, flattening the colour gradient.

With this interpretation, if steep age gradients are established in galaxies by the time they reach the transition mass, and if their stars are not redistributed after quenching, then the trend we observe implies that star formation is suppressed relatively earlier in relatively more massive galaxies. This seems plausible, and would be equally consistent with differences between the colour gradient trends for red and blue galaxies. It is interesting that the analysis of IFU spectra by \citet{Parikh2021SDSS-IVGalaxies} implies flat age gradients and steep metallicity gradients for massive ETGs, and much steeper age gradients for late types in the same mass range. That result appears more consistent with a merger-driven explanation for the colour gradient trends. However, as we discuss further below, at least one other study based on the same MaNGA data explains flat age gradients in massive early types through a combination of steep age gradients and a rapid increase in dust reddening at larger radii.

In summary, although we cannot attribute trends in colour gradients directly to metallicity or age gradients based on our data alone, our results appear at least qualitatively consistent with inferences from spectroscopic age and metallicity observations \citep[][]{Tortora:2010,2Goddard2017SDSS-IVType,Parikh2021SDSS-IVGalaxies} and hence with the galaxy formation scenarios that have been proposed to explain them. As we discuss in the following section, however, the latest generation of IFU surveys now allows for more robust constraints on the connection between optical colour gradients and underlying population gradients.

\begin{figure*}
    \centering
    \includegraphics[width=\textwidth, trim=0 30 0 0]{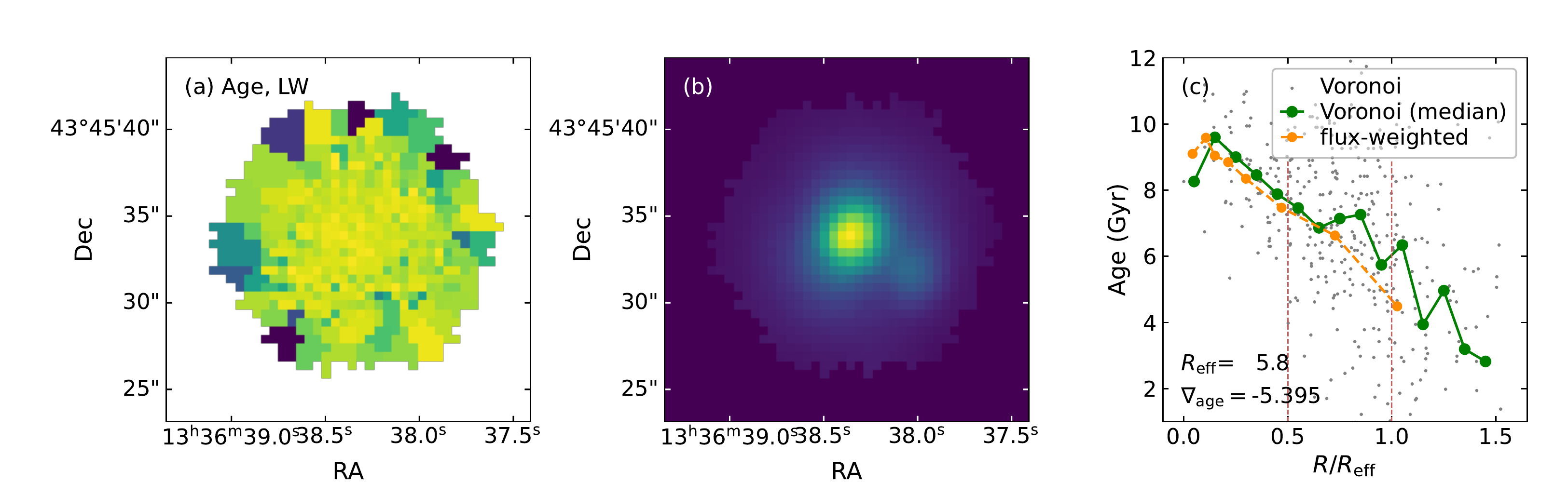}
    \caption{An example of the data used in our comparison with the FIREFLY reduction of MaNGA, for an early type galaxy with $M_\star = 10^{11.1} \, M_\odot$  and redshift $z=0.062$: (a) Light-weighted age map from a FIREFLY MaStar  datacube, (b) corresponding $r$-band image fromt the MaNGA datacube, and (c) age profile. The grey dots show the age and corresponding radius of each individual Voronoi cell.
    The green dots are the medians of the Voronoi datapoints in bins of 0.1 $R/R_\mathrm{eff}$. The orange dots show mean ages dervied from individual pixels, flux-weighted according to the $r$-band image, using the LS photometric apertures. The vertical dashed lines label 0.5 and 1$R_\mathrm{eff}$.}
    \label{fig:manga_example}
\end{figure*}

\subsection{Comparisons to integral field spectroscopy}
\label{sec:manga}

Template fits to spectra \citep[or to well-sampled broad-band SEDs, e.g.][]{Abdurrouf2022Pixedfit} can break degeneracies between the effects of age, metallicity and dust on the colour of stellar populations. In principle, spatially resolved spectroscopy can be used to measure stellar population gradients directly, and therefore help both to explain the colour gradient trends we observe and to enable more straightforward comparisons with simulations. Such data are currently only available for a relatively sparse sampling of the galactic CMD. The current state of the art in resolved spectroscopy surveys is MaNGA \citep[][]{Bundy2015Manga,Yan2016MangaOverview}, datacubes from which have been used to construct resolved age and metallicity maps for a sample of nearby galaxies  \citep[e.g.][]{Zheng2017SDSS-IVGalaxies,Li2018SDSS-IVPlane,Neumann2011manga, Neumann2022firefly}. The MaNGA sample was selected to be broadly representative of the galaxy population by requiring a flat number density with respect to $M_i$. The MaNGA sample spans a mass range of $5\times10^8 \leq M_\star \leq 3\times10^{11 \, M_\odot}$. It provides spectroscopic coverage to 1.5 $R_\mathrm{eff}$ for the whole sample and 2.5 $R_\mathrm{eff}$ for a subset \citep{Law2015MangaObserve,Wake2017MangaDesign}.

A thorough comparison between these data and our results is beyond the scope of this paper. However, we can make a simple face-value comparison between the colour gradients we measure for galaxies in the MaNGA sample and the resolved ages and metallicities for the same galaxies provided by the public MaNGA data products. This provides some quantitative insight into the qualitative discussion in the previous section. For this purpose, we have matched our dataset to the publicly available value-added data products \footnote{\url{hhttps://data.sdss.org//datamodel/files/MANGA_FIREFLY/FIREFLY_VER/manga_firefly.html}} based on the \firefly{} spectral energy distribution fitting code (\citealt{Wilkinson2015Manga,Wilkinson2017firefly}, see also \citealt{2Goddard2017SDSS-IVType} for a summary of the application of \firefly{} to MaNGA spectra). \firefly{} is a $\chi$-squared minimisation code that iteratively seeks a best-fitting combination of single-burst stellar population templates, attempting to map degeneracies between parameters robustly. Dust extinction is included in fitting process, using a novel approach that does not assume a specific form for the attenuation curve \citep{Wilkinson2017firefly}.

We find 3,139 galaxies in common (from a total of 10,010 unique galaxies in MaNGA overall and 9,816 with suitable \firefly{} data). The majority of MaNGA targets that are not included in our matched sample (5048) have $R_\mathrm{eff} > 6$\arcsec. A further 1097 have $R_\mathrm{eff} < 3\arcsec$, 408 have $z<0.008$ and 694 have $z>0.1$ (these excluded sets have some overlap). \firefly{} results are available for two alternative choices of stellar population libraries: the M11-MILES models \citep{Maraston2011Miles} and the MaStar  SSP models \citep{Maraston2020Hill}. The M11-MILES models are based on MILES stellar libraries \citep{Sanchez2006miles} with wavelength coverage of 3500 to 4730\angstrom, age ranges of 6.5 Myr to 15 Gyr and  [Z/H] metallicity ranges of $-2.25$ to 0.35. 
The MaStar  SSP models, based on the MaStar stellar libraries \citep{Yan2019Mastar}, adopt the same stellar population models and physical assumptions as M11-MILES, but with larger parameter ranges (3 Myr to 15 Gyr for age and $-2.25$ to 0.35 for metallicity [Z/H]) and broader wavelength coverage (3600 to 10,300\angstrom).

The \firefly{} catalogues report gradients measured between zero radius and 1.5 $R_\mathrm{eff}$, rather than 0.5 to 1 $R_\mathrm{eff}$. We therefore recompute the gradients over the latter range, to match our colour gradient measurements. Fig.~\ref{fig:manga_example} illustrates our approach. We use the \firefly{} age and metallicity maps, which are provided on the same pixel grid as the reconstructed $r$-band image. The parameter values were fit from spectra co-added in Voronoi cells, which combine multiple pixels in these maps, such that blocks of neighbouring pixels may have identical stellar population parameters. We derive age and metallicity profiles (hence gradients) from the maps using the same LS apertures we used to compute colour gradients. In each aperture, we compute the $r$-band flux-weighted average age or metallicity. We have confirmed this procedure results in gradients consistent with those reported in the \firefly{} data products \citep[see also e.g. section 3.2 of][]{2Goddard2017SDSS-IVType} when we use the same range of radii\footnote{We note that \citet{2Goddard2017SDSS-IVType} and \citet{Neumann2022firefly} correct their gradients for the ellipticity of the source, whereas we report values for circular apertures.}.

\begin{figure}
    \centering
    \includegraphics[width=\columnwidth, trim= 0 50 0 0]{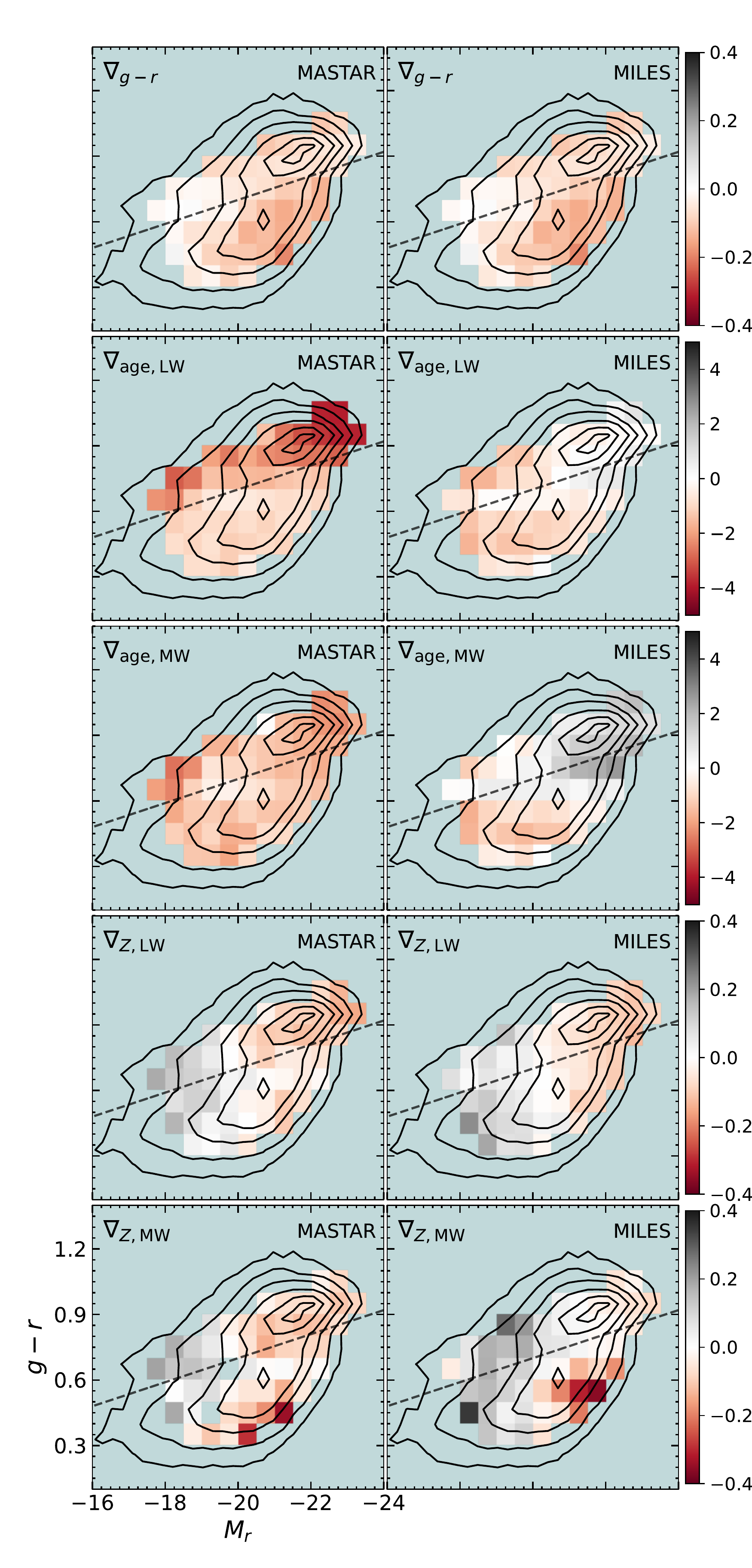}
    \caption{Average gradients binned in the plane of $M_r$, $g-r$ for the MaStar (left) and MILES (right) results for the subset of our sample matched to the \firefly{} dataset. From top to bottom, rows show our $\nabla_{g-r}$ measurements from LS photometry, light-weighted $\nabla_\mathrm{age}$, mass-weighted $\nabla_\mathrm{age}$,  light-weighted $\nabla_Z$ and mass-weighted $\nabla_Z$. The contours show the joint density distribution of galaxies in our \textit{sz} sample.}
    \label{fig:firefly_gradients}
\end{figure}

\begin{figure}
    \centering
    \includegraphics[width=\columnwidth, trim=0 35 0 0]{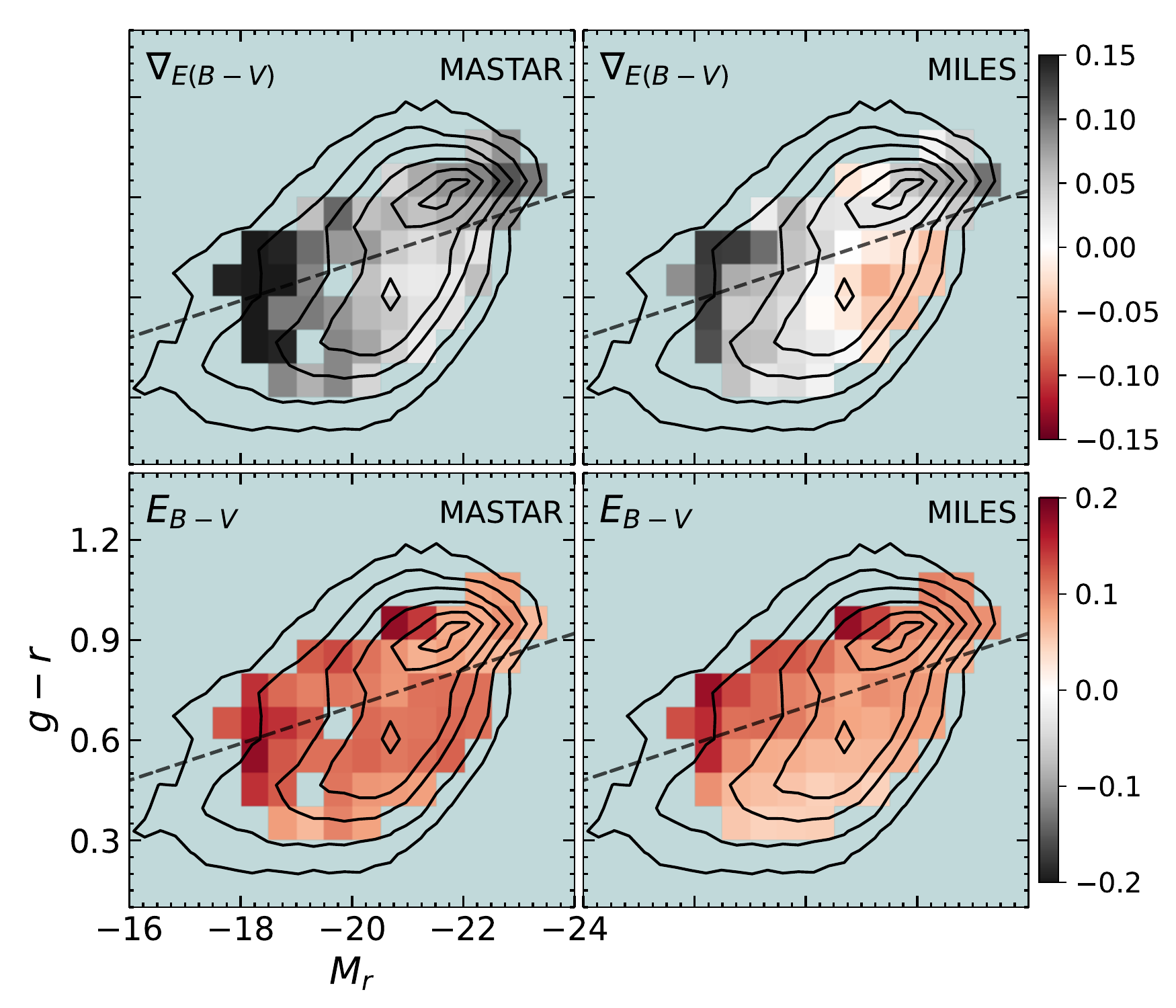}
    \caption{Average $\nabla_{E(B-V)}$ and $E_{B-V}$ at $R_\mathrm{eff}$ binned in the plane of $M_r$, $g-r$ for MaStar (left) and MILES (right) results fo the subset of our sample matched to the \firefly{} dataset. The contours show the joint density distribution of galaxies in our \textit{sz} sample.}
    \label{fig:firefly_dust}
\end{figure}

In Fig.~\ref{fig:firefly_gradients} we plot average $\nabla_{g-r}$, age gradients ($\nabla_\mathrm{age}$), and metallicity gradients ($\nabla_Z$) across the colour magnitude diagram for the subset of our sample matched to the \firefly{} catalogues.  Fig.~\ref{fig:firefly_dust} shows equivalent plots of the inferred dust reddening gradient $\nabla_{E(B-V)}$ as well as the absolute value of $E_{B-V}$ at 1$R_\mathrm{eff}$. There are clear differences between MILES and MaStar  results, especially for age gradients and reddening gradients. The trends within each model and the differences between models are discussed by \citet{2Goddard2017SDSS-IVType} and \citet{Neumann2022firefly}, but Fig.~\ref{fig:firefly_gradients} emphasizes the ways in which they may lead to somewhat different interpretations of colour gradient trends.

The light-weighted MaStar results suggest a very different explanation for the flat colour gradients along the red sequence, compared to the hypothesis of intrinsically flat age gradients sketched in the previous section. They infer steeper negative age gradients for bright red sequence galaxies, together with weakly negative age gradients. At face value, these results suggest massive early types should have steep negative (red to blue) colour gradients. However, the MaStar  fits appear to compensate for this by inferring strong positive dust reddening gradients, much stronger than for late types of similar magnitude (which are found to have more reddening overall, but a flatter reddening gradient). The mass-weighted MaStar  results differ only in that the inferred age gradients are shallower, suggesting that the steep colour gradients arise from intrinsically young populations.

The MILES results for massive early types, in contrast, imply flat light-weighted age gradients (which arise from steeper \textit{positive} mass-weighted gradients, i.e. older stars at larger radii) together with weak negative metallicity gradients and more moderate positive reddening gradients. These findings are closer to the interpretation sketched in the previous section, although the degeneracy between age and dust still seems to be significant for the brightest galaxies. 

There is less difference between the stellar populations inferred by the two variants for blue cloud galaxies. In both variants, light-weighted age gradients are weakly negative or flat. The MaStar results imply the mass-weighted age gradients are slightly steeper (and have no apparent trend with magnitude or colour) whereas the MILES results imply inherently flat age gradients, except fainter and bluer galaxies. The steepening of colour gradients with magnitude across the blue cloud is therefore implied by the MILES fits to be the consequence of a metallicity gradient combined with \textit{positive} reddening gradients in brighter galaxies, and by the MaStar fits to be the consequence of intrinsic negative age gradients modulated by \textit{negative} reddening gradients in fainter galaxies.

Finally, the faintest galaxies in the matched sample ($M_{r} \gtrsim -19$) show interesting and perhaps unexpected trends in both the MaStar and MILES results, for both the red sequence and blue cloud. On the red sequence, faint galaxies are found to have steep negative age gradients and very steep dust gradients in both variants. Unlike the massive early types in the MaStar results, both variants infer positive metallicity gradients at low mass. This would reinforce the negative colour gradient implied by the steep age gradient, perhaps explaining the need for even steeper dust reddening gradiens and higher extinction overall. A similar effect is seen for the faintest blue cloud galaxies. We note that our matched MaNGA sample does not extend to the faintest galaxies in our full sample, for which we see, on average, stronger positive colour gradients.

From our exploration of the two \firefly{} variants, together with the separate study by \citet{Parikh2021SDSS-IVGalaxies}, we conclude that state-of-the-art inferences of stellar population parameters from MaNGA spectroscopy do not yet provide a unique interpretation of the large scale trends we find in galaxy colour gradients.

\subsection{Positive colour gradients in field dwarf galaxies}
\label{sec:dw}

Fig.~\ref{fig:fig2cghist} shows that the majority of galaxies in our $z<0.1$ sample have negative colour gradients.
However, positive colour gradients are increasingly common below $M_{\star}\lesssim10^{9}\,\mathrm{M_{\odot}}$ ($M_V \gtrsim -17$) and typical at $M_{\star}\lesssim10^{8}\,\mathrm{M_{\odot}}$, as shown in Fig.~\ref{fig:cg_optcolor} and \ref{fig:cg_sm_sfr}. These magnitudes and stellar masses correspond to conventional dividing lines between `dwarf' galaxies and other late types, which also (loosely) correspond to a transition between coherent thin disks and irregular or spheroidal morphology \citep[e.g.][]{Tammann:1994eso,Tolstoy2009Star-FormationGroup}. It is well established that many dwarf galaxies in the Local Group and other very nearby samples have positive colour gradients \citep[e.g.][]{Tully:199612,Jansen2000}. Our results follow those of \citet{Tortora:2010} in confirming this phenomenon for a large, relatively unbiased sample of field dwarfs, and in showing that it is consistent with the extrapolation of the well-defined trend evident at higher stellar mass. To put this result in context, we give a brief summary of expectations for population gradients in dwarf galaxies.

Long-lived thin discs in dwarf galaxies are disfavoured by recent halo assembly times, short cooling timescales, the high efficiency of supernova-driven gas expulsion in shallow gravitational potentials, and circular speeds comparable to random velocities in the dense star-forming gas \citep[e.g.][]{Mo1998TheDiscs}. These factors also act to limit the overall star formation efficiency in dwarfs and so favour sporadic and relatively extended star formation histories, with star formation rates rising gradually towards the present day. 
The lack of centrifugal support means stars form close to the centre of the galaxy at all epochs; the less gas is involved in a starburst, the more concentrated the resulting population is likely to be. 

Resolved stellar population studies of dwarf galaxies have found evidence for a trend towards positive age gradients at lower luminosity \citep[][]{Koleva2011AgeSequence}. Although Local Group dwarfs have diverse stellar populations, it is clear that younger stars are typically more centrally concentrated in most systems \citep[e.g.][]{Harbeck2001, Hidalgo2013, Albers2019}. These galaxies are particularly significant because their internal structure and star formation histories can be studied in detail \citep[e.g.][]{Tolstoy2009Star-FormationGroup}. They provide an archaeological record of the baryon cycle in dwarf galaxies that is vitally important for tests of the nature of dark matter on sub-galactic scales \citep[e.g.][]{Walker2011, Bullock2017Small-ScaleParadigm, McGaugh2017TheGalaxies}. 

These observations above imply that the typical concentration of new star formation increases with time, or that older populations become more diffuse; in practice both seem likely. The early history of these systems may include an initial phase of rapid cooling and intense star formation that leaves significant mass in a relatively extended population (reminiscent of the \citet{Eggen1962} model). 
Older stars may diffuse to larger radii  s the host halo potential evolves \citep[e.g.][]{LeBret2017ParticleDynamics,Cooper2017ComparingHalo}. This diffusion may be enhanced by mergers \citep[][]{Benitez-Llambay:2016vx}. Negative metallicity gradients may result in negative colour gradients if early starbursts result in stronger self-enrichment compared to less intensive, more diffuse star formation at later epochs \citep[e.g.][]{Larson1974DynamicalGalaxies, Mori1997TheSupershell, Tortora:2010, Mercado2021AGalaxies}. 

The current generation of simulations in this regime support this theoretical picture. For example, \citet{Graus:2019} found that positive age gradients were ubiquitous in a sample of 26 isolated dwarf galaxies ($10^{5} < M_\star < 10^{9} \, \mathrm{M_{\odot}}$) in the FIRE-2 hydrodynamical simulation suite. \citet{Mercado2021AGalaxies} found the same simulated galaxies have generally negative metallicity gradients, as the result of expansion of older metal-poor populations (driven by gas expulsion) and fall-back of metal-rich ejecta to fuel central star formation at late times. This mixture of positive age gradients and negative metallicity gradients suggests relatively flat colour gradients. Other studies have explored how environmental effects may give rise to gradients in satellite galaxies, as observed around the Milky Way and M31, and in galaxy clusters \citep[e.g.][]{Genina2019,Pfeffer2022}. Our results, and those of \citet{Tortora:2010}, imply that strong stellar population gradients in dwarf galaxies can be established in isolation, as the generic outcome of the baryon cycle in shallow potentials.

\begin{figure}
    \centering
    \includegraphics[width=0.8\columnwidth, trim=30 15
    0 0]{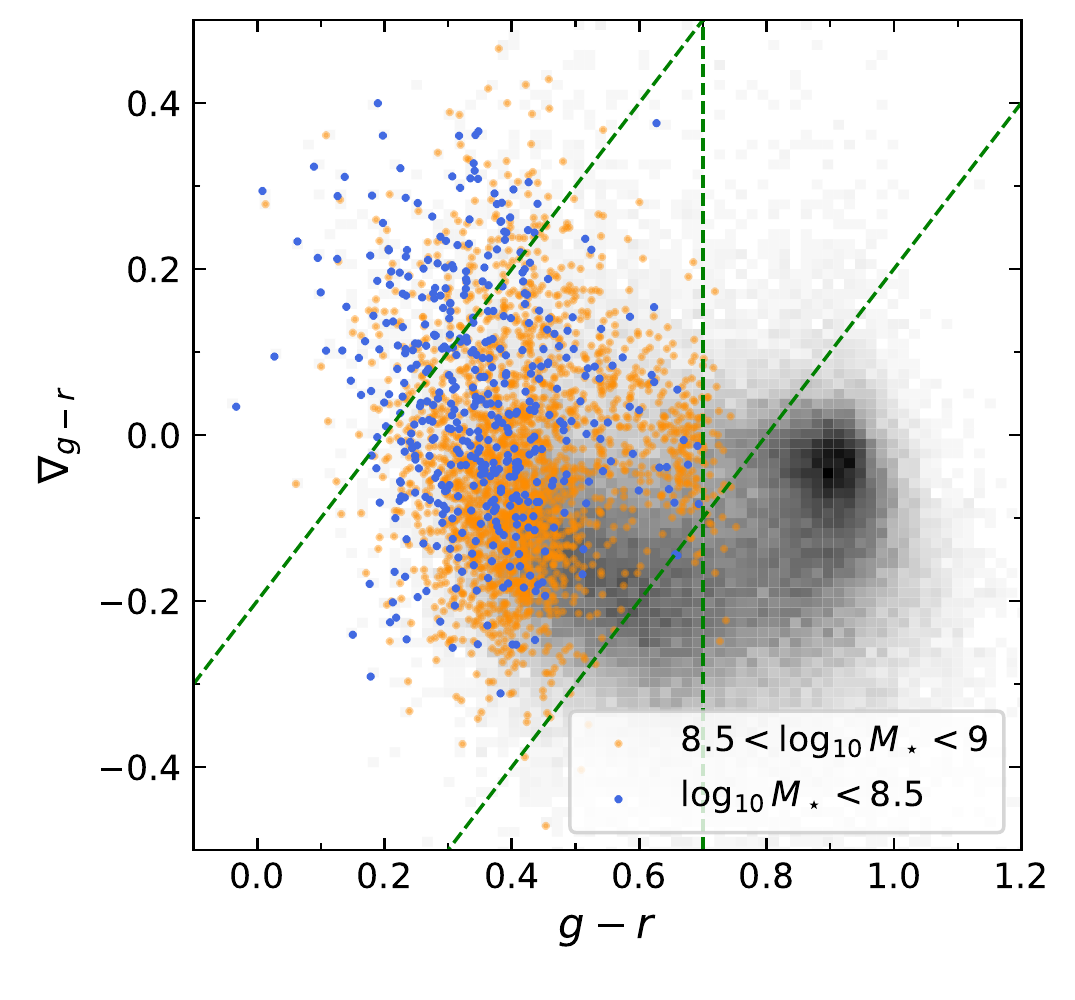}
    \caption{Distribution of galaxies in the plane of average colour and $\nabla_{g-r}$, (without $1/V_\mathrm{max}$ weighting).
    The orange dots correspond to galaxies with $10^{8.5} < M_\star < 10^9 \, \mathrm{M_\odot}$ and the blue dots to galaxies with $M_\star < 10^{8.5} \, \mathrm{M_\odot}$.
    The vertical green dashed line is drawn at $g-r=0.7$.
    The diagonal lines show cuts at $\nabla_{g-r} = (g-r) - a$, with $a=0.2$ (upper line)  and $a=0.8$ (lower line).}
    \label{fig:dw_distribution}
\end{figure}

\begin{figure*}
    \centering
    \includegraphics[width=\textwidth, trim= 0 40 0 0]{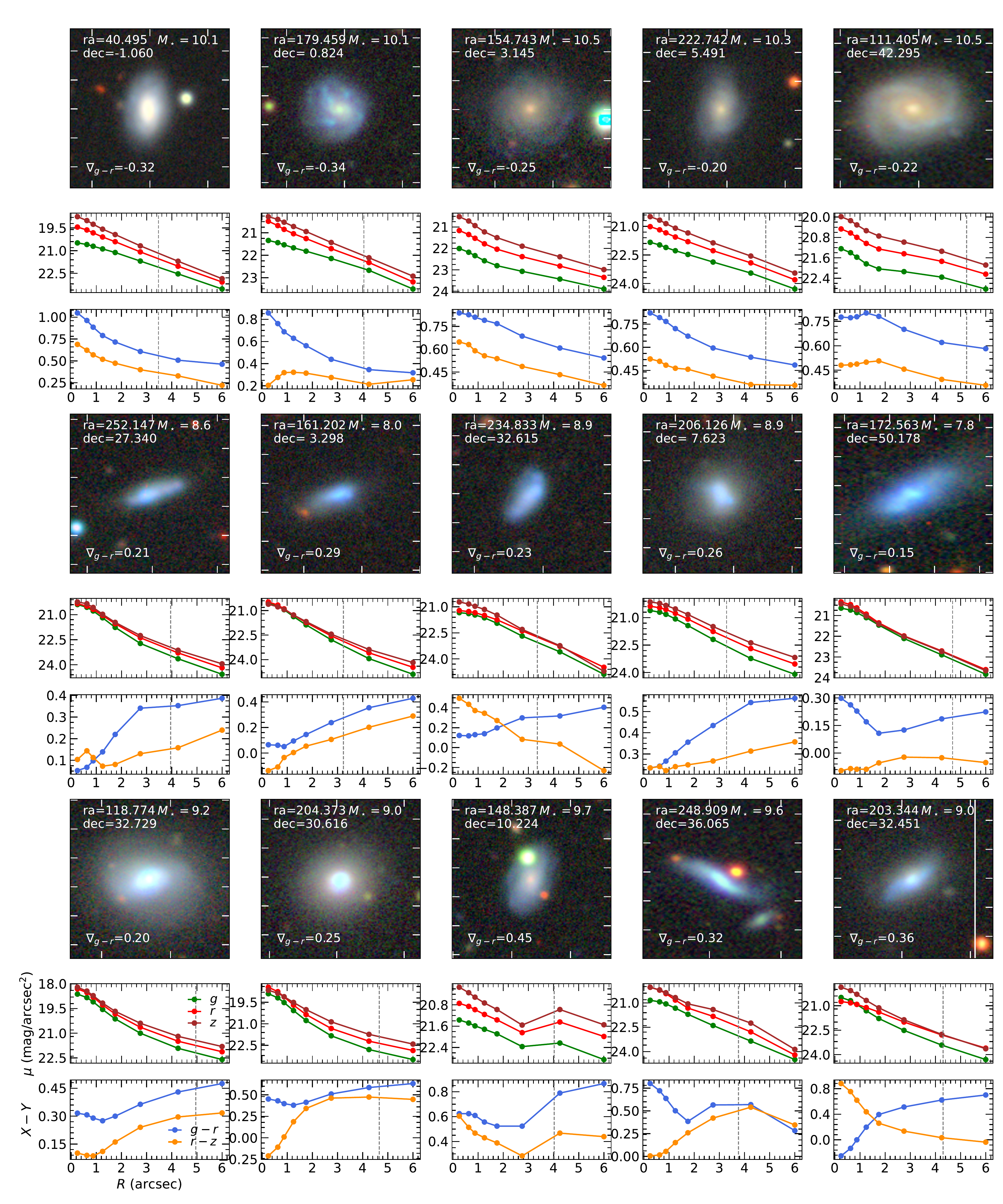}
    \caption{The top row shows examples of galaxies (with their corresponding surface brightness and colour profiles) that fail a colour gradient cut $\nabla_{g-r} < (g-r)-0.8$.
    The middle and bottom rows show galaxies that pass a cut at $\nabla_{g-r} > (g-r)-0.2$. The middle row shows examples of such galaxies with $M_\star < 10^9 \, \mathrm{M_\odot}$ and the bottom row examples with $M_\star > 10^9 \, \mathrm{M_\odot}$.
    In each panel, a white dashed circle shows $R_{\mathrm{eff}}$. Each panel is $36\arcsec\times36\arcsec$.
    For the profiles, the units of distance are arcseconds and the units of surface brightness are mag arcsec$^{-2}$. The dashed lines show $R_\mathrm{eff}$.}
    \label{fig:exdw_example}
\end{figure*}

To provide a broader context for Local Group observations, several recent large-scale surveys of dwarf galaxies have focused on the satellites of nearby $L^{\star}$ galaxies \citep[e.g.][]{Tanaka2018}, of which SAGA is currently the most extensive \citep{Geha2017TheAnalogs, Mao2021SAGA}. SAGA defines a catalogue of candidate satellites near Milky Way analogues using photometric morphology and colour-magnitude criteria, which are then confirmed with multi-object spectroscopy. The initial photometric selection aims for a high ratio of true dwarfs to candidate targets to maximise the efficiency of the spectroscopic follow-up. This is aided by the restriction to regions around brighter galaxies.

In principle, positive colour gradients in dwarfs provide an additional means of distinguishing them from more distant galaxies with similar blue colours, which could be particularly useful for photometric selection in the field. As a simple illustration of this idea, Fig.~\ref{fig:dw_distribution} shows the distribution of galaxies in our \textit{sz} sample in the space of LS $g-r$ colour and colour gradient, effectively an alternative projection of Fig.~\ref{fig:binned_map}. This figure follows \citet[][]{Park2005MorphologySpace}, who showed that morphological early and late types can be separated effectively in the space of SDSS $u-r$ colour and $\nabla(g-i)$ gradient. Clearly, a cut in average colour at $g-r\approx 0.7$ is an effective means of separating massive galaxies (predominantly early types, per \citealt{Park2005MorphologySpace}) from intrinsically fainter galaxies. However, blue cloud galaxies with $M_\star\sim10^{10} \, \mathrm{M_\odot}$ have similar average colours to dwarfs. The shallow slope of the size-mass relation means the apparent size distributions of bright and faint galaxies also have significant overlap at these modest distances.

Within the subset of blue field dwarf candidates, we show two further cuts on colour gradient as a function of average colour. The more generous, $\nabla_{g-r} = (g-r)-0.8$, retains most $M_\star <10^9 \, \mathrm{M_\odot}$ galaxies (orange points) while excluding a relatively small but significant number of more massive galaxies with similar average colours. The top row of Fig.~\ref{fig:exdw_example} shows five randomly chosen galaxies with $g-r < 0.7$ colour for which we measure colour gradients more negative than this cut.
These clearly have red centres and high stellar masses, as expected. As evident from Fig.~\ref{fig:dw_distribution}, a sample of $M_\star < 10^9 \, \mathrm{M_\odot}$ galaxies selected in this way would have high completeness, but only a modest improvement in purity compared to a simple colour selection.

Stricter colour gradient cuts lower both the mass to which the selection is complete and the typical mass of the `contaminants'. For example, we show a second cut at $\nabla_{g-r} = (g-r)-0.2$.
The middle and the bottom rows in Fig.~\ref{fig:exdw_example} show ten galaxies which meet the average colour cut and this stricter gradient criterion. The five galaxies in the middle row of Fig.~\ref{fig:exdw_example} have $M_\star < 10^9 \, \mathrm{M_\odot}$ and the five galaxies in bottom row have $M_\star > 10^9 \, \mathrm{M_\odot}$. They all have similar morphology to the SAGA satellites \citep[][]{Geha2017TheAnalogs}. The similarity of the galaxies in the middle and bottom rows reflects both the dispersion in colour gradient and the uncertainty in the measured stellar masses. More generally, it may reflect the limited significance of any hard division by stellar mass in this regime, in which empirical models and simulations predict that stellar mass has only a weak correlation with dark matter halo mass. 

Further investigation is beyond the scope of this paper. Although the limitations of a simple gradient-cut selection alone are clear from Fig.~\ref{fig:dw_distribution}, galaxy colour profiles may nevertheless be useful to improve the purity of photometrically selected samples of field dwarfs in combination with other selection methods.

\section{Conclusions}
\label{sec:conclusion}

In this paper, our goal was to study the distribution of galaxy colour gradients and their correlation with other photometric observables at $z<0.1$. We measured colour gradients between $R_{\mathrm{eff}}/2$ and $R_{\mathrm{eff}}$ for 93,680 galaxies with spectroscopic redshifts and 574,493 galaxies with photometric redshifts, using images and photometric catalogs from the DESI Legacy Imaging Survey DR9. We obtained $M_\star$ and SFR for our spectroscopic redshift sample using the GSWLC-X2 catalogue \citep{Salim2016GALEX-SDSS-WISEGalaxies,Salim2018DustAnalogs}. Our main results are as follows:
\begin{itemize}
    \item The colour gradients of most galaxies in our sample are negative (redder at the centre and bluer in the outskirts), consistent with previous work (Figs.~\ref{fig:fig2cghist}, \ref{fig:cghist_pz}).
    
    \item Colour gradients vary across the galaxy population as a function of both magnitude and average galaxy colour (Fig.~\ref{fig:binned_map}). Gradients in $g-r$ and $r-z$ show broadly similar trends in all parameter combinations we have examined (although see Figs.~\ref{fig:cg_optcolor}, \ref{fig:cg_sm_sfr}, \ref{fig:cg_color_pz}).
    
    \item For blue cloud galaxies, colour gradients are positive at the faint limit of our sample ($M_r\sim-16$) and become flatter then increasingly negative with luminosity, up to $M_r\sim-21$. For the small number of blue cloud galaxies brighter than this, colour gradients become slightly flatter with increasing luminosity (Figs.~\ref{fig:cggr_color_rsbc_sz}, \ref{fig:cg_color_pz}).
    
    \item For red sequence galaxies, the trend of colour gradient with luminosity is almost identical in slope and amplitude to that for blue cloud galaxies up to $M_r\sim-21$. At higher luminosity, the gradients of red sequence galaxies flatten much more rapidly and the galaxy-to-galaxy scatter diminishes (Figs.~\ref{fig:cggr_color_rsbc_sz}, \ref{fig:cg_color_pz}). 
    
    \item Similar trends are seen as a function of stellar mass; gradients become increasingly negative up to $M_\star \sim 10^{10.5} \, \mathrm{M_\odot}$ (Fig.~\ref{fig:cg_sm_sfr}). This mass is consistent with the characteristic `transition' stellar mass seen (for example) in galaxy colours, sSFR, and morphology \citep{Kauffmann2002TheGalaxies,Croton2006TheGalaxies,Cattaneo2008DownsizingGalaxies}.
    
    \item High- and low-sSFR subsets of our sample show similar trends below the transition mass; at higher masses gradients in the low sSFR subset flatten more rapidly. Similar results are seen if our sample is divided into blue and red subsets (Figs.~\ref{fig:cggr_sm_rsbc_sz}, \ref{fig:cg_color_pz}). This suggests negative colour gradients are persevered by whatever process is responsible for transition to the red sequence at lower masses.
    
    \item Below the transition mass, at fixed $M_\star$, low-sSFR (red) galaxies have slightly flatter gradients than high-sSFR (blue) galaxies, particularly at higher masses. This separation is less apparent at fixed $M_r$. This result implies that the presence or absence of active star formation is not, on its own, the main effect responsible for colour gradients (Fig.~\ref{fig:cggr_color_rsbc_sz}, \ref{fig:cggr_sm_rsbc_sz}).
    
    \item Below the transition mass, the above results support the consensus in the literature that strong negative colour gradients in the blue cloud are established by `inside out' formation, and hence can be attributed to stellar age gradients (see also below). 
    In this picture, more massive galaxies have a greater difference in age and radius between their oldest and younger populations, leading to stronger colour gradients. 
    
    \item Below the transition mass, the very similar trends of colour gradient with luminosity for blue and red subsets of our sample suggest that galaxies are fading passively to the red sequence without a significant change in the radial distribution of their stellar populations.
    
    \item Above the transition mass, the rapid flattening of colour gradients with increasing mass for quiescent galaxies in comparison to star-forming galaxies suggests that the evolution of stellar population gradients in this mass range is driven by merging, although the passive fading of recently star-forming outer regions may also contribute. 
    
    \item Positive colour gradients are typical at the low mass and high sSFR limit of our sample (Fig.~\ref{fig:cg_sm_sfr}). The prevalence of positive gradients in field dwarf galaxies is consistent with the theoretical expectation that these systems have extended star formation histories, in which new star formation is centrally concentrated and older populations diffuse outwards over time. Our results imply that, whatever the mechanism responsible for generating positive gradients, it must operate in the bulk of field dwarfs (i.e.\ cannot be an environmental effect alone) and must transition smoothly to `inside-out' formation at higher stellar masses.

    \item The use of stellar population studies based on resolved spectroscopy from the MaNGA survey to interpret trends in colour gradients across the CMD is limited by an apparent lack of consensus between the two \firefly{} datasets with which we compare (Figs.~\ref{fig:firefly_gradients}, \ref{fig:firefly_dust}). Both datasets imply that dust is a significant factor giving rise to shallow colour gradients along the red sequence and for low-mass galaxies.
    
    \item The separation between the colour gradients of dwarf galaxies and massive late types suggests that gradients could be a useful addition to other photometric dwarf galaxy selection criteria (Fig.~\ref{fig:dw_distribution}).
    
\end{itemize}

Our study suggests several opportunities for further work. The angular resolution and photometric accuracy of LS and Hyper SuprimeCam (and in future, LSST) are high enough to study colour profiles, instead of simple linear gradients. Doing so would make better use of the full radial range of the data. Exploring trends in subsets of galaxies defined by properties beyond stellar mass and star formation rate (for example environment, gas content or nuclear activity) may help to understand the origin of the modest scatter in the trends we report. Further work to better understand the contribution of dust and contamination from nearby sources to that scatter, would, in turn, better quantify the nature of galaxies with extreme colour gradients. It is likely, for example, that there are subsets of galaxies for which dust dominates colour gradients in the region we consider, even if this is not the case for the galaxy population as a whole.

It is somewhat surprising that we do not find a clearer correspondence between the colour gradients we measure and the age, dust and metallicity gradients reported by different studies based on the MaNGA spectra. At face value, however, the two \firefly{} datasets we consider in Section~\ref{sec:manga} apparently agree that neither age, metallicity nor dust trends alone are likely to dominate the observed variation of colour gradients across the CMD. This, and other inferences from those models such as steep positive age gradients in massive early types, seem to be in tension with the `standard picture' invoked by earlier work on colour gradients (outlined in \ref{sec:origin_cg}) and with naive expectations based on galaxy formation models, in which age gradients dominate. It is striking that the MaNGA results imply a significant contribution to colour variations from dust reddening gradients, particularly towards the extremes of the red sequence. We believe a more extensive comparison with these results would be valuable. If IFU studies can connect readily observed trends in colour gradients to underlying trends in age, metallicity and dust, then the much larger samples available from photometric studies would provide strong and complementary statistical constraints on galaxy formation models.

On the theoretical side, the current generation of cosmological galaxy formation simulations may already be accurate and detailed enough to explain how the processes that regulate quiescent star formation (and redistribute stars in mergers) give rise to the precise slopes of the trends we find. At lower stellar masses, a detailed comparison to observed colour gradients may constrain the relationship between disk assembly, quenching and halo mass. The flattening of gradients at high masses may be due to a combination of merging and passive evolution following quenching; simulations would help to disentangle these effects. Separately, we plan to compare our results to an equivalent analysis of mock images from cosmological simulations to illustrate how colour gradients may help to constrain models in future (Liao et al. in prep).

To assist with comparisons to theoretical predictions, we provide in Appendix \ref{appendix:catalog} a table of our fits to colour gradient trends in the LS data. We also make available a catalogue containing all our measurements for individual galaxies in the LS DR9 (described in Appendix \ref{appendix:catalog}) at the following URL: \dataurl{}. We note that the DESI Bright Galaxy survey \citep{Hahn2022BGS} will soon obtain spectroscopic redshifts for a significant fraction of galaxies in our photometric redshift sample.

\section*{Acknowledgements}

The authors thank Lihwai Lin and Abdurro'uf for useful discussions and insightful advice. We are grateful to the anonymous referee for their constructive comments and suggestions, which helped to improve the paper, and in particular for encouraging us to pursue the comparison to MaNGA. We acknowledge support from a Taiwan Ministry of Education Yushan Fellowship awarded to APC and the Taiwan National Science and Technology Council grant 109-2112-M-007-011-MY3.
This work used high-performance computing facilities operated by the
Center for Informatics and Computation in Astronomy (CICA) at National
Tsing Hua University. This equipment was funded by the Taiwan Ministry of
Education, the Taiwan National Science and Technology Council,
and National Tsing Hua University.

The Legacy Surveys consist of three individual and complementary projects: the Dark Energy Camera Legacy Survey (DECaLS; Proposal ID \#2014B-0404; PIs: David Schlegel and Arjun Dey), the Beijing-Arizona Sky Survey (BASS; NOAO Prop.\ ID \#2015A-0801; PIs: Zhou Xu and Xiaohui Fan), and the Mayall z-band Legacy Survey (MzLS; Prop.\ ID \#2016A-0453; PI: Arjun Dey). DECaLS, BASS and MzLS together include data obtained, respectively, at the Blanco telescope, Cerro Tololo Inter-American Observatory, NSF’s NOIRLab; the Bok telescope, Steward Observatory, University of Arizona; and the Mayall telescope, Kitt Peak National Observatory, NOIRLab. The Legacy Surveys project is honored to be permitted to conduct astronomical research on Iolkam Du’ag (Kitt Peak), a mountain with particular significance to the Tohono O’odham Nation.

NOIRLab is operated by the Association of Universities for Research in Astronomy (AURA) under a cooperative agreement with the National Science Foundation.

This project used data obtained with the Dark Energy Camera (DECam), which was constructed by the Dark Energy Survey (DES) collaboration. Funding for the DES Projects has been provided by the U.S. Department of Energy, the U.S. National Science Foundation, the Ministry of Science and Education of Spain, the Science and Technology Facilities Council of the United Kingdom, the Higher Education Funding Council for England, the National Center for Supercomputing Applications at the University of Illinois at Urbana-Champaign, the Kavli Institute of Cosmological Physics at the University of Chicago, Center for Cosmology and Astro-Particle Physics at the Ohio State University, the Mitchell Institute for Fundamental Physics and Astronomy at Texas A\&M University, Financiadora de Estudos e Projetos, Fundacao Carlos Chagas Filho de Amparo, Financiadora de Estudos e Projetos, Fundacao Carlos Chagas Filho de Amparo a Pesquisa do Estado do Rio de Janeiro, Conselho Nacional de Desenvolvimento Cientifico e Tecnologico and the Ministerio da Ciencia, Tecnologia e Inovacao, the Deutsche Forschungsgemeinschaft and the Collaborating Institutions in the Dark Energy Survey. The Collaborating Institutions are Argonne National Laboratory, the University of California at Santa Cruz, the University of Cambridge, Centro de Investigaciones Energeticas, Medioambientales y Tecnologicas-Madrid, the University of Chicago, University College London, the DES-Brazil Consortium, the University of Edinburgh, the Eidgenossische Technische Hochschule (ETH) Zurich, Fermi National Accelerator Laboratory, the University of Illinois at Urbana-Champaign, the Institut de Ciencies de l’Espai (IEEC/CSIC), the Institut de Fisica d’Altes Energies, Lawrence Berkeley National Laboratory, the Ludwig Maximilians Universitat Munchen and the associated Excellence Cluster Universe, the University of Michigan, NSF’s NOIRLab, the University of Nottingham, the Ohio State University, the University of Pennsylvania, the University of Portsmouth, SLAC National Accelerator Laboratory, Stanford University, the University of Sussex, and Texas A\&M University.

BASS is a key project of the Telescope Access Program (TAP), which has been funded by the National Astronomical Observatories of China, the Chinese Academy of Sciences (the Strategic Priority Research Program “The Emergence of Cosmological Structures” Grant \#XDB09000000), and the Special Fund for Astronomy from the Ministry of Finance. The BASS is also supported by the External Cooperation Program of Chinese Academy of Sciences (Grant \# 114A11KYSB20160057), and Chinese National Natural Science Foundation (Grant \#11433005).

The Legacy Survey team makes use of data products from the Near-Earth Object Wide-field Infrared Survey Explorer (NEOWISE), which is a project of the Jet Propulsion Laboratory/California Institute of Technology. NEOWISE is funded by the National Aeronautics and Space Administration.

The Legacy Surveys imaging of the DESI footprint is supported by the Director, Office of Science, Office of High Energy Physics of the U.S. Department of Energy under Contract No.\ DE-AC02-05CH1123, by the National Energy Research Scientific Computing Center, a DOE Office of Science User Facility under the same contract; and by the U.S.\ National Science Foundation, Division of Astronomical Sciences under Contract No.\ AST-0950945 to NOAO. 

The Photometric Redshifts for the Legacy Surveys (PRLS) catalogue used in this paper was produced thanks to funding from the U.S. Department of Energy Office of Science, Office of High Energy Physics via grant DE-SC0007914.

Funding for the Sloan Digital Sky Survey IV has been provided by the Alfred P. Sloan Foundation, the U.S. Department of Energy Office of Science, and the Participating Institutions. SDSS acknowledges support and resources from the Center for High-Performance Computing at the University of Utah. The SDSS web site is www.sdss.org.

SDSS is managed by the Astrophysical Research Consortium for the Participating Institutions of the SDSS Collaboration including the Brazilian Participation Group, the Carnegie Institution for Science, Carnegie Mellon University, Center for Astrophysics | Harvard \& Smithsonian (CfA), the Chilean Participation Group, the French Participation Group, Instituto de Astrofísica de Canarias, The Johns Hopkins University, Kavli Institute for the Physics and Mathematics of the Universe (IPMU) / University of Tokyo, the Korean Participation Group, Lawrence Berkeley National Laboratory, Leibniz Institut für Astrophysik Potsdam (AIP), Max-Planck-Institut für Astronomie (MPIA Heidelberg), Max-Planck-Institut für Astrophysik (MPA Garching), Max-Planck-Institut für Extraterrestrische Physik (MPE), National Astronomical Observatories of China, New Mexico State University, New York University, University of Notre Dame, Observatório Nacional / MCTI, The Ohio State University, Pennsylvania State University, Shanghai Astronomical Observatory, United Kingdom Participation Group, Universidad Nacional Autónoma de México, University of Arizona, University of Colorado Boulder, University of Oxford, University of Portsmouth, University of Utah, University of Virginia, University of Washington, University of Wisconsin, Vanderbilt University, and Yale University.

This work made use of the following Python software: Numpy \citep[][]{Harris2020ArrayNumPy}, matplotlib \citep[][]{Hunter2007Matplotlib:Environment}, Astropy \citep[][]{Collaboration2013Astropy:Astronomy,Collaboration2018ThePackage}.

\section*{Data Availability} 
The data underlying this article were accessed from the DESI Legacy Imaging Survey (\url{https://www.legacysurvey.org/dr9/files/}). The FIREFLY datacubes are provided by SDSS-MaNGA (\url{https://data.sdss.org/sas/dr17/manga/spectro/firefly/v3_1_1/}). The derived data generated in this research can be found online at \dataurl{}.




\bibliographystyle{mnras}
\bibliography{references}




\appendix

\section{Extreme colour gradients}
\label{appendix:extcg}

\begin{figure}
    \centering
    \includegraphics[width=\columnwidth, trim=0 70 0 0]{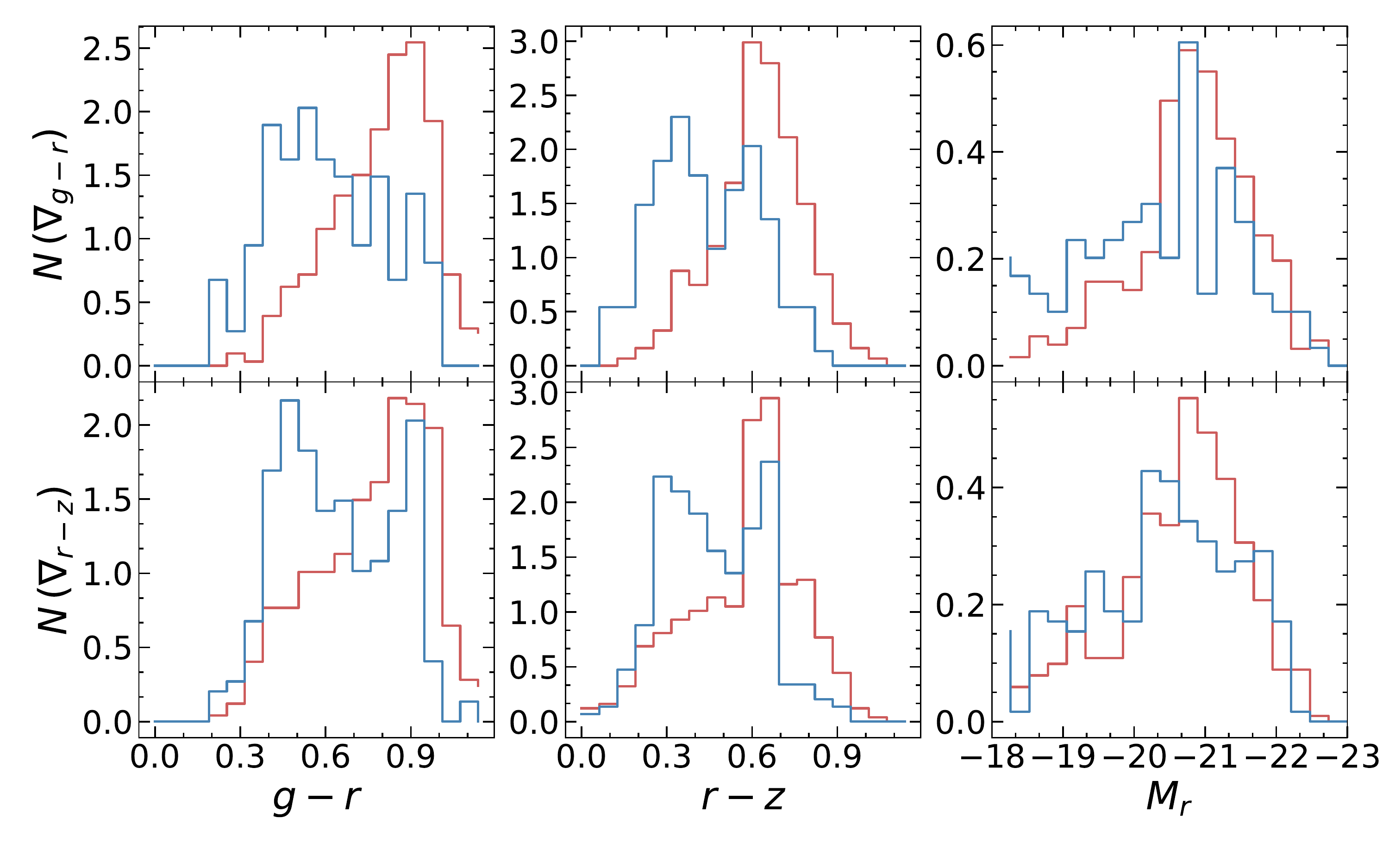}
    \caption{Histograms of colour and magnitude for subsets of our sample that have extreme positive (blue) and negative (red) colour gradients. The upper row corresponds to $\nabla_{g-r}$ outliers and the lower row to $\nabla_{r-z}$ outliers.}
    \label{fig:ext_histogram}
\end{figure}

\begin{figure*}
    \centering
    \includegraphics[width=0.95\textwidth, trim=0 10 0 0]{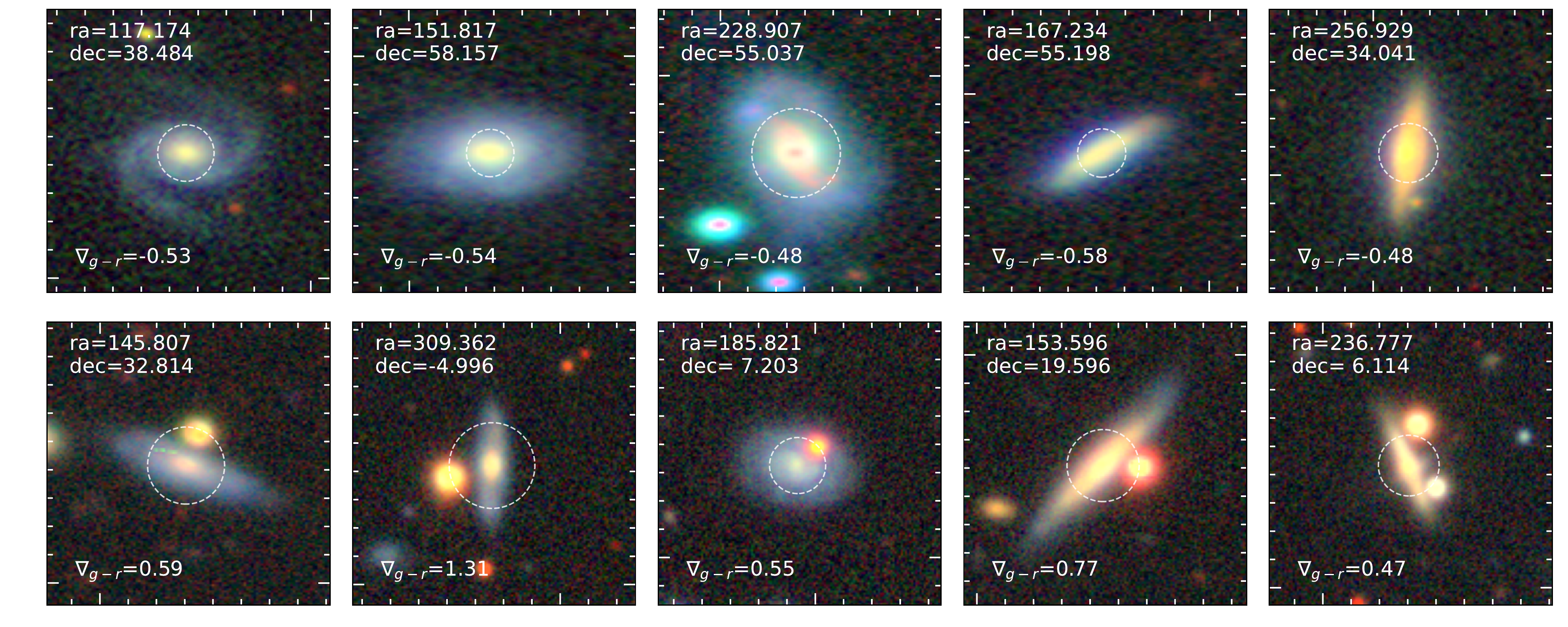}
    \caption{Randomly selected examples of galaxies with extreme negative colour gradients (upper panels) and galaxies with extreme positive colour gradients (lower panels).
    White dashed circles indicate $R_{\mathrm{eff}}$.
    Each panel is $36\arcsec\times36\arcsec$.}
    \label{fig:ext_map}
\end{figure*}

\begin{figure}
    \centering
    \includegraphics[width=0.9\columnwidth, trim=0 25 0 0]{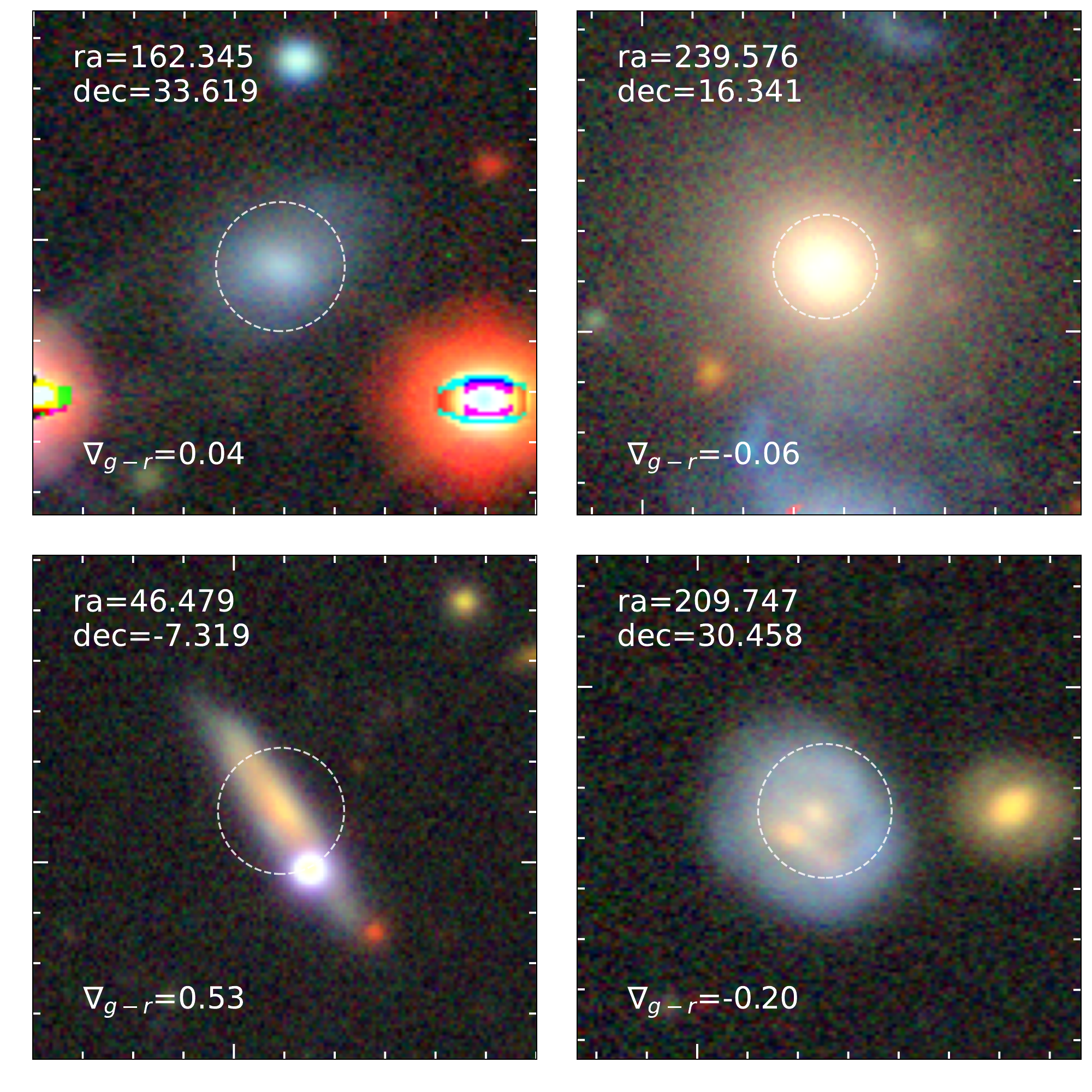}
    \caption{Example galaxies with high \texttt{FRACFLUX}. White dashed circles indicate $R_\mathrm{eff}$. Each panel is $36\arcsec\times36\arcsec$.}
    \label{fig:highfrac}
\end{figure}

\begin{figure}
    \centering
    \includegraphics[width=0.7\columnwidth, trim=0 20 0 0]{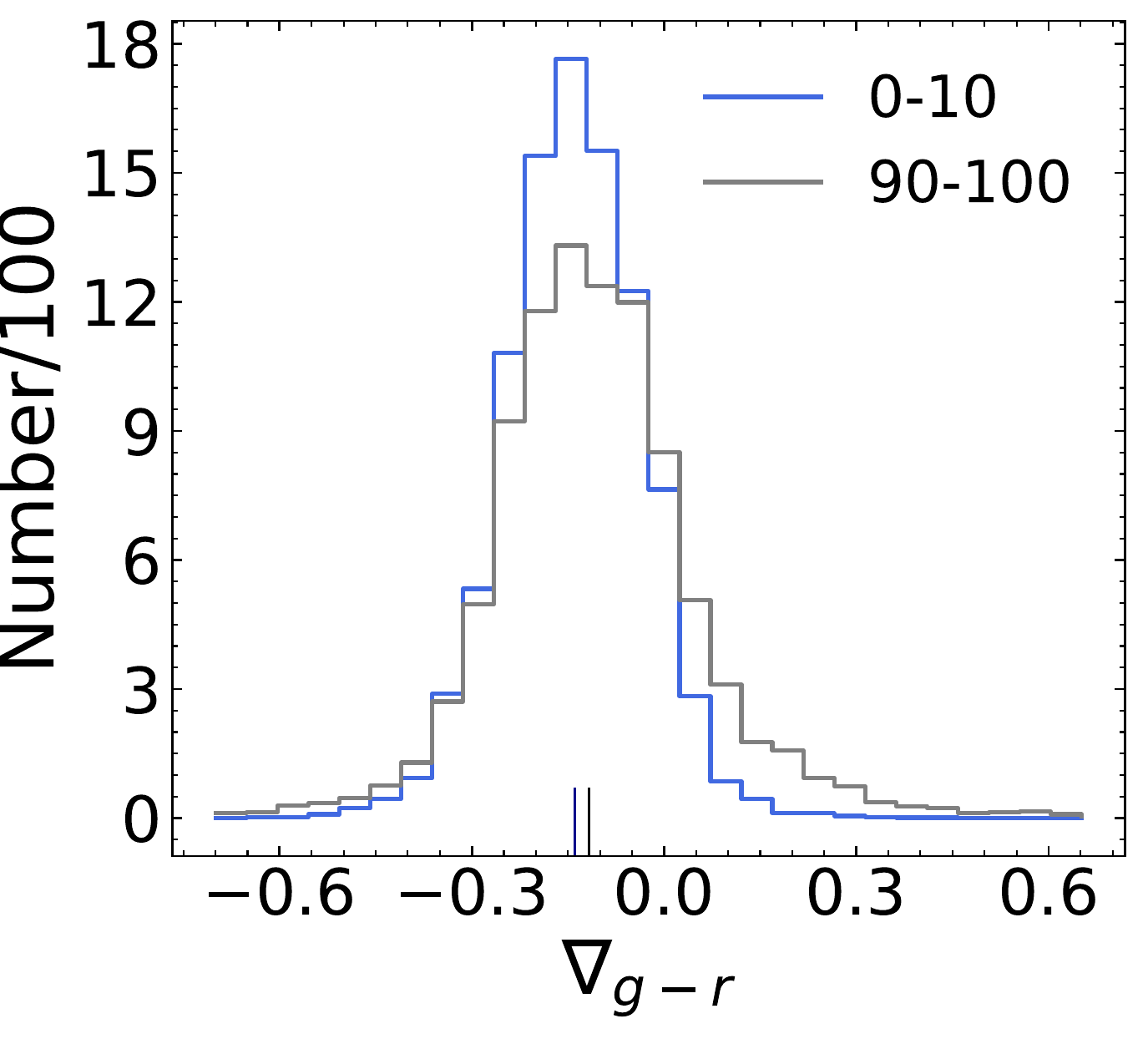}
    \caption{Number distribution of galaxies with lower $10^\mathrm{th}$ \texttt{FRACFLUX\_r} (blue)  and upper $90^\mathrm{th}$ \texttt{FRACFLUX\_r} (grey). The short vertical bars label the median of the distribution, $-0.06$ for lower $10^\mathrm{th}$ galaxies and $-0.03$ for upper $90^\mathrm{th}$ galaxies. The standard deviation are 0.13 and 0.23 for lower and upper percentile galaxies.}
    \label{fig:hist_fracfux}
\end{figure}

Colour gradient distributions show significant dispersion around the mean \citep{Gonzalez:2011}. This reflects a mix of physical variation and systematics in the photometry. We briefly explore the properties of galaxies with gradients further than $2\sigma$ from the mean in our \textit{sz} sample. Fig.~\ref{fig:ext_histogram} shows histograms of $g-r$, $r-z$, and $M_r$ for these galaxies.
Galaxies with extreme negative colour gradient are systematically brighter and redder (in both $g-r$ and $r-z$) than galaxies with extreme positive gradients.
In Fig.~\ref{fig:ext_map} we show images of randomly chosen examples. We see that galaxies with extreme negative colour gradients often have a prominent red centre, are dusty, or have potential contamination from blue stars.
Galaxies with extreme positive colour gradients are mainly late types with bluer centres and contamination from red stars. Overlapping (predominantly stellar) sources clearly contribute to extreme gradients, but do not appear to be their only cause.

We use the \texttt{FRACFLUX\_r} measurements in the LS catalogue to explore how these contaminated galaxies may affect the distribution of colour gradients in our sample. This parameter quantifies the fractional contribution of contaminating flux to the pixels associated with a particular source model, weighted by the source profile; a high value of \texttt{FRACFLUX\_r} for a source indicates that other sources contribute a significant fraction of the $r$-band flux in its pixels. As noted in Section~\ref{sec:LS}, our selection criteria already exclude galaxies with $\texttt{FRACFLUX\_x} > 5$ in any of the LS bands. Galaxies that remain in our sample after these cuts may nevertheless have nearby sources that contribute significant flux in the region we study, even if their total flux much less than that of the galaxy overall.

Fig.~\ref{fig:highfrac} shows examples of galaxies with high \texttt{FRACFLUX\_r} that remain in our sample. These galaxies do have one or more nearby sources. Visual inspection of a larger number of images confirms that that high \texttt{FRACFLUX\_r} is a reasonable predictor of visual contamination, and conversely, low \texttt{FRACFLUX\_r} selects isolated objects with clean photometry.

In Fig.~\ref{fig:hist_fracfux}, we plot the colour gradient distributions of galaxies in the $10^\mathrm{th}$ and $90^\mathrm{th}$ percentile tails of \texttt{FRACFLUX\_r} for our sample. These percentiles correspond to \texttt{FRACFLUX\_r} values of $5.4\times10^{-5}$ and $4.8\times10^{-2}$ respectively. The medians of the low- and high-\texttt{FRACFLUX\_r} subsets are $-0.06$ and $-0.03$ and their standard deviations are $0.13$ and $0.23$, respectively. The two distributions are similar, except for a small excess of galaxies with extreme positive gradients in the high-\texttt{FRACFLUX\_r} subset. This is consistent with the visual impression given by Fig.~\ref{fig:ext_map}. If we examine instead the lower $25^\mathrm{th}$ and upper $75^\mathrm{th}$ percentile subsets of \texttt{FRACFLUX\_r}, we find their distributions are almost identical, with an even less significant excess of extreme positive gradients.

We conclude that the mean colour gradient in our sample is not significantly affected by contaminated galaxies. The fact that the galaxies with the lowest \texttt{FRACFLUX\_r} in our sample do not have a significantly narrower colour gradient distribution suggests that contamination from nearby sources does not dominate the scatter either. In the case of galaxies with extreme negative gradients, our measurements appear to reflect faithfully the intrinsic variation of colour in the images. Contamination may contribute more to the tail of positive gradients (perhaps because red stars and blue galaxies in our sample have higher density on the sky than blue stars and red galaxies). More rigorous removal of contaminated sources may therefore reduce the scatter of colour gradients in our sample, but is unlikely to have a significant effect on the average trends we report.

\section{Effects of total dust reddening on the colour magnitude diagram}
\label{appendix:dust}

\begin{figure}
    \centering
    \includegraphics[width=\columnwidth, trim=0 40 0 0]{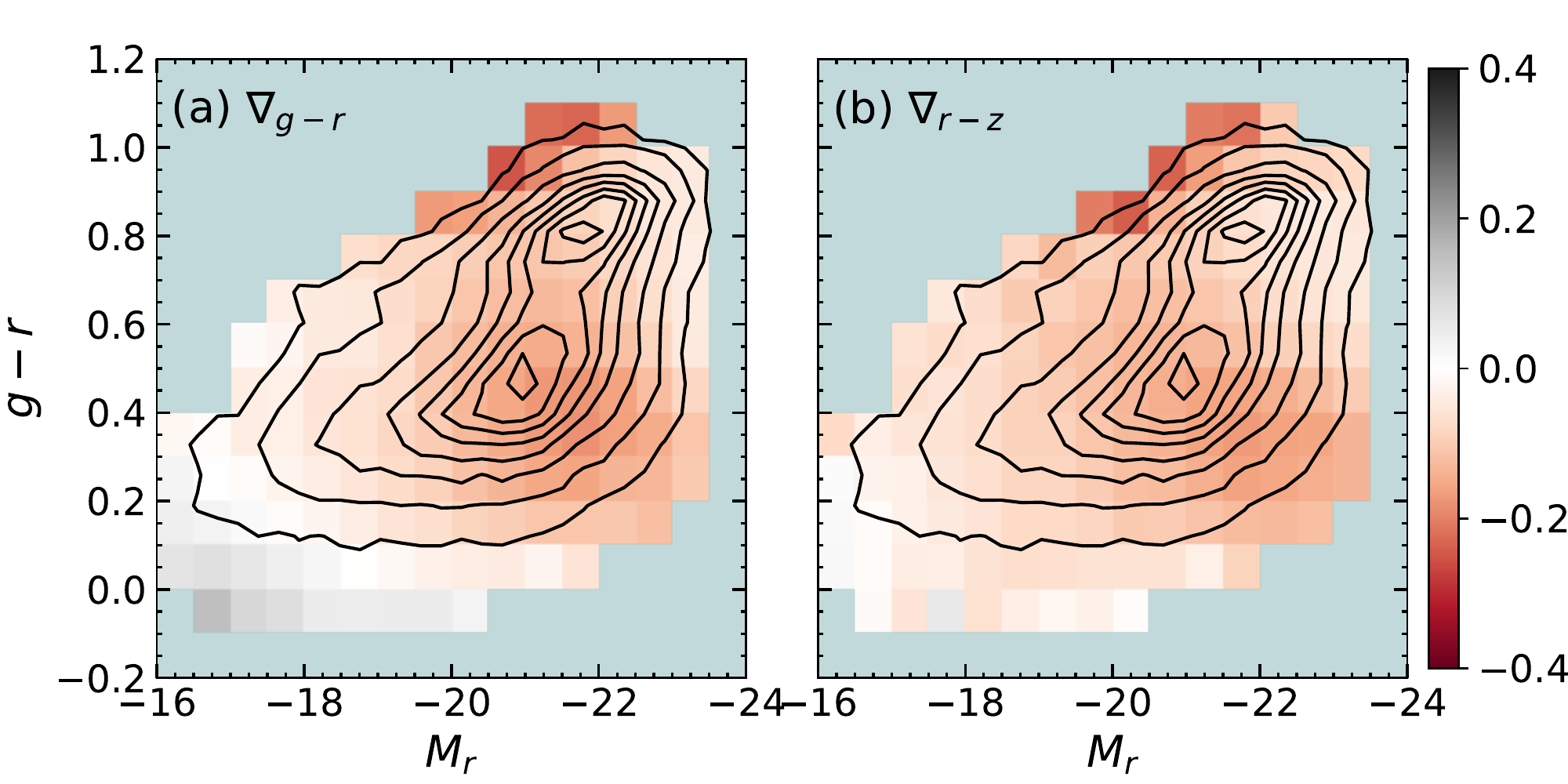}
    \caption{Equivalent to     Fig.~\ref{fig:binned_map}, showing the distribution of mean colour gradients $\nabla_{g-r}$ (left) and $\nabla_{r-z}$ (right) in bins of $M_r$ and $g-r$. In this version of the figure, we correct for the effects of dust on the average $g-r$ colour (see text). The contours show the density of galaxies in each distribution.}
    \label{fig:cgmap_Mrgr_dust}
\end{figure}

In the main text, we do not correct the average colours of galaxies for reddening by dust. Here we apply a correction using $A_B$ and $A_V$ from the GSWLC catalogue \citep{Salim2018DustAnalogs}.
These values are derived from CIGALE \citep{Noll2009AnalysisSample} using a modified \citet{Calzetti2000TheGalaxies} extinction law  \citep{Stecher1965InterstellarUltraviolet.,Fitzpatrick1986AnBump,Conroy2010TheEvaluation}:  $k(\lambda)_\mathrm{mod} = k(\lambda)_\mathrm{Cal}(\lambda/5500\angstrom)^\delta$ where $k(\lambda)_\mathrm{Cal}$ is the original \citet{Calzetti2000TheGalaxies} law and the exponent $\delta=-0.5$ or $-1.0$ accounts for the UV bump \citep{Noll2009AnalysisSample, Salim2016GALEX-SDSS-WISEGalaxies}.
We use the DES $grz$ bands as proxies for the LS bands and calculate $A_x = R_x E(B-V)$, where $E(B-V) = A_B-A_V$ and $R_x=3.237, 2.176$ and $1.217$ for $g$, $r$, and $z$ bands, respectively, with $R_V = 3.1$ \citep[from the appendix in][]{Schlafly2010MeasuringSFD}. We compute the dust-corrected colour $(X-Y)_{c}=(X-Y)-E(X-Y)$ and absolute magnitude $M_{x,c} = M_x-A_x$.

Fig.~\ref{fig:cgmap_Mrgr_dust} shows the equivalent of Fig.~\ref{fig:binned_map} after de-reddening. Comparing Fig.~\ref{fig:cgmap_Mrgr_dust} to Fig.~\ref{fig:binned_map}, we see a much lower density of galaxies around $M_r\sim-21$ and $g-r\sim1.2$, where mean colour gradients are very steep.
The trends across the blue cloud are preserved; gradients become steeper as galaxies become brighter. The variation of gradients across red sequence is reduced. These results match our expectation that the steepest colour gradients correspond to extremely dusty and luminous late-type galaxies, which appear above the red sequence in an uncorrected CMD. Their steep negative gradients stand out from the flat gradients of morphologically early-type red sequence galaxies with similar luminosity. Our reddening correction redistributes most (but not all) of this population to the high-luminosity edge of the blue cloud, where steep negative colour gradients are typical.

\section{catalogue and fitting results}
\label{appendix:catalog}

Our full set of colour gradient measurements for both \sz\/ and \pz\/ samples are available online at \dataurl{}. 
The tables contain all galaxies selected from DR9 of the Legacy Survey according to the criteria described in Section~\ref{sec:szsample} and Section~\ref{sec:pzsample} for the \sz\/ and \pz\/ samples, respectively.
Table~\ref{tab:sz_catalog} describes the contents of the \sz\/ table.
The \pz\/ has almost the same structure, but does not include spectroscopic redshifts, stellar masses or sSFRs.

Table~\ref{tab:fitresults} presents the coefficients of our linear fits to trends of colour gradient with average galaxy properties.

\begin{table*}
    \caption{\label{tab:sz_catalog}Contents of our \sz\/ colour gradient catalogue. We tabulate all 93,680 galaxies selected from DR9 of the Legacy Survey according to the criteria in Section~\ref{sec:szsample}. Columns 1 to 6 are taken directly from the Legacy Survey DR9 catalogue. Columns 28 to 31 are taken from the GSWLC catalogue of \citet{Salim2016GALEX-SDSS-WISEGalaxies}. Column 9 and 10 are taken from \citet{Zhou2020TheRedshifts}.}
    \centering
    \begin{tabular}{l l l l}
    \hline\hline
    No. & Column Name & Units & Description \\
    \hline
    1 & brickname & ... & Name of brick in the Legacy Survey, encoding the brick sky position\\
    2 & objid & ... & Catalogue object number in the corresponding brick of the Legacy Survey\\
    3 & ra & deg & Right ascension at equinox J2000 \\
    4 & dec & deg & Declination at equinox J2000 \\
    5 & ra\_ivar & 1/deg$^2$ & Inverse variance of RA, excluding astrometric calibration errors\\
    6 & dec\_ivar & 1/deg$^2$ & Inverse variance of DEC, excluding astrometric calibration errors\\
    7 & z & ... & Spectroscopic redshift from SDSS DR14 catalogue \\
    8 & z\_err & ... & Error of the redshift \\
    9 & z\_phot\_median & ... & Photometric redshift from \citet{Zhou2020TheRedshifts}\\
    10 & z\_phot\_std & ... & Error of photometric redshift \\
    11 & r\_size & arcsec & $R_\mathrm{eff}$\\
    12 & r\_err & arcsec & Error of $R_\mathrm{eff}$ \\
    13 & M\_r & mag & $M_r$ \\
    14 & mag\_g & mag & $g$ band magnitude \\
    15 & mag\_g\_err & mag & Error of $g$ \\
    16 & mag\_r & mag & $r$ band magnitude \\
    17 & mag\_r\_err & mag & Error of $r$ \\
    18 & mag\_z & mag & $z$ band magnitude \\
    19 & mag\_z\_err & mag & Error of $z$ \\
    20 & g\_r & mag & $g-r$ \\
    21 & g\_r\_err & mag & Error of $g-r$ \\
    22 & r\_z & mag & $r-z$\\
    23 & r\_z\_err & mag & Error of $r-z$ \\
    24 & cgr & $\cgunit$ &  $g-r$ gradient of galaxies measured at 0.5 and $1R_\mathrm{eff}$, $\nabla_{g-r} = 100$ are galaxies with no corresponding colour gradient \\
    25 & cgr\_err & $\cgunit$ & Error of $\nabla_{g-r}$ \\
    26 & crz & $\cgunit$ & $r-z$ gradient of galaxies measured at 0.5 and $1R_\mathrm{eff}$, $\nabla_{r-z} = 100$ are galaxies with no corresponding colour gradient\\
    27 & crz\_err & $\cgunit$ & Error of $\nabla_{r-z}$\\
    28 & stellarmass & $\mathrm{M_\odot}$ & $\log_{10} \, M_*$ from the GSWLC catalogue \\
    29 & stellarmass\_err & $\mathrm{M_\odot}$ & Error of the stellar mass \\
    30 & SFR & $\mathrm{M_\odot} \, \mathrm{yr}^{-1}$ & $\log_{10} \, \mathrm{SFR_{SED}}$ from the GSWLC catalogue\\
    31 & SFR\_err & $\mathrm{M_\odot} \, \mathrm{yr}^{-1}$ & Error of SFR \\
    \hline
    \end{tabular}
\end{table*}

\begin{landscape}
\begin{table}
 \caption{\label{tab:fitresults} Fits to colour gradient trends with physical parameters as described in the main text, either a straight line (see Eq.\ref{eq:linefit}) or a broken line (inflection points are indicated by subscript $i$). The values in parentheses are 1$\sigma$ errors.}
 \begin{tabular}{p{1.7cm}
 C{0.75cm}C{0.75cm}C{0.75cm}C{0.75cm}
 C{0.75cm}C{0.75cm}C{0.75cm}C{0.75cm}C{0.75cm}
 C{0.75cm}C{0.75cm}C{0.75cm}C{0.75cm}C{0.75cm}
 C{0.75cm}C{0.75cm}C{0.75cm}C{0.75cm}C{0.75cm}}
  \hline \hline
  $\nabla_{g-r}$ & \multicolumn{2}{c}{$g-r$} & \multicolumn{2}{c}{$r-z$} &
  $M_{r,i}$ &
  \multicolumn{2}{c}{$M_r > -21$} & \multicolumn{2}{c}{$M_r < -21$}  &
  $M_{*,i}$ &
  \multicolumn{2}{c}{$M_\star < M_{*,i}$} &
  \multicolumn{2}{c}{$M_\star > M_{*,i}$} &
  $\mathrm{sSFR}_i$ &
  \multicolumn{2}{c}{$\mathrm{sSFR} < \mathrm{sSFR}_i$} &
  \multicolumn{2}{c}{$\mathrm{sSFR} > \mathrm{sSFR}_i$}\\
  & $a$ & $b$ & $a$ & $b$
  & & $a$ & $b$ & $a$ & $b$ 
  & & $a$ & $b$ & $a$ & $b$ 
  & & $a$ & $b$ & $a$ & $b$\\
  \hline 
  \sz\/ sample & 
  $-0.215$ & $0.080$ & 
  $-0.394$ & $0.135$ & 
  $-21.082$ & $0.041$ & $0.698$ & 
  $-0.080$ & $-1.859$ & 
  $10.479$ & $-0.085$ & $0.724$ & 
  $0.131$ & $-1.546$ & 
  $-10.081$ & $-0.030$ & $-0.402$ & 
  $0.165$ & $1.567$ 
  \\
  & $(0.017)$ & $(0.010)$ & 
  $(0.032)$ & $(0.025)$ & 
  $(0.051)$ & $(0.001)$ & $(0.016)$ & 
  $(0.008)$ & $(0.149)$ & 
  $(0.076)$ & $(0.004)$ & $(0.039)$ & 
  $(0.046)$ & $(0.474)$ & 
  $(0.064)$ & $(0.004)$ & $(0.038)$ & 
  $(0.013)$ & $(0.126)$ 
  \\
  \sz\/ (red) & 
  $-0.353$ & $0.212$ & 
  $-0.457$ & $0.192$ & 
  $-20.966$ & $0.017$ & $0.223$ & 
  $-0.054$ & $-1.260$ & 
  $10.505$ & $-0.073$ & $0.638$ & 
  $0.087$ & $-1.038$ & 
  $-10.412$ & $-0.036$ & $-0.479$ & 
  $0.195$ & $1.932$ 
  \\
  & $(0.066)$ & $(0.044)$ & 
  $(0.010)$ & $(0.008)$ & 
  $(0.223)$ & $(0.004)$ & $(0.068)$ & 
  $(0.016)$ & $(0.317)$ & 
  $(0.067)$ & $(0.002)$ & $(0.022)$ & 
  $(0.028)$ & $(0.290)$ & 
  $(0.063)$ & $(0.004)$ & $(0.041)$ & 
  $(0.023)$ & $(0.236)$ 
  \\
  \sz\/ (blue) & 
  $-0.463$ & $0.143$ & 
  $-0.467$ & $0.075$ & 
  $-21.631$ & $0.046$ & $0.779$ & 
  $-0.050$ & $-1.297$ & 
  $10.612$ & $-0.101$ & $0.857$ & 
  $0.044$ & $-0.680$ & 
  $-9.977$ & $0.130$ & $1.194$ & 
  $0.235$ & $2.242$ 
  \\
  & $(0.015)$ & $(0.008)$ & 
  $(0.025)$ & $(0.012)$ & 
  $(0.151)$ &$(0.003)$ & $(0.059)$ & 
  $(0.060)$ & $(1.288)$ & 
  $(0.091)$ & $(0.006)$ & $(0.057)$ & 
  $(0.049)$ & $(0.510)$ & 
  $(0.223)$ & $(0.018)$ & $(0.184)$ & 
  $(0.038)$ & $(0.377)$ 
  \\  
  \sz\/ (low-sSFR) & 
  $-0.011$ & $0.072$ & 
  $-0.309$ & $0.064$ & 
  $-20.973$ & $0.044$ & $0.751$ & 
  $-0.072$ & $-1.666$ & 
  $10.578$ & $-0.079$ & $0.690$ & 
  $0.123$ & $-1.450$ & 
  -- & $-0.036$ & $-0.468$ & 
  -- & -- 
  \\
  & $(0.035)$ & $(0.024)$ & 
  $(0.017)$ & $(0.008)$ & 
  $(0.056)$ & $(0.004)$ & $(0.079)$ & 
  $(0.008)$ & $(0.154)$ & 
  $(0.108)$ & $(0.009)$ & $(0.090)$ & 
  $(0.145)$ & $(1.532)$ & 
  -- & $(0.003)$ & $(0.032)$ & 
  -- & --
  \\
  \sz\/ (high-sSFR) & 
  $-0.287$ & $0.093$ & 
  $-0.368$ & $0.069$ & 
  $-21.505$ & $0.044$ & $0.744$ & 
  $-0.042$ & $-1.106$ & 
  -- & $-0.087$ & $0.728$ & 
  -- & -- & 
  -- & -- & -- & 
  $0.174$ & $1.665$ 
  \\
  & $(0.024)$ & $(0.019)$ & 
  $(0.014)$ & $(0.006)$ &
  $(0.146)$ & $(0.002)$ & $(0.040)$ & 
  $(0.027)$ & $(0.563)$ & 
  -- & $(0.004)$ & $(0.035)$ & 
  -- & -- & 
  -- & -- & -- & 
  $(0.024)$ & $(0.235)$ 
  \\
  \pz\/ sample & 
  $-0.276$ & $0.096$ & 
  $-0.157$ & $0.047$ & 
  $-20.672$ & $0.042$ & $0.687$ & 
  $-0.067$ & $-1.561$ & 
  -- & -- & -- & 
  -- & -- & 
  -- & -- & -- & 
  -- & -- 
  \\
  & $(0.050)$ & $(0.027)$ & 
  $(0.026)$ & $(0.013)$ & 
  $(0.108)$ & $(0.003)$ & $(0.059)$ & 
  $(0.010)$ & $(0.194)$ & 
  -- & -- & -- & 
  -- & -- & 
  -- & -- & -- & 
  -- & -- 
  \\ 
  \pz\/ (red) & 
  $-0.233$ & $0.117$ & 
  $-0.340$ & $0.109$ & 
  $-20.280$ & $0.053$ & $0.920$ & 
  $-0.047$ & $-1.111$ & 
  -- & -- & -- & 
  -- & -- & 
  -- & -- & -- & 
  -- & -- 
  \\
  & $(0.088)$ & $(0.064)$ & 
  $(0.004)$ & $(0.003)$ & 
  $(0.023)$ & $(0.003)$ & $(0.047)$ & 
  $(0.003)$ & $(0.057)$ & 
  -- & -- & -- & 
  -- & -- & 
  -- & -- & -- & 
  -- & -- 
  \\
  \pz\/ (blue) & 
  $-0.526$ & $0.181$ & 
  $-0.389$ & $0.018$ & 
  $-21.102$ & $0.040$ & $0.650$ & 
  $-0.015$ & $-0.516$ & 
  -- & -- & -- & 
  -- & -- & 
  -- & -- & -- & 
  -- & -- 
  \\
  & $(0.033)$ & $(0.016)$ & 
  $(0.034)$ & $(0.016)$ & 
  $(0.097)$ & $(0.002)$ & $(0.037)$ & 
  $(0.007)$ & $(0.129)$ & 
  -- & -- & -- & 
  -- & -- & 
  -- & -- & -- & 
  -- & -- 
  \\
  \hline
  \hline
  $\nabla_{r-z}$ & \multicolumn{2}{c}{$g-r$} & \multicolumn{2}{c}{$r-z$} &
  $M_{r,i}$ &
  \multicolumn{2}{c}{$M_r > -21$} & \multicolumn{2}{c}{$M_r < -21$}  &
  $M_{*,i}$ &
  \multicolumn{2}{c}{$M_\star < M_{*,i}$} &
  \multicolumn{2}{c}{$M_\star > M_{*,i}$} &
  $\mathrm{sSFR}_i$ & 
  \multicolumn{2}{c}{$\mathrm{sSFR} < \mathrm{sSFR}_i$} &
  \multicolumn{2}{c}{$\mathrm{sSFR} > \mathrm{sSFR}_i$}\\
  & $a$ & $b$ & $a$ & $b$ 
  & & $a$ & $b$ & $a$ & $b$ 
  & & $a$ & $b$ & $a$ & $b$ 
  & & $a$ & $b$ & $a$ & $b$\\
  \hline
  \sz\/ sample & 
  $-0.138$ & $0.009$ &
  $-0.202$ & $0.004$ & 
  $-20.924$ & $0.028$ & $0.439$ & 
  $-0.064$ & $-1.480$ & 
  $10.420$ & $-0.072$ & $0.596$ & 
  $0.131$ & $-1.526$ & 
  $-10.185$ & $-0.051$ & $-0.626$ & 
  $0.119$ & $1.096$ 
  \\
  & $(0.015)$ & $(0.010)$ & 
  $(0.040)$ & $(0.019)$ & 
  $(0.062)$ & $(0.002)$ & $(0.031)$ & 
  $(0.008)$ & $(0.156)$ & 
  $(0.054)$ & $(0.001)$ & $(0.011)$ & 
  $(0.028)$ & $(0.287)$ & 
  $(0.064)$ & $(0.004)$ & $(0.036)$ & 
  $(0.047)$ & $(0.473)$ 
  \\
  \sz\/ (red) & 
  $-0.188$ & $0.062$ & 
  $-0.339$ & $0.114$ & 
  $-20.330$ & $0.036$ & $0.605$ & 
  $-0.048$ & $-1.103$ & 
  $10.240$ & $-0.074$ & $0.632$ & 
  $0.073$ & $-0.865$ & 
  $-10.576$ & $-0.043$ & $-0.541$ & 
  $0.081$ & $0.769$ 
  \\
  & $(0.126)$ & $(0.101)$ & 
  $(0.031)$ & $(0.016)$ & 
  $(0.050)$ & $(0.005)$ & $(0.089)$ & 
  $(0.003)$ & $(0.061)$ & 
  $(0.036)$ & $(0.005)$ & $(0.043)$ & 
  $(0.011)$ & $(0.106)$ & 
  $(0.120)$ & $(0.005)$ & $(0.052)$ & 
  $(0.045)$ & $(0.475)$ 
  \\
  \sz\/ (blue) & 
  $-0.334$ & $0.073$ & 
  $-0.336$ & $0.024$ & 
  $-21.511$ & $0.032$ & $0.505$ & 
  $-0.018$ & $-0.561$ & 
  $10.567$ & $-0.069$ & $0.547$ & 
  $0.034$ & $-0.535$ & 
  $-10.180$ & $0.027$ & $0.153$ & 
  $0.140$ & $1.306$ 
  \\
  & $(0.025)$ & $(0.016)$ & 
  $(0.030)$ & $(0.015)$ & 
  $(0.108)$ & $(0.002)$ & $(0.031)$ & 
  $(0.024)$ & $(0.500)$ & 
  $(0.055)$ & $(0.003)$ & $(0.026)$ & 
  $(0.021)$ & $(0.219)$ & 
  $(0.035)$ & $(0.011)$ & $(0.105)$ & 
  $(0.008)$ & $(0.080)$ 
  \\
  \sz\/ (low-sSFR) & 
  $-0.037$ & $-0.061$ & 
  $-0.202$ & $0.012$ & 
  $-21.040$ & $0.012$ & $0.135$ & 
  $-0.062$ & $-1.433$ & 
  $10.357$ & $-0.065$ & $0.543$ & 
  $0.059$ & $-0.741$ & 
  -- & $-0.044$ & $-0.554$ & 
  -- & -- 
  \\
  & $(0.018)$ & $(0.014)$ & 
  $(0.052)$ & $(0.030)$ & 
  $(0.108)$ & $(0.001)$ & $(0.021)$ & 
  $(0.020)$ & $(0.418)$ & 
  $(0.027)$ & $(0.003)$ & $(0.029)$ & 
  $(0.007)$ & $(0.065)$ & 
  -- & $(0.003)$ & $(0.033)$ & 
  -- & -- 
  \\
  \sz\/ (high-sSFR) & 
  $-0.149$ & $0.009$ & 
  $-0.439$ & $0.120$ & 
  $-21.500$ & $0.031$ & $0.497$ & 
  $-0.001$ & $-0.201$ & 
  -- & $-0.071$ & $0.575$ & 
  -- & -- & 
  -- & -- & -- & 
  $0.109$ & $1.000$ 
  \\
  & $(0.034)$ & $(0.020)$ & 
  $(0.109)$ & $(0.051)$ & 
  $(0.215)$ & $(0.002)$ & $(0.032)$ & 
  $(0.014)$ & $(0.296)$ & 
  -- & $(0.004)$ & $(0.035)$ & 
  -- & -- & 
  -- & -- & -- & 
  $(0.002)$ & $(0.019)$ 
  \\
  \pz\/ sample & 
  $-0.149$ & $-0.011$ & 
  $-0.198$ & $-0.004$ & 
  $-20.638$ & $0.037$ & $0.605$ & 
  $-0.058$ & $-1.344$ & 
  -- & -- & -- & 
  -- & -- & 
  -- & -- & -- & 
  -- & -- 
  \\
  & $(0.025)$ & $(0.020)$ & 
  $(0.024)$ & $(0.008)$ & 
  $(0.077)$ & $(0.001)$ & $(0.012)$ & 
  $(0.011)$ & $(0.209)$ & 
  -- & -- & -- & 
  -- & -- & 
  -- & -- & -- & 
  -- & -- 
  \\
  \pz\/ (red) & 
  $-0.304$ & $0.144$ & 
  $-0.389$ & $0.145$ & 
  $-20.215$ & $0.046$ & $0.800$ & 
  $-0.051$ & $-1.176$ & 
  -- & -- & -- & 
  -- & -- & 
  -- & -- & -- & 
  -- & -- 
  \\
  & $(0.264)$ & $(0.200)$ & 
  $(0.030)$ & $(0.014)$ & 
  $(0.018)$ & $(0.004)$ & $(0.079)$ & 
  $(0.004)$ & $(0.066)$ & 
  -- & -- & -- & 
  -- & -- & 
  -- & -- & -- & 
  -- & -- 
  \\
  \pz\/ (blue) & 
  $-0.369$ & $0.094$ & 
  $-0.355$ & $0.033$ & 
  $-21.067$ & $0.037$ & $0.609$ & 
  $-0.007$ & $-0.306$ & 
  -- & -- & -- & 
  -- & -- & 
  -- & -- & -- & 
  -- & -- 
  \\
  & $(0.007)$ & $(0.004)$ & 
  $(0.018)$ & $(0.009)$ & 
  $(0.090)$ & $(0.001)$ & $(0.010)$ & 
  $(0.013)$ & $(0.257)$ & 
  -- & -- & -- & 
  -- & -- & 
  -- & -- & -- & 
  -- & -- 
  \\
  \hline
  \multicolumn{13}{l}{$^*$This fit is less secure because specific star formation rates have substantial uncertainty for $\log_{10}$\,sSFR < -11 \citep[see][]{Salim2016GALEX-SDSS-WISEGalaxies}.}
 \end{tabular}
\end{table}
\end{landscape}


\bsp	
\label{lastpage}
\end{document}